\documentclass[10pt,preprint]{aastex}
\usepackage{epstopdf,natbib}
\usepackage{lscape}
\usepackage[usenames,dvipsnames]{color}
\usepackage{color}
\usepackage{amssymb}
\usepackage[caption=false]{subfig}
\usepackage{natbib}
\usepackage{amsmath}

\slugcomment{To be submitted to AJ; draft \today}

\shorttitle{Resolving Spectral Binaries}
\shortauthors{Bardalez Gagliuffi et al.}

\begin{document}

\title{High Resolution Imaging of Very Low Mass Spectral Binaries: Three Resolved Systems and Detection of Orbital Motion in an L/T Transition Binary\altaffilmark{*}}
\author{Daniella C.\ Bardalez Gagliuffi\altaffilmark{1,2,3}, Christopher R.\ Gelino\altaffilmark{2,4}, Adam J. Burgasser\altaffilmark{1}}

\altaffiltext{1}{Center for Astrophysics and Space Sciences, University of California, San Diego, 9500 Gilman Dr., Mail Code 0424, La Jolla, CA 92093, USA.}
\altaffiltext{2}{Infrared Processing and Analysis Center, California Institute of Technology, Pasadena, CA 91125, USA.}
\altaffiltext{3}{IPAC Visiting Graduate Student Fellow 2014.}
\altaffiltext{4}{NASA Exoplanet Science Institute, California Institute of Technology, Pasadena, CA 91125.}
\altaffiltext{*}{Some of the data presented herein were obtained at the W.M. Keck Observatory, which is operated as a scientific partnership among the California Institute of Technology, the University of California, and the National Aeronautics and Space Administration. The Observatory was made possible by the generous financial support of the W.M. Keck Foundation.}

\begin{abstract}
We present high resolution Laser Guide Star Adaptive Optics imaging of 43 late-M, L and T dwarf systems with Keck/NIRC2.  These include 17 spectral binary candidates, systems whose spectra suggest the presence of a T dwarf secondary. We resolve three systems: 2MASS J1341$-$3052, SDSS J1511+0607 and SDSS J2052$-$1609; the first two are resolved for the first time. All three have projected separations $<8$ AU and estimated periods of $14-80$ years. We also report a preliminary orbit determination for SDSS J2052$-$1609 based on six epochs of resolved astrometry between 2005$-$2010. Among the 14 unresolved spectral binaries, 5 systems were confirmed binaries but remained unresolved, implying a minimum binary fraction of $47^{+12}_{-11}\%$ for this sample. Our inability to resolve most of the spectral binaries, including the confirmed binaries, supports the hypothesis that a large fraction of very low mass systems have relatively small separations and are missed with direct imaging.
\end{abstract}


\section{Introduction}\label{sec:intro}

Observational studies of field brown dwarfs indicate that only $\sim10-20\%$ are found in very low mass (VLM) binary systems~\citep{2003AJ....126.1526B, 2003IAUS..211..249C, 2006AJ....132..663B, 2007ApJ...668..492A, 2007ApJ...659..655B, 2012ApJ...757..141K}. In contrast, the binary fraction for G stars is $\sim40\%$~\citep{1991A&A...248..485D} and $\sim30\%$ for M dwarfs~\citep{1992ApJ...396..178F}. These statistics suggest a steady decline of binary fraction with mass. The peak in the observed projected separation distribution also decreases with mass, going from $30$AU for G dwarfs~\citep{1991A&A...248..485D}, $4-30$AU for M dwarfs~\citep{2010ApJS..190....1R,1992ApJ...396..178F} to $6-8$AU for VLM stars and brown dwarfs~\citep{2007ApJ...668..492A,2007prpl.conf..427B,2012ApJ...757..141K}. 


The observed peak in the projected separation distribution for VLM dwarfs is largely based on direct imaging studies, which have discovered $>80\%$ of the VLM binary systems to date~(\citealt{2007prpl.conf..427B,2014ApJ...794..143B} [hereafter: BG14]). Angular resolution limits impose a bias on the separations observed. For ground-based telescopes with Adaptive Optics (AO) and the Hubble Space Telescope (HST), this resolution limit is roughly $0\farcs05-0\farcs1$, which at the typical distances of known VLM dwarfs, $20-40$pc\footnote{In order to achieve S/N$\gtrsim$25 with low and high resolution spectroscopy.}, corresponds to the observed peak in the projected separation distribution. Tighter systems are unresolvable. Measurements of radial velocity (RV) and astrometric variability more adequately probe the small projected separation regime, but such measurements are resource-intensive and introduce their own set of geometric biases. An alternative approach to identifying closely-separated VLM binaries is as \emph{spectral binary} systems. Spectral binaries exhibit peculiarities in blended-light spectra that arise from the superposition of two components with distinct spectral morphologies. This method has been used to disentangle the spectra of white dwarf/M dwarf binaries~\citep{2007AJ....134..741S} and more recently, VLM stars and brown dwarfs, especially those with a T dwarf component~(e.g.~\citealt{2004ApJ...604L..61C,2008ApJ...676.1281M,2010ApJ...710.1142B} [hereafter: B10];~\citealt{2011ApJ...732...56G,2013MNRAS.430.1171D}; BG14). The identification of spectral binaries is independent of their projected separation, allowing the identification of binaries with very tight separations. The selection biases for this method (small separation, distinct component masses) are different from those of direct imaging, RV and astrometric variability, and overluminosity, providing a complementary approach to finding VLM binary systems.


While many brown dwarf spectral binaries have been discovered serendipitously~\citep{2004ApJ...604L..61C,2007AJ....134.1330B,2010AJ....140..110G}, recent systematic searches (B10, BG14) have increased the number of known spectral binaries to $\sim$50. Follow-up of candidates is necessary to confirm their binary nature since the spectral peculiarities that signal binarity may instead be the result of atmospheric variability, as in the case of the T1.5 2MASS J21392676+0220226\footnote{Hereafter targets observed in this study are referred to by shorthand notation: J$hhmm+ddmm$, where $h$ is hour, $d$ is degree and $m$ is minute. Full coordinates are listed in Table~\ref{tab:obs}.}~\citep{2012ApJ...750..105R,2013AJ....145...71K}. Only 12 spectral binaries have been confirmed by direct imaging, radial velocity, astrometric variability or overluminosity (See Table~\ref{tab:confSB};~\citealt{2006ApJS..166..585B,2008ApJ...678L.125B,2010AJ....140..110G,2011ApJ...739...49B,2011A&A...525A.123S,2012ApJ...757..110B,2012ApJS..201...19D,2012ApJ...752...56F,2013A&A...560A..52M}) and many of these have turned out to be close separation systems. The M9 dwarf SDSS J0006$-$0852AB~\citep{2012ApJ...757..110B} and the M8.5 dwarf 2MASS J0320$-$0446AB~\citep{2008ApJ...678L.125B,2008ApJ...681..579B} were confirmed as binaries by RV variability and found to have projected separations $<1$AU. The L4 dwarf SDSS J0805+4812~\citep{2007AJ....134.1330B}, confirmed as a binary through astrometric variability, has a semi-major axis $0.9-2.3$AU~\citep{2012ApJS..201...19D}. Even with the high resolution images provided by the Keck II Laser Guide Star Adaptive Optics (LGS-AO) system, none of these binaries can be resolved.

Nevertheless, high resolution imaging remains an efficient first test for binarity. In this article, we present high resolution LGS-AO observations of 43 late-M, L and T dwarfs, 17 of which are spectral binary candidates. Section~\ref{sec:obs} describes the target selection and observation procedures using the LGS-AO system and Keck II/NIRC2~\citep{2006PASP..118..310V,2006PASP..118..297W}. For the unresolved spectral binaries (visual and index-selected) we determine detection and separation limits in Section~\ref{sec:fakebinary}. We discuss in detail each of the known, unresolved binaries in Section~\ref{sec:unresbin}. We report three resolved sources and describe their properties in Section~\ref{sec:binaries}. In Section~\ref{sec:2052orbit} we analyze multi-epoch AO images of SDSS J2052$-$1609 and determine a first astrometric orbit for this L/T transition system. For the other two resolved systems, we estimate orbital parameters with Monte Carlo methods in Section~\ref{sec:orbit_est}. We discuss the broader implications of our results in the context of small separation VLM binaries in Section~\ref{sec:discussion}. Our results are summarized in Section~\ref{sec:summary}.


\section{Target Selection and Observations}\label{sec:obs}

\subsection{Spectral Binary Identification}
The 43 sources observed in our study (Table~\ref{tab:obs}) were selected from known late-M, L and T dwarfs in the vicinity of the Sun with a suitable tip-tilt star for LGS-AO correction. These include 33 M9-T3 dwarfs initially classified as spectral binaries by visual inspection, before the B10 and BG14 selection criteria had been defined. We re-examined their binary candidacy by dividing them into two groups according to spectral type: 15 objects in the M7-L7 range analyzed with the BG14 method, and 22 objects in the L5-T3 range, analyzed with the B10 method. The four objects overlapping in these spectral type ranges were analyzed by both methods. Ten other low mass stars and brown dwarfs were also observed as back-up targets, but were excluded from the analysis because visual inspection rejected them as spectral binary candidates.

Indices were measured from low resolution ($\lambda/\Delta\lambda = 75-120$), near-infrared IRTF/SpeX spectra~\citep{2003PASP..115..362R} covering $0.9-2.4\mu$m, accessed from the SpeX Prism Libraries~\citep{2014ASInC..11....7B}. One of our targets, 2MASS J2126+7617, has a declination outside the observable range of SpeX/IRTF ($-50^{\circ}<\delta<+67^{\circ}$), so a smoothed Keck/NIRSPEC spectrum was used instead~\citep{2010ApJS..190..100K}. Spectral indices given in B10 and BG14 were calculated for each spectrum, and regions of interest (ROI) in index-index spaces were delineated using confirmed binaries. Slight modifications to the limits of some ROIs in both B10 and BG14 were made to include known binaries WISEP J0720$-$0846 and 2MASS J1209$-$1004~(\citealt{2015AJ....149..104B,2010ApJ...722..311L}, respectively), which had not been detected at the time the index selection criteria were defined. Table~\ref{tab:indices} shows the updated limits of the index selection ROIs for both sets of criteria. Strong and weak candidates are selected by the number of times they fall within the ROIs, as described in B10 and BG14. 


From the BG14 set, 8 sources were selected as candidates from spectral indices (4 as strong, 4 as weak). Single and binary templates were fit to these index-selected sources, ranked by a $\chi^2$ statistic. The best fit single and binary templates were compared to each other with an F-test to assess the percentage confidence that the binary fit is statistically better than the single fit. The primary types were constrained to $\pm$3 subtypes from the combined optical spectral type or, in its absence, near infrared type, and the secondary types were allowed to vary between T1 and T8. After template fitting, 6 sources remained as candidates. From the B10 set, 16 sources were selected as index candidates (11 as strong, 5 as weak) and after fitting, 12 sources remained as candidates. 2MASS J1711+2232 was selected as a candidate on both sets. In all, we classify 17 sources as true spectral binary candidates (Table~\ref{tab:obs}), close to half of the visually-selected spectral binaries.

\subsection{NIRC2 High Resolution Imaging and Reduction}
High angular resolution images of our targets were obtained using the Keck II LGS-AO system with NIRC2 on nine nights between August 2009 and January 2014. Tip-tilt reference stars within 60$\arcsec$ of the targets were selected from the USNO-B catalog~\citep{2003AJ....125..984M}. A 3-point dither pattern was used to avoid the noisy lower left quadrant of the array, and was repeated as needed with different dither offsets to build up long exposures. Total integration times were between 60s and 720s, depending on the brightness of the source and the atmospheric conditions. All objects were observed with the Mauna Kea Observatories (MKO) $H$ filter~\citep{2002PASP..114..169S,2002PASP..114..180T} and narrow plate scale (9.970$\pm$0.012 $mas$/pixel for a single-frame field-of-view of $10\arcsec\times10\arcsec$;~\citealt{2006ApJ...649..389P}).  The MKO $J$ and/or $K_s$ filters were also used for targets with apparent companions.


The images were reduced in a standard fashion using Interactive Data Language (IDL) scripts. First, a dark frame was subtracted from each science frame.  For each science exposure a sky frame was constructed from the median average of all images acquired for the target, exclusive of the frame being reduced. The sky-subtracted frames were then divided by a normalized dome flat. A bad pixel mask was applied to smooth over bad pixels using the average of the neighboring pixels. All images in a given epoch and common filter were shifted to align the target to a common location, and the stack was median-combined to create the final mosaics.  

\section{Analysis}

\subsection{Image Characterization and Companion Detection Limits for Unresolved Sources}\label{sec:fakebinary}

The reduced image mosaics around each target are shown in Figures~\ref{fig:binaries} and~\ref{fig:unresbin}. Strehl and signal-to-noise ratios (S/N) are reported in Table~\ref{tab:obs}. The Strehl ratio was calculated by comparing each point source to a theoretical, diffraction-limited, monochromatic, NIRC2 point spread function (PSF) with the \texttt{NIRC2Strehl} IDL routine\footnote{Retrieved from~\url{https://www2.keck.hawaii.edu/optics/lgsao/software/}}. The S/N was computed assuming Poisson statistics:
\begin{equation}
S/N = \frac{N_{star}}{\sqrt{n_{sky}~\sigma_{sky}^2 + \frac{N_{star}}{g}}}
\end{equation}
where $N_{star}$ is the total counts from the star at a radius of 1.5 times the full width at half maximum, $n_{sky}$ is the number of pixels used for the standard deviation of the sky counts, $\sigma_{sky}$, which encompasses noise from several sources (read out, dark current, image reduction, etc.) and $g$ is the gain in DN/$e^{-}$ (data number per electron). 

\begin{figure}
\centering
\includegraphics[scale=0.6]{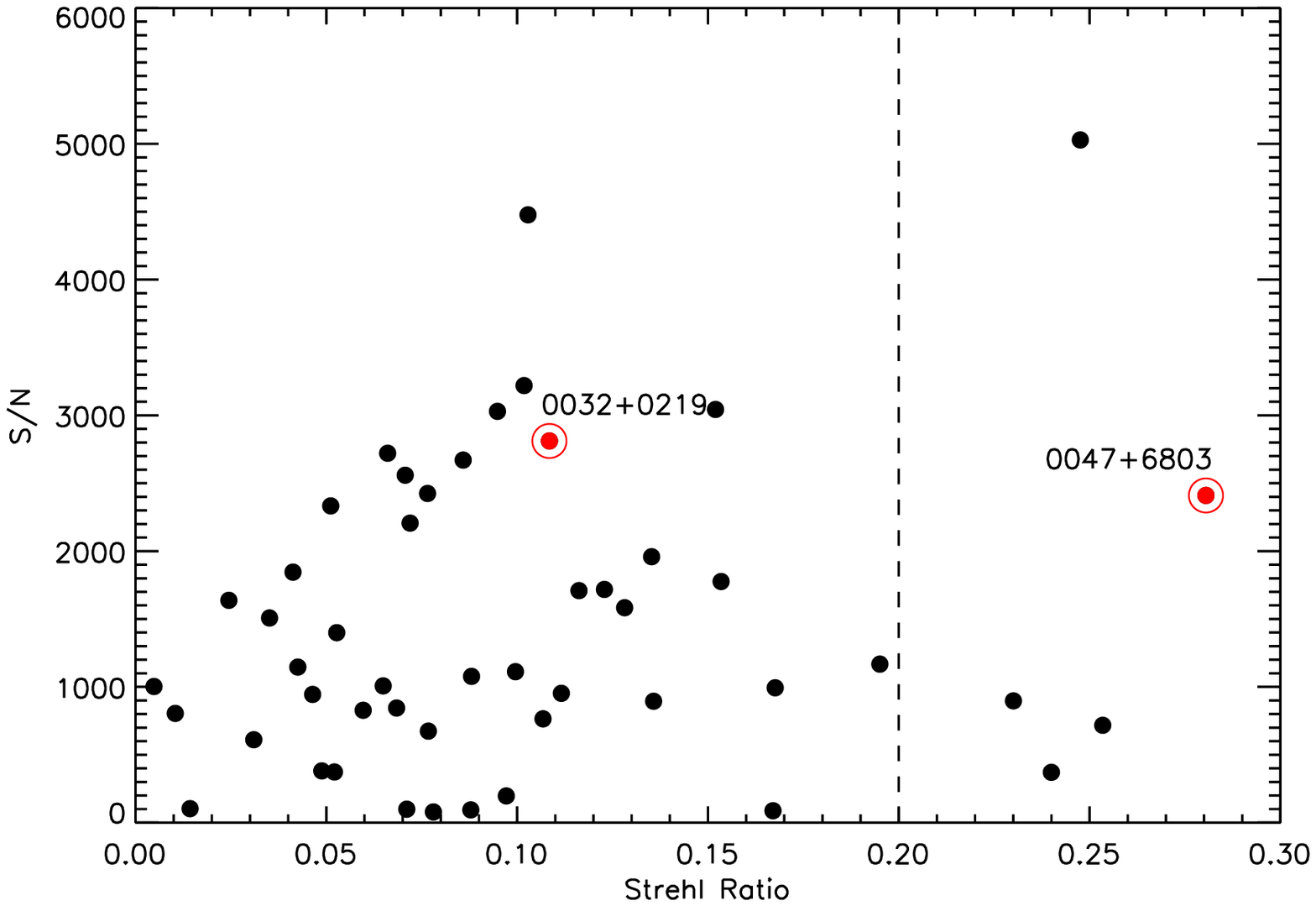}
\caption{Strehl ratio vs. signal-to-noise ratio of all $H$ band observations (black dots). The two representative sources used in the empirical sensitivity curves are marked in red. The three groups are separated by the dashed lines. See Table~\ref{tab:obs} for the list of objects.\label{fig:Strehl_vs_SNR}}
\end{figure}

Three of our sources are resolved: 2MASS 1341$-$3052, SDSS 1511+0607 and SDSS 2052$-$1609; these are shown in Figure~\ref{fig:binaries} and discussed further in Section~\ref{sec:binaries}. One source, 2MASS J1733$-$1654, has a feature that we cannot distinguish between bona fide source and PSF structure, so we consider this to be a ``source of interest''. The remaining sources are unresolved at the limits of our sensitivity and image quality. Because the PSF of the images vary considerably, we determined detection limits through a source implantation simulation of representative images. We organized the targets by Strehl and S/N and selected two representative sources of high Strehl (WISE J0047+6308) and low Strehl (2MASS J0032+0219), as shown in Figure~\ref{fig:Strehl_vs_SNR}. For these sources, we simulated binary companions by scaling down the brightness of each image, and then shifting and superimposing it onto the original image. The implanted image was scaled down by a maximum of 6 magnitudes, which was the largest magnitude difference inferred from the template fitting of spectral binary candidates, and shifted by up to 50 pixels or $\sim0\farcs5$ in any angle. Magnitude difference, separation and position angle were all drawn from a uniform random distribution.

\begin{figure}[htp]
\centering
\includegraphics[width=.5\textwidth]{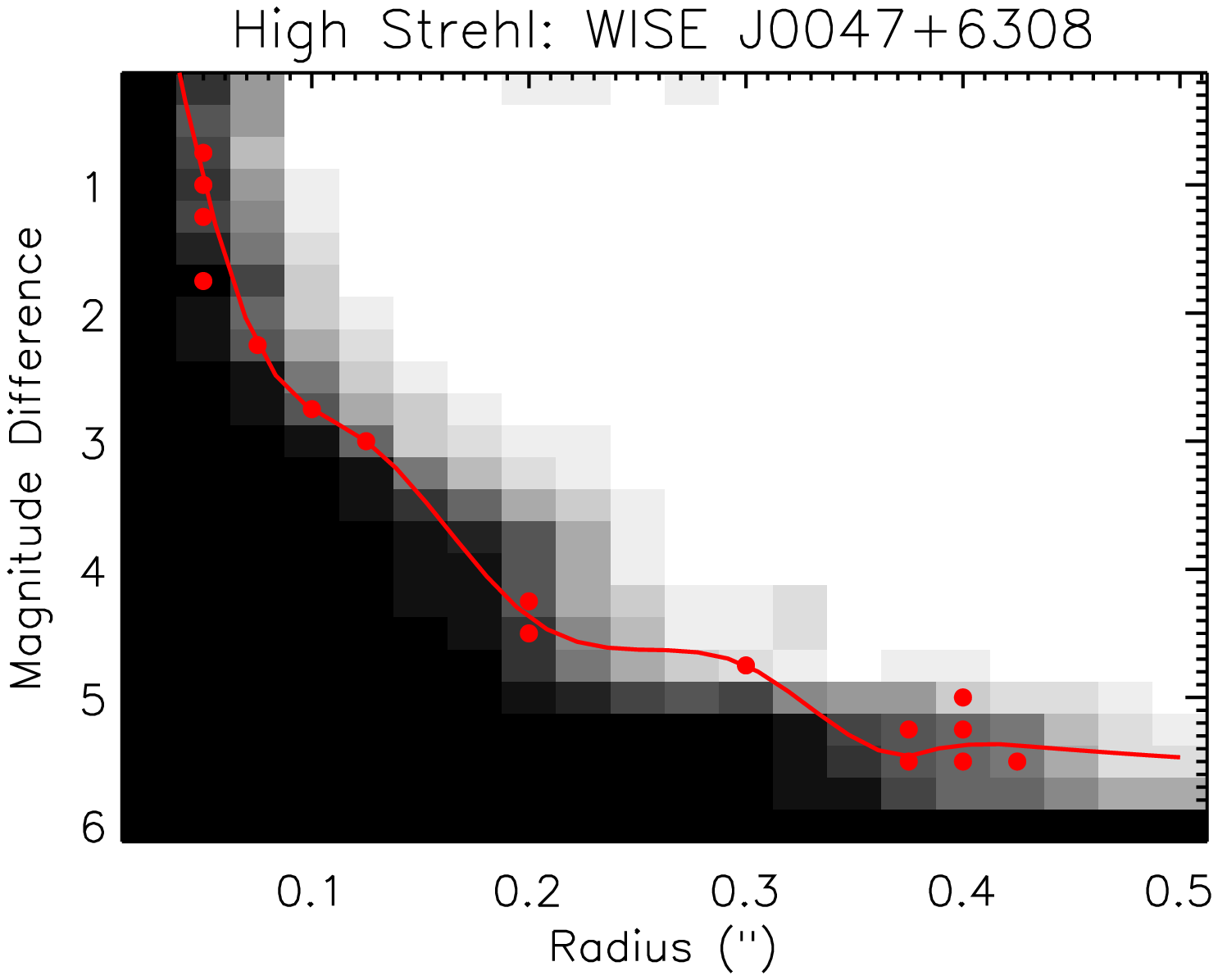}\includegraphics[width=.5\textwidth]{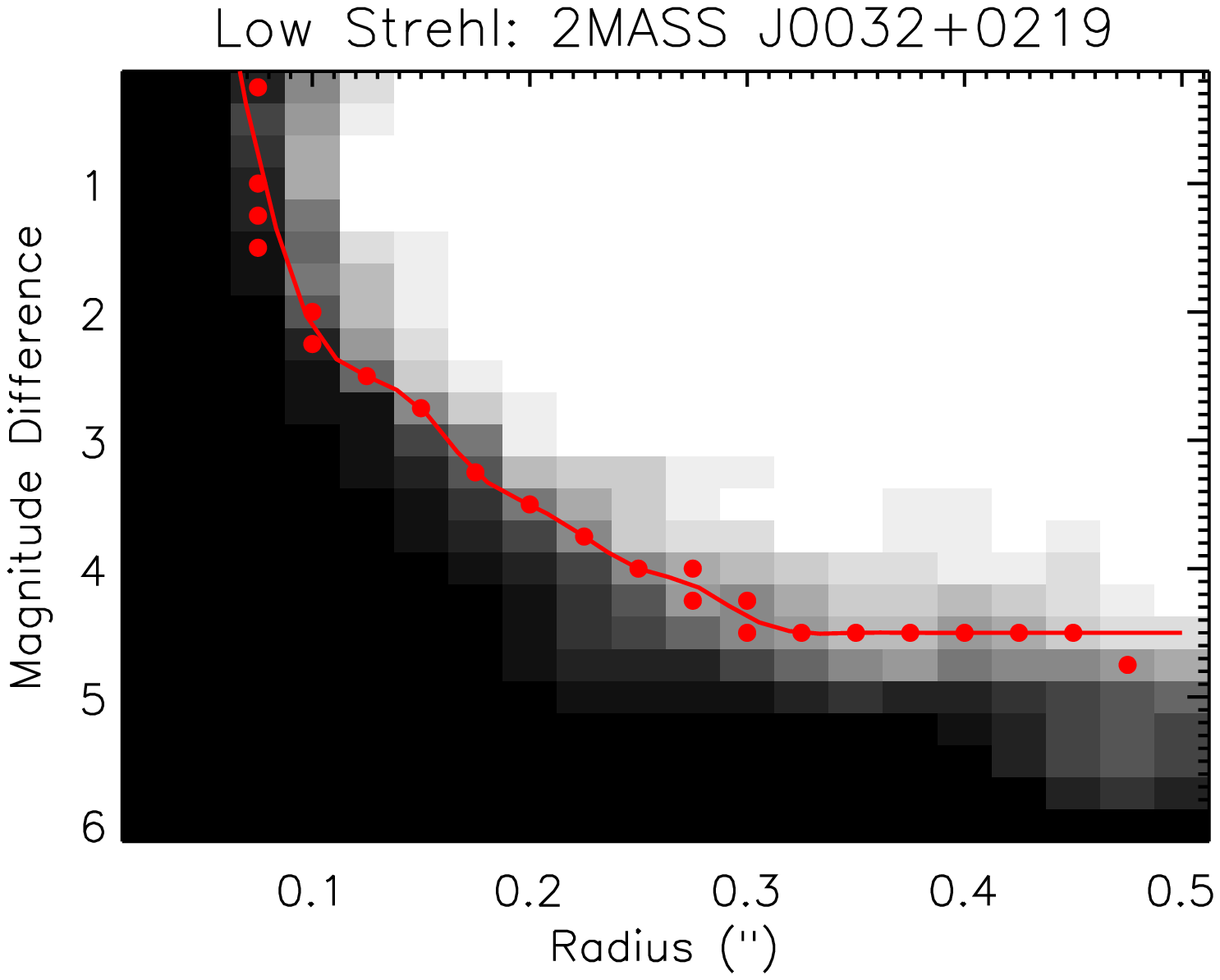}
\caption{Empirical sensitivity curves from simulated binaries for two representative sources based on their Strehl and signal-to-noise ratios (see Table~\ref{tab:obs}). The red line delimits the detections of secondaries (white) from non-detections (black).}
\label{fig:fakebin}
\end{figure}

We visually examined each image at multiple contrast ratios to search for the implanted companion. This experiment was performed N$\gtrsim$12,000 times per source, varying the target, scale factor and offset. A ``detection'' required clicking within 15 pixels of the implanted secondary, with the option to decide if an implanted companion was visually undetectable. We determined the maximum relative magnitude as a function of separation for which the detection fraction exceeded 50\%. The detection fraction was calculated in steps of 0.5 mag and $0\farcs05$, sliding by half a step along both axes for a total of $\sim$400 overlapping bins. Figure~\ref{fig:fakebin} shows that the PSF dominates the sensitivity close to the star centroid. For the case of low Strehl ratio, detections reach a minimum at $\Delta H\approx 5$ mag, $0\farcs3$ away from the center of the PSF, beyond which our sensitivity is limited by sky noise. For the high Strehl ratio case the floor lies around 5.5 magnitude difference at radii greater than $0\farcs4$. 



\begin{figure}
\centering
\includegraphics[scale=0.4]{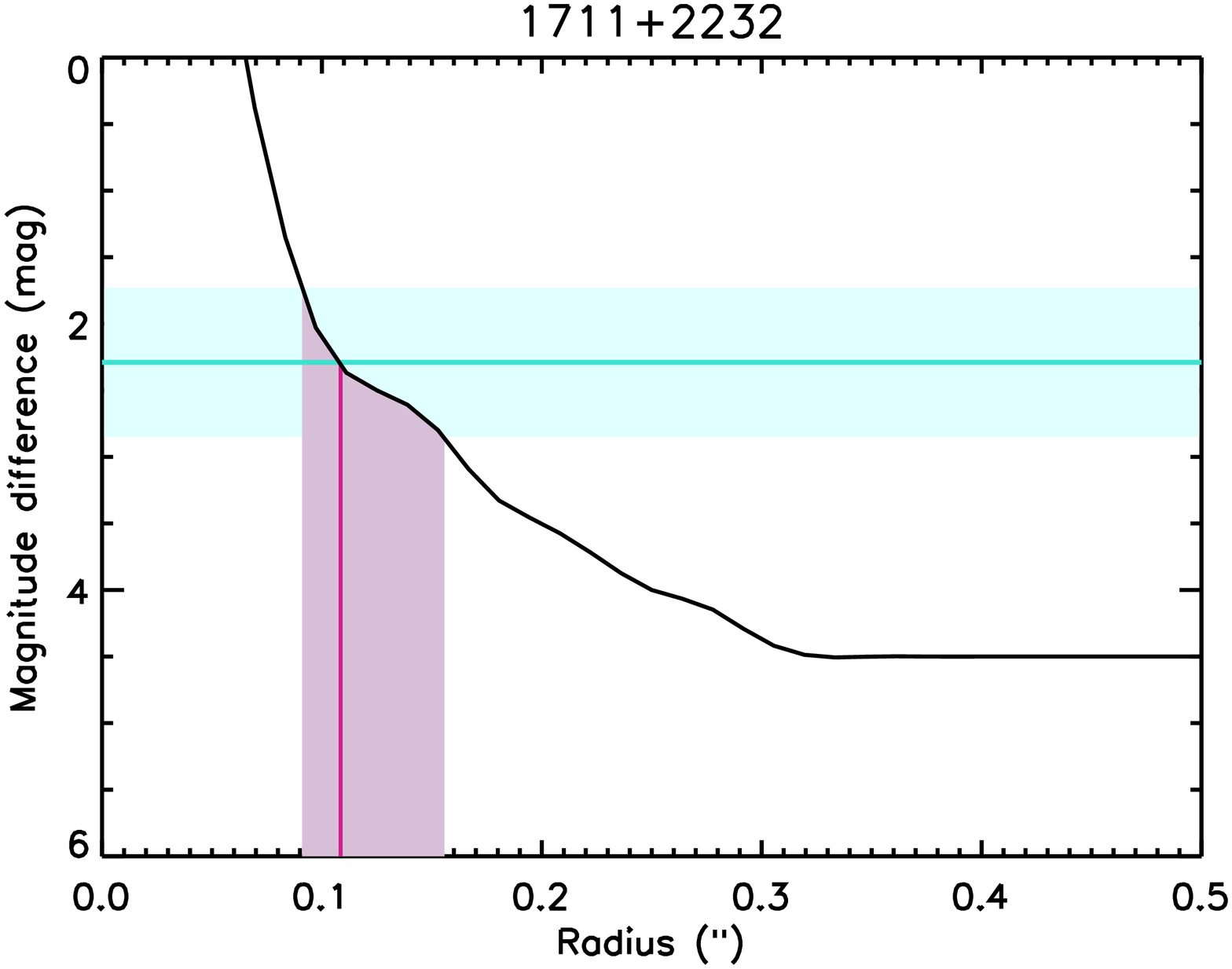}
\caption{Example separation constraint for 2MASS J1711+2232. Given the empirical sensitivity curve for its Strehl ratio vs. S/N group (low Strehl-low S/N, representative source: 2MASS J0032+0219) in black, and the estimated magnitude difference in cyan (uncertainties are shown as the shaded region), we can set an upper limit for an undetected secondary at the intersection (magenta line). The values for the separation constraints for all observations are reported in Table~\ref{tab:seplim}.\label{fig:seplim}}
\end{figure}

We applied the sensitivity curves of our representative sources to systems with similar Strehl ratios. For the 30 unresolved spectral binary candidates, we compared these sensitivity limits to the magnitude differences predicted from template fitting to determine separation limits (Table~\ref{tab:seplim}). Figure~\ref{fig:seplim} shows an example of the sensitivity curve and separation constraint for the spectral binary candidate 2MASS J1711+2232. Five of our unresolved spectral binaries have been previously confirmed as true binaries (See Section~\ref{sec:unresbin}), but our separation limits are up to $40\%$ greater, i.e. these binaries can not be resolved with our observations. For the case of 2MASS J1209-1004, our estimated separation limit is smaller than the measured separation, suggesting that the secondary has moved to a closer configuration. Similarly, for our three resolved systems the calculated separation limit is always smaller than the measured separation, which means that our separation limits correctly constrain the PSF of the primary. The remaining 9 unresolved spectral binaries have angular separation limits between $0\farcs04-0\farcs28$.



\subsection{Unresolved Known Binaries}\label{sec:unresbin}

\subsubsection{2MASS J05185995-2828372}
2MASS J0518$-$2828 was the first source to be identified as a spectral binary of L6 and T4 components~\citep{2004ApJ...604L..61C} and was marginally resolved with HST~\citep{2006ApJS..166..585B} with an angular separation of $0\farcs051\pm0\farcs012$. ~\citet{2012ApJS..201...19D} find a small astrometric perturbation for this source that cannot be clearly attributed to orbital motion.~\citet{2010ApJ...711.1087K} observed this source with LGS-AO at Keck in 2006 and were not able to resolve it.  Its parallactic distance has been measured to be 22.9$\pm$0.4 pc~\citep{2012ApJS..201...19D}, implying a projected separation of 1.17$\pm$0.28 AU from the HST measurement. Our LGS-AO observations also fail to resolve this system to a limit of $98~mas$ or 2.2 AU, which is consistent with the HST observations. This system appears to be a very tight binary whose separation is just below the limits of ground-based AO imaging.


\subsubsection{WISEP J072003.20$-$084651.2}
WISEP J0720$-$0846 was discovered by~\citet{2014A&A...561A.113S} and confirmed by~\citet{2015AJ....149..104B} as an M9 at a distance of 6.0$\pm$1.0 pc. The latter study identified  a candidate companion at an angular separation of 139$\pm14~mas$ in NIRC2 LGS-AO observations, which has been confirmed at a slightly wider offset (angular separation $197\pm3~mas$, projected separation $1.18\pm0.21$AU) with $\Delta{H}$ = 3.85$\pm$0.11 mag in follow-up observations~\citep{2015arXiv150806332B}. Our analysis does not resolve the companion to limits of $500~mas$ and 3 AU, beyond the separation reported in that study.


\subsubsection{SDSS J080531.84+481233.0}
SDSS J0805+4812 is a blue L dwarf discovered by~\citet{2002AJ....123.3409H}, and a spectral binary of L4.5 and T5 components~\citep{2007AJ....134.1330B}. This source shows astrometric variability with an amplitude of 15 $mas$~\citep{2012ApJS..201...19D}. Our LGS-AO observations show an elongated PSF that we attribute to tip-tilt correction errors, but no resolved companion.~\citet{2012ApJS..201...19D} estimate a semi-major axis of $40-100~mas$ assuming a mass ratio $q = 0.55-0.88$ and from the measured parallactic distance of $22.9\pm0.6$pc, they infer a projected separation of $0.9-2.3$AU. Our observations do not resolve this system to limits of $164~mas$ and 3.8 AU, both consistent with the~\citet{2012ApJS..201...19D} estimates.



\subsubsection{2MASS J11061197+2754225}
The T2.5 2MASS J1106+2754 was first discovered by~\citet{2007AJ....134.1162L} and later observed with NIRC2 with LGS-AO in June 2006, but was unresolved~\citep{2008ApJ...686..528L}. B10 identified it as a spectral binary of T0.0$\pm$0.2 and T4.5$\pm$0.2 components due to its CH$_4$ absorption feature in the $H$ band, and ruled out a separation greater than 1.5 AU based on Keck imaging.~\citet{2013A&A...560A..52M} finds that this source is $\sim$1 mag overluminous and determined a parallactic distance of $20.6^{+1.0}_{-1.2}$ pc. Our LGS-AO observations were unable to resolve this source, implying upper limits of $74~mas$ and 1.5 AU, the same constraint as that reported by~\citet{2010ApJ...710.1142B}.


\subsubsection{2MASS J12095613-1004008}
2MASS J1209$-$1004 was first discovered by~\citet{2004AJ....127.2856B} and is the T3 spectral standard~\citep{2006ApJ...639.1095B}.~\citet{2010ApJ...722..311L} resolved the system with NIRC2 and LGS-AO in the $J$ band, and estimated component types of T2.0$\pm$0.5 and T7.5$\pm$0.5 based on photometry. The mass ratio of this binary is estimated to be $q = 0.5$, which is unusually small for brown dwarf binaries~(\citealt{2003AJ....126.1526B,2007ApJ...659..655B}).~\citet{2010ApJ...722..311L} found an angular separation of $151\pm13~mas$ at a position angle of $314^{\circ}\pm5^{\circ}$ with a magnitude difference of $\Delta H = 2.8\pm0.3$ mag. Its parallactic distance $d = 21.8\pm$0.5 pc~\citep{2012ApJS..201...19D} leads to a projected separation of 3.3$\pm$0.3 AU. This source was not resolved in our $H$ band LGS-AO image, with limits of 95~$mas$ and 2.1 AU. In this case, our observation should have detected the companion, suggesting that orbital motion may have moved into closer projected alignment, or that the companion could be the source of the elongation of the PSF to the South East.

\subsection{Resolved Binaries}\label{sec:binaries}


\subsubsection{2MASS J13411160$-$30525049}
The L3 2MASS 1341$-$3052 was first discovered by~\citet{2008AJ....136.1290R} in the Two Micron All Sky Survey (2MASS;~\citealt{2003yCat.2246....0C}) and later identified as a spectral binary candidate with L1.0$\pm$0.5 and T6.0$\pm$1.0 components (BG14). Our NIRC2 observations resolve the source with an angular separation of $279\pm17~mas$ at a position angle of $317.9^{\circ}\pm0.6^{\circ}$. The original template matching analysis assumed relative spectral fluxes on the~\citet{2008ApJ...685.1183L} absolute magnitude to spectral type relation. We repeated this analysis using the relative photometry in all three $JHK_s$ bands to scale and select binary templates in a similar fashion as described in~\citet{2011AJ....141...70B}.  Our revised template fit analysis resulted in component spectral types of L2.5$\pm$1.0 and T6.0$\pm$1.0; i.e. we infer a primary classification more consistent with the combined-light optical classification.


2MASS J1341$-$3052 is the only one of the three resolved systems that does not have a parallax measurement. We estimated its distance using the calculated component apparent magnitudes with the~\citet{2012ApJS..201...19D} spectral type to absolute magnitude relation and component spectral types from template fitting. Uncertainties were propagated from spectral type, apparent magnitudes and spectral type relation, accordingly. We calculated one distance per filter and then obtained the weighted average distance from the MKO $JHK_s$ filters. The distances for both components (29$\pm$3 pc for the primary and 31$\pm$6 pc for the secondary were consistent across filters. The uncertainty-weighted average yields a distance estimate of 29$\pm$3 pc. From this we infer a projected separation of $8.1\pm0.5$ AU.

\begin{figure}[htp]
\centering
\includegraphics[width=.5\textwidth]{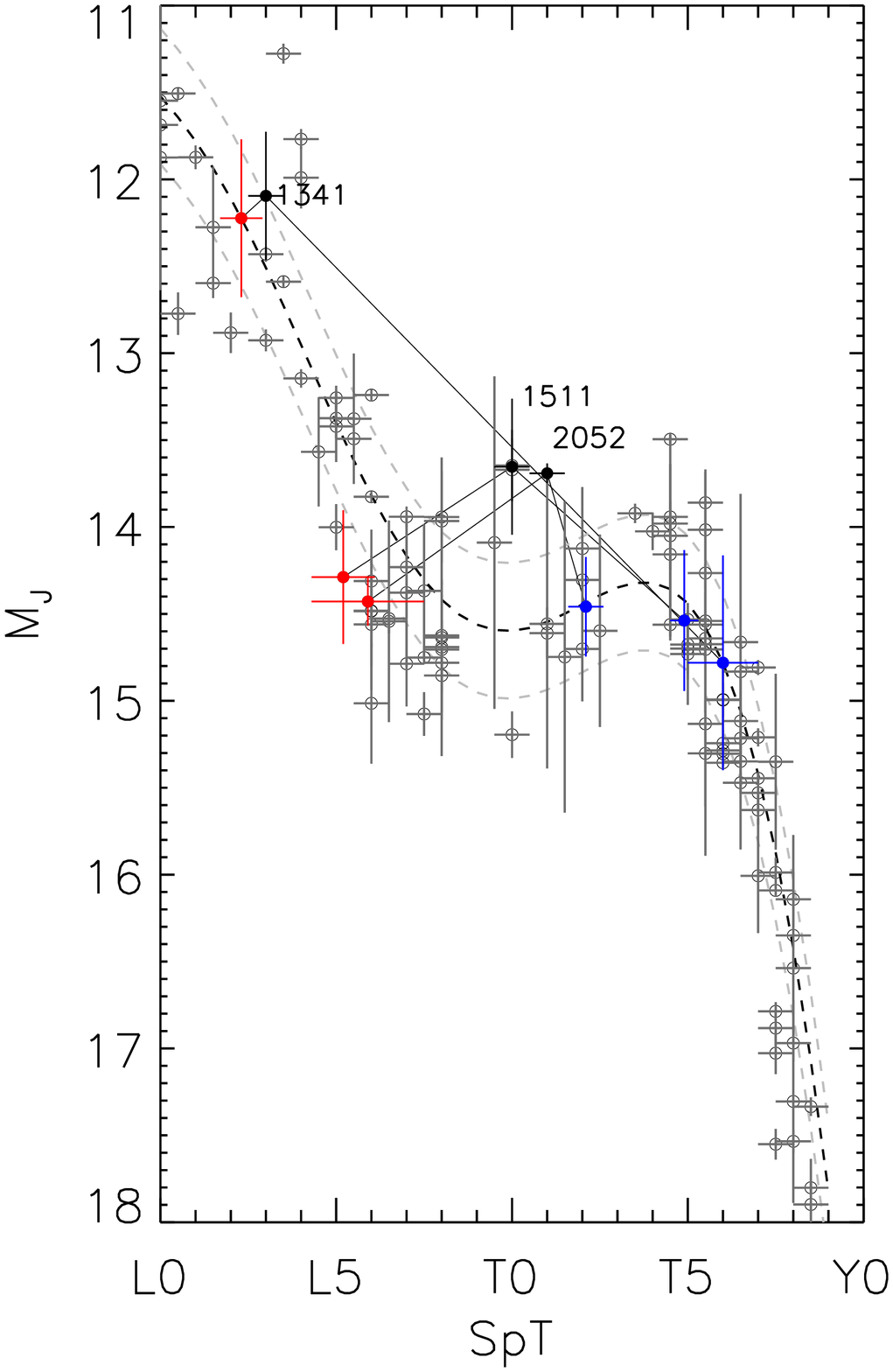}\hfill
\includegraphics[width=.5\textwidth]{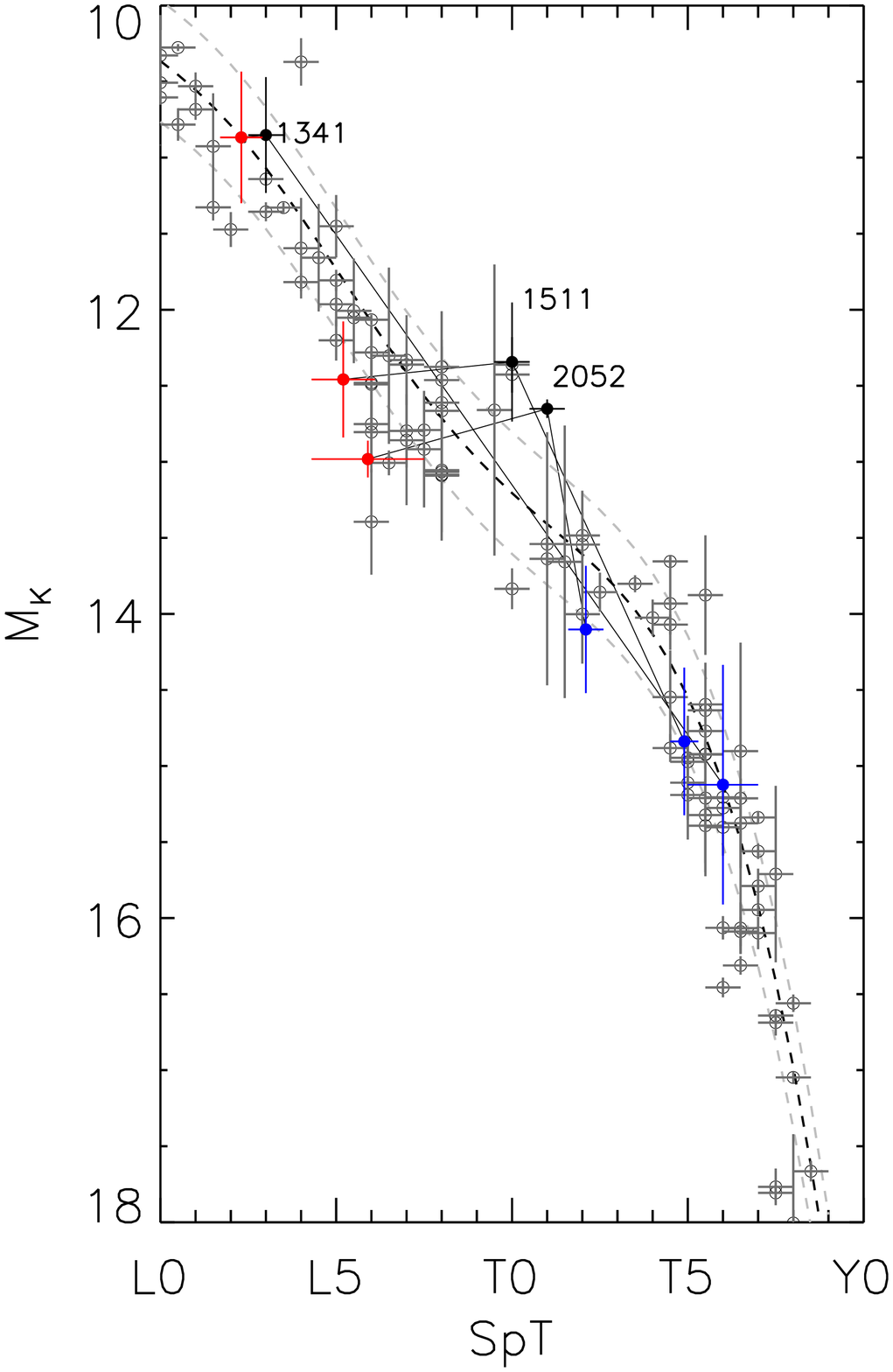}
\caption{Absolute magnitudes in $J$ and $K_s$ for the three resolved binaries (black dots) and the~\citet{2012ApJS..201...19D} parallax sample (grey circles) shown with the~\citet{2012ApJS..201...19D} spectral type to absolute magnitude relation (black dashed line) and its 1-$\sigma$ outliers (grey dashed lines). Only SDSS J1511+0607 and SDSS~J2052$-$1609 have parallaxes reported in the literature. The three binaries are split into component spectral types, where the primaries for the two objects with parallaxes lie on the faint end of the absolute magnitude relation for their spectral type (red dots) , and the secondaries (blue dots) are all within 1-$\sigma$ from the relation.}
\label{fig:absmagspt}
\end{figure}

A spectral type versus absolute magnitude plot is shown in Figure~\ref{fig:absmagspt}, using the~\citet{2012ApJS..201...19D} parallax sample of 259 objects as a reference. The primary absolute magnitude is anchored to the~\citet{2012ApJS..201...19D} relation due to the lack of a trigonometric distance, while the secondary absolute magnitude was derived from the primary's using the measured relative magnitude. In the $J$ band, the secondary adds 0.09 mags to the primary, barely enough to make the combined absolute magnitude look like an outlier. In the $K$ band, the secondary adds 0.02 mags to the primary, so it appears as if the secondary absolute magnitude also falls on the relation and the combined magnitude is within the $+1\sigma$ curve. This source could not have been detected as an overluminous binary candidate because of the late type of the secondary.

\subsubsection{SDSS J151114.66+060742.9}
The T0 SDSS J1511+0607 was discovered by~\citet{2006AJ....131.2722C} in the Sloan Digital Sky Survey (SDSS;~\citealt{2007ApJS..172..634A}). This source was identified as a ``strong'' binary candidate in B10, and found to be 1 mag overluminous for its spectral type in M$_{JHK}$ by~\citet{2012ApJ...752...56F}. Multi-band imaging with NIRC2 on 2009 Aug 15 resolved the system into two components separated by $108\pm11~mas$ at a position angle of 335$^{\circ}\pm4^{\circ}$. The parallax measurement of 37$\pm7~mas$ by~\citet{2012ApJ...752...56F} implies a distance of 28$\pm$5 pc, which in turn corresponds to a projected separation of 2.9$\pm$0.3 AU. Including the measured relative $JHK_s$ magnitudes in our template fitting gives updated component spectral types of L5.0$\pm$1.0 and T5.0$\pm$0.5.

Using the combined light magnitude, the relative magnitudes and the parallactic distance, we determined the absolute magnitudes for the components (Figure~\ref{fig:absmagspt}). While the combined absolute magnitude of SDSS J1511+0607 clearly stands out as an outlier in spectral type to absolute magnitude plots, its components look typical. Indeed, its primary lies slightly below the~\citet{2012ApJS..201...19D} spectral type to absolute magnitude relation. 

\subsubsection{SDSS J205235.31$-$160929.8}
Also discovered by~\citet{2006AJ....131.2722C}, SDSS 2052$-$1609 was classified as an T1$\pm$1 brown dwarf. B10 identified it as a spectral binary candidate with component types of L7.5$\pm$1.0 and T2$\pm$0.5, noting that the best fit primary was unusually blue when compared to the median $J-K_s$ colors of~\citet{2009AJ....137....1F}.~\citet{2011A&A...525A.123S} was able to resolve the components with VLT/NACO, and also reported archival HST/NICMOS data which confirmed common proper motion and indicated some orbital motion.~\citet{2011A&A...525A.123S} determined component spectral types by comparing the objects' $JHK_s$ colors to mean colors from Dwarf Archives\footnote{\url{http://www.dwarfarchives.org}} and obtained divergent results for the primary component (T0.5$\pm$0.5 and T2.5$\pm$0.5) as compared to those from B10. Using our NIRC2 photometry to constrain spectral template fitting, we find component spectral types of L6.0$\pm$2.0 and T2.0$\pm$0.5.~\citet{2012ApJS..201...19D} find a parallactic distance of 29.5$\pm$0.7 pc. The angular separation between the components was measured to be 103$\pm2~mas$, leading to a projected separation of 3.0$\pm$0.1 AU. 



Absolute magnitudes of this source and its components were calculated from its parallactic distance and measured magnitude differences. As for the case of SDSS J1511+0607, the primary of SDSS J2052$-$1609 appears to be underluminous, while its secondary falls comfortably within $1\sigma$ from the~\citet{2012ApJS..201...19D} relation (Figure~\ref{fig:absmagspt}).

\section{Discussion}

\subsection{A Preliminary Orbit for SDSS~J2052$-$1609AB}\label{sec:2052orbit}

\begin{figure}
\epsscale{0.6}
\plotone{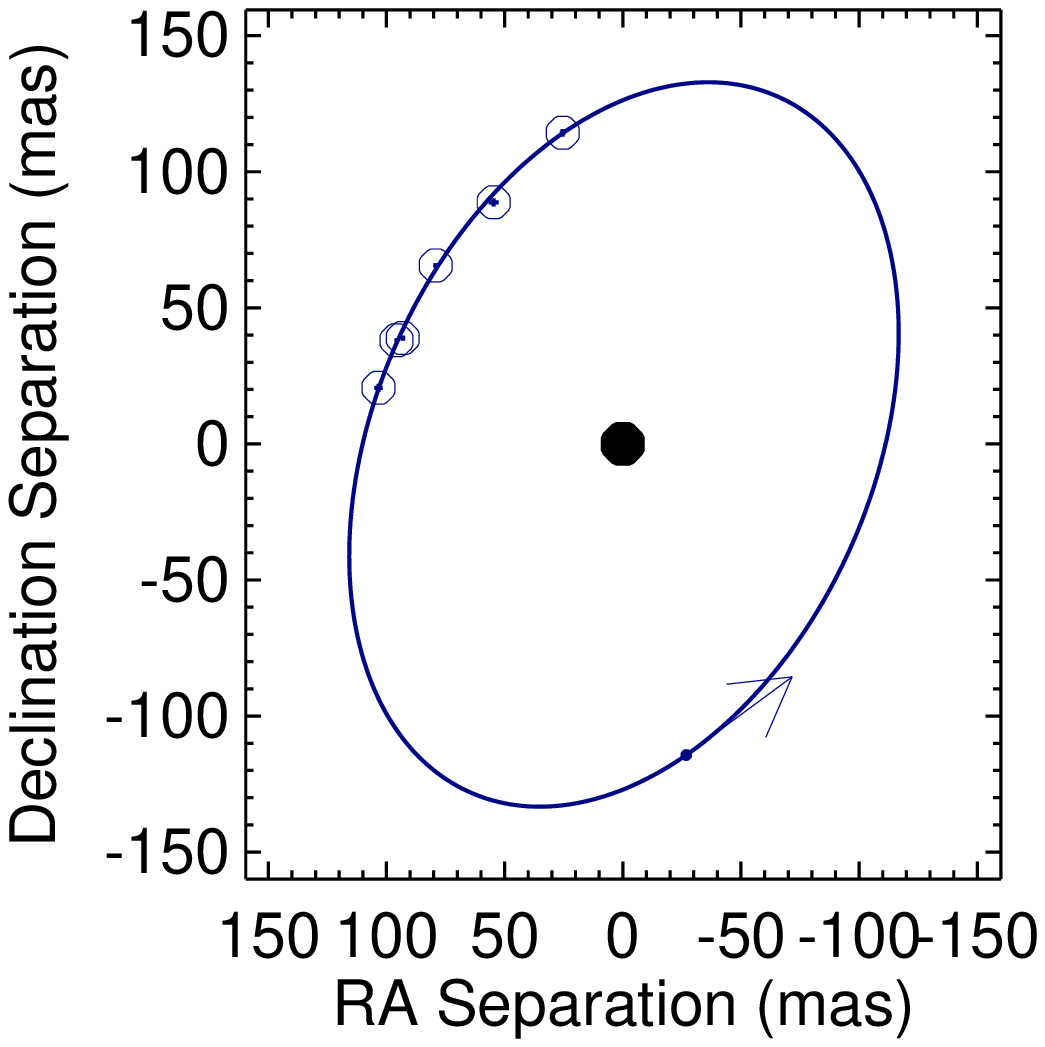}
\caption{Visual orbit of SDSS~J2052$-$1609AB based on MCMC analysis of separation measurements reported here and in \citet[open circles]{2011A&A...525A.123S}. The orbital motion of the secondary (blue line) relative to the primary (black dot at the origin) is shown projected on the sky, with the arrow indicating the direction of orbital motion at periaps ($M$ = 0$\degr$).  Error bars are plotted but indiscernible on this scale.
\label{fig:orbit2052}}
\end{figure}

Our observations of SDSS~J2052$-$1609AB confirm prior results by \citet{2011A&A...525A.123S} and adds to coverage of its orbital motion first detected in that study.  We identified additional archival NIRC2 + LGSAO images of the system taken on 2005 October 11, 2007 April 23 (PI M.\ Liu) and 2010 May 1 (PI B.\ Biller), and analyzed these data in the same manner as described above. The resulting six epochs of relative astrometry spanning just over 4.5~yr are listed in Table~\ref{tab:astrometry2052} and displayed in Figure~\ref{fig:orbit2052}.  These measurements confirm the direction of motion previously identified and cover a significant fraction of the system's orbit.  

To more tightly constrain the orbit of this system, we used an Markov Chain Monte Carlo (MCMC) routine with Metropolis-Hasting algorithm \citep{1953JChPh..21.1087M,hastings1970} to iteratively fit a seven-parameter orbit model to the twelve astrometric measurements (six each in relative Right Ascension and declination) and parallax distance measurement ($d$ = 29.5$\pm$0.7~pc; \citealt{2012ApJS..201...19D}).  
The methodology used is described in detail in~\citet{2015arXiv150806332B}.
The parameter vector is
\begin{equation}
\vec{\theta} =  \left(P,a,e,i,\omega,\Omega,M_0,d\right)
\end{equation}
where $P$ is the period of the orbit in years, $a$ the semi-major axis in AU, $e$ the eccentricity, $i$ the inclination, $\omega$ the argument of periastron, $\Omega$ the longitude of nodes, $M_0$ the mean anomaly at epoch $\tau_0$ =  2453654.31 (Julian Date), and $d$ is the distance in pc.  
We computed an MCMC chain of 10$^7$ parameter sets, at each step varying parameters using a normal distribution with scale factors that were allowed to vary dynamically to improve convergence.\footnote{We used an initial scale factor set $\vec{\beta}$ = (5~yr, 0.5~AU, 0.2, 20$\degr$, 20$\degr$, 20$\degr$, 20$\degr$, 0.7~pc), but if a parameter $\theta_j$ did not change in 1000 iterations, $\beta_j$ was changed to 3 times the standard deviation of the previous (up to 100) changed values.} We applied additional constraints of 1~yr $< P <$ 100~yr and $0 < e < 0.6$ to eliminate improbable regions of parameter space, and constained the distance to lie within 28~pc $< d <$ 31~pc; our parameter chain was largely insensitive to these limits. Convergence of the chain was monitored through autocorrelation of parameters and evolution of divergence in sequential subchains, and acceptance rates were typically 0.5--1\%. The first 10\% of the MCMC chain was removed from subsequent analysis.

\begin{figure}
\epsscale{1.0}
\centering
\plotone{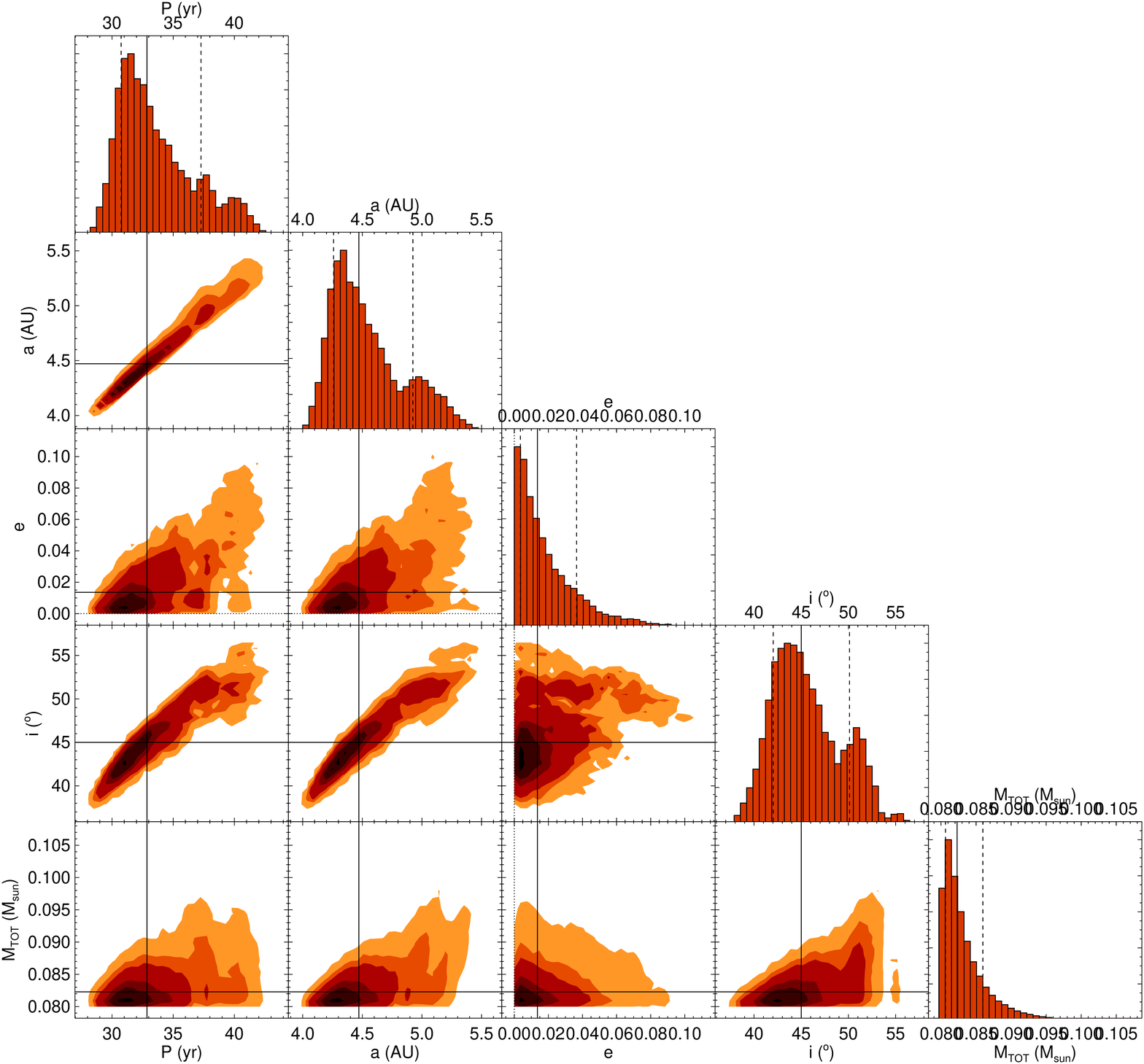}
\caption{Parameter distributions and correlations (triangle plot) for period ($P$), semi-major axis ($a$), eccentricity ($e$), 
inclination ($i$), and total system mass ($M_{tot}$) based on our MCMC orbital analysis of SDSS~J2052$-$1609AB.
Contour plots show $\chi^2$ distributions as a function of parameter pairs, highlighting correlations.
Normalized histograms at the ends of rows are marginalized over all other parameters.
Median values are indicated by solid lines in all panels, and 16\% and 84\% quantiles are indicated by dashed lines in the histograms.  
\label{fig:orbit2052_parameters}}
\end{figure}

Figure~\ref{fig:orbit2052} shows the best-fit relative visual orbit compared to the measurements, which is an acceptable fit ($\chi^2$ = 12.05 for 6 degrees of freedom).  Table~\ref{tab:orbit2052} lists the best-fit orbital parameters, as well as median values and 16\% and 84\% quartiles, while 
Figure~\ref{fig:orbit2052_parameters} displays the distributions and correlations of $P$, $a$, $e$, $i$ and M$_{tot}=a^3/P^2$, the total system mass in Solar units. All of the parameters are reasonably well-determined despite the limited phase coverage, although there are strong correlations between $P$, $a$, $i$ and $e$ and a hint of a secondary solution (double-peaked distributions). From the primary solution, we estimate an orbit period of 33$^{+4}_{-2}$~yr and total system mass of 0.0823$^{+0.0037}_{-0.0017}$~M$_{\odot}$, which is consistent with the lower limit of 0.074~M$_{\odot}$ proposed by \citet{2011A&A...525A.123S} assuming a circular orbit.  Indeed, the orbit of SDSS~J2052$-$1609AB appears to be fairly circular (0.014$^{+0.023}_{-0.010}$) and significantly inclined to the line of sight (45$\degr^{+4\degr}_{-2\degr}$). The best fit template fitting results suggest component spectral types of L5.8$\pm$1.8 and T2.1$\pm$0.5. Assuming effective temperatures corresponding to these spectral types (1544$\pm$181K for the primary and 1248$\pm$101K for the secondary;~\citealt{2009ApJ...702..154S}), and using the evolutionary models of~\citet{2008ApJ...689.1327S} to estimate age-dependent component masses, we estimate an age of 0.4--1.4~Gyr for this system, where the range accounts for the total mass uncertainty, effective temperature uncertainties, and cloud effects on brown dwarf evolution. Observations over the next decade should greatly improve the mass and orbit constraints on this system, and resolved spectroscopy should make it possible to critically test evolutionary models (e.g., \citealt{2010ApJ...711.1087K,2014ApJ...790..133D}).


\subsection{Estimated orbital parameters for 2MASS J1341$-$3052 and SDSS J1511+0607}\label{sec:orbit_est}

For 2MASS J1341$-$3052 and SDSS J1511+0607 only a single epoch of astrometry is available. Hence, we performed a simple Monte Carlo simulation to find the distributions of likely semi-major axes and periods for these systems. Following the procedure described in~\citet{2015AJ....149..104B}, we created random uniformly-distributed vectors for eccentricity $0<\epsilon<0.6$~\citep{2011ApJ...733..122D},  inclination $0< \mathrm{sin}~i< 1$, longitude of ascending node $0<\Omega<2\pi$, argument of periapse $0<\omega<2\pi$, and mean anomaly angle $0<M<2\pi$ for $10^5$ hypothetical orbits with a fixed semi-major axis of $a = 1$ AU. We numerically solved the Kepler equation to find the eccentric anomaly, calculated the Thiele-Innes constants~(\citealt{1907Obs....30..310I,1927BAN.....3..261V}), and found the $x$ and $y$ projected positions on the sky leading to the total projected separation, $r_{tot}$. The distributions of semi-major axes for the resolved systems were inferred by transforming variables:

\begin{equation}
a = (1 \mathrm{AU})\times\frac{\rho d}{r_{tot}}
\end{equation}

where $a$ is the semi-major axis in AU, $\rho$ is the angular separation in arc seconds, $d$ is the distance to the system in parsecs and $r_{tot}$ is in AU. The observed projected separation, $r_{obs} = \rho d$, constrains the array of allowed orbits, $r_{tot}$, and as a result we arrive at a distribution of probable semi-major axes, $a$.

\begin{figure}
\centering
\subfloat[][\emph{2MASS J1341$-$3052}.]
{\includegraphics[width=.5\textwidth]{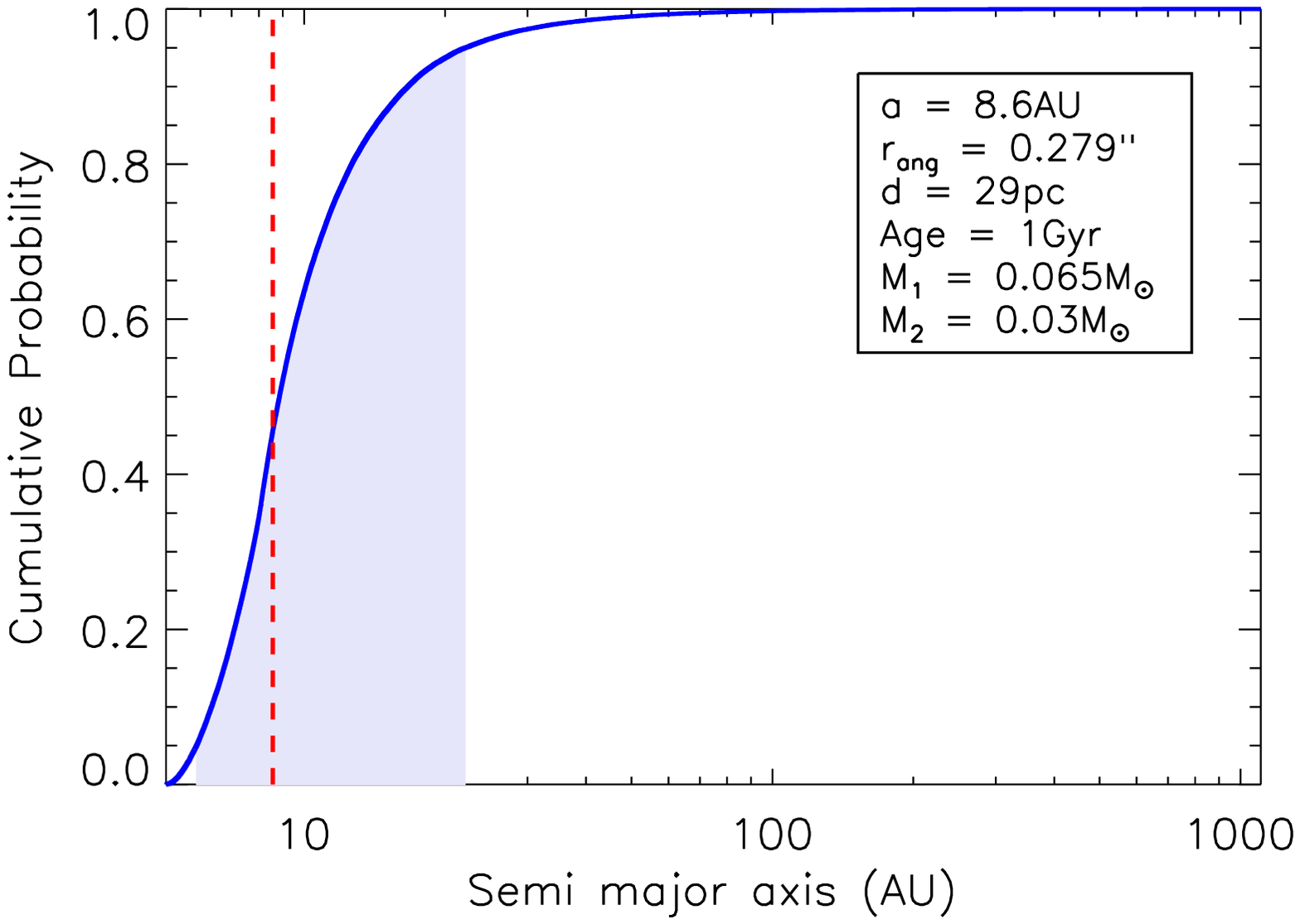}\includegraphics[width=.5\textwidth]{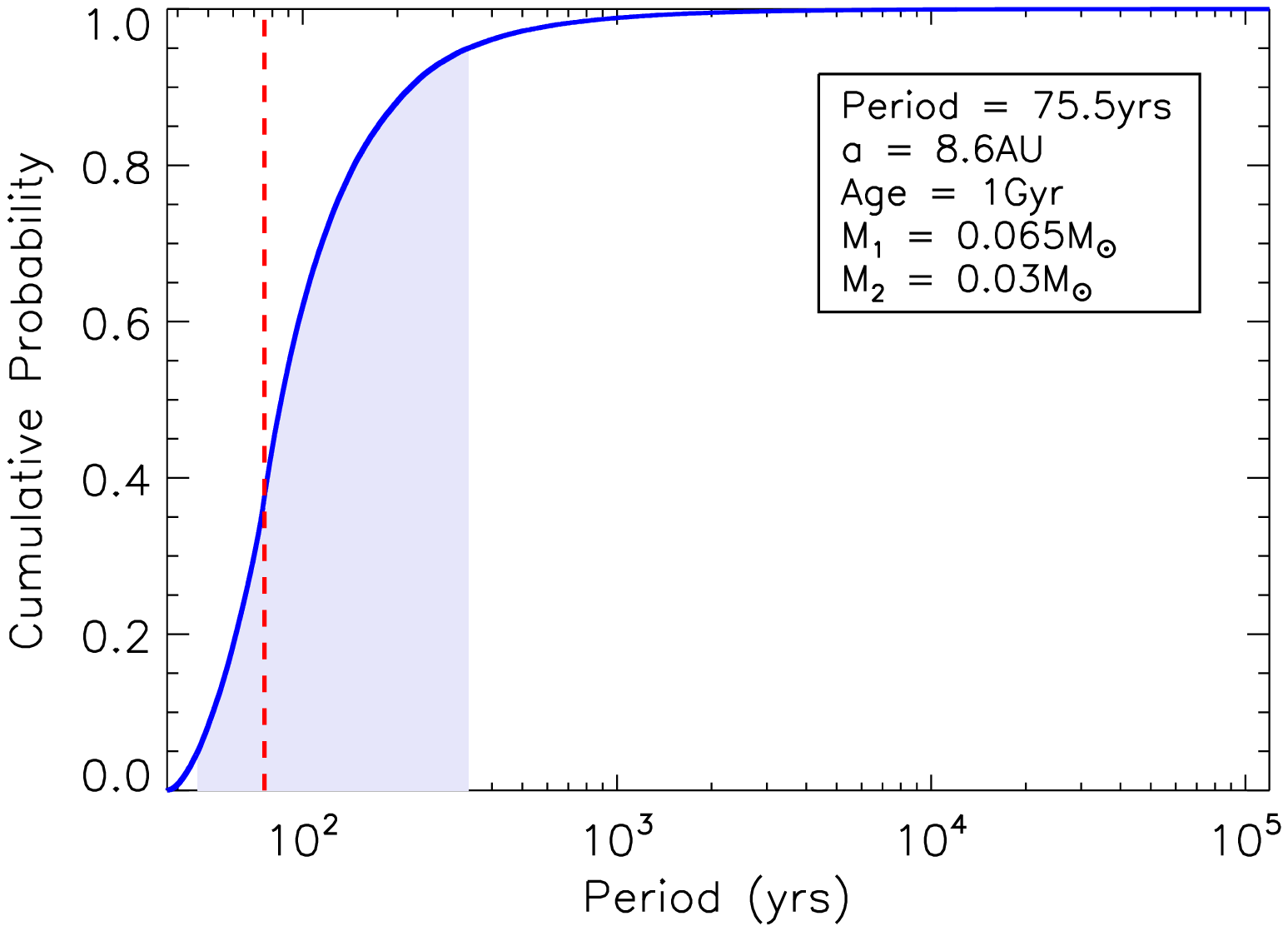}}\\
\subfloat[][\emph{SDSS J1511+0607}.]
{\includegraphics[width=.5\textwidth]{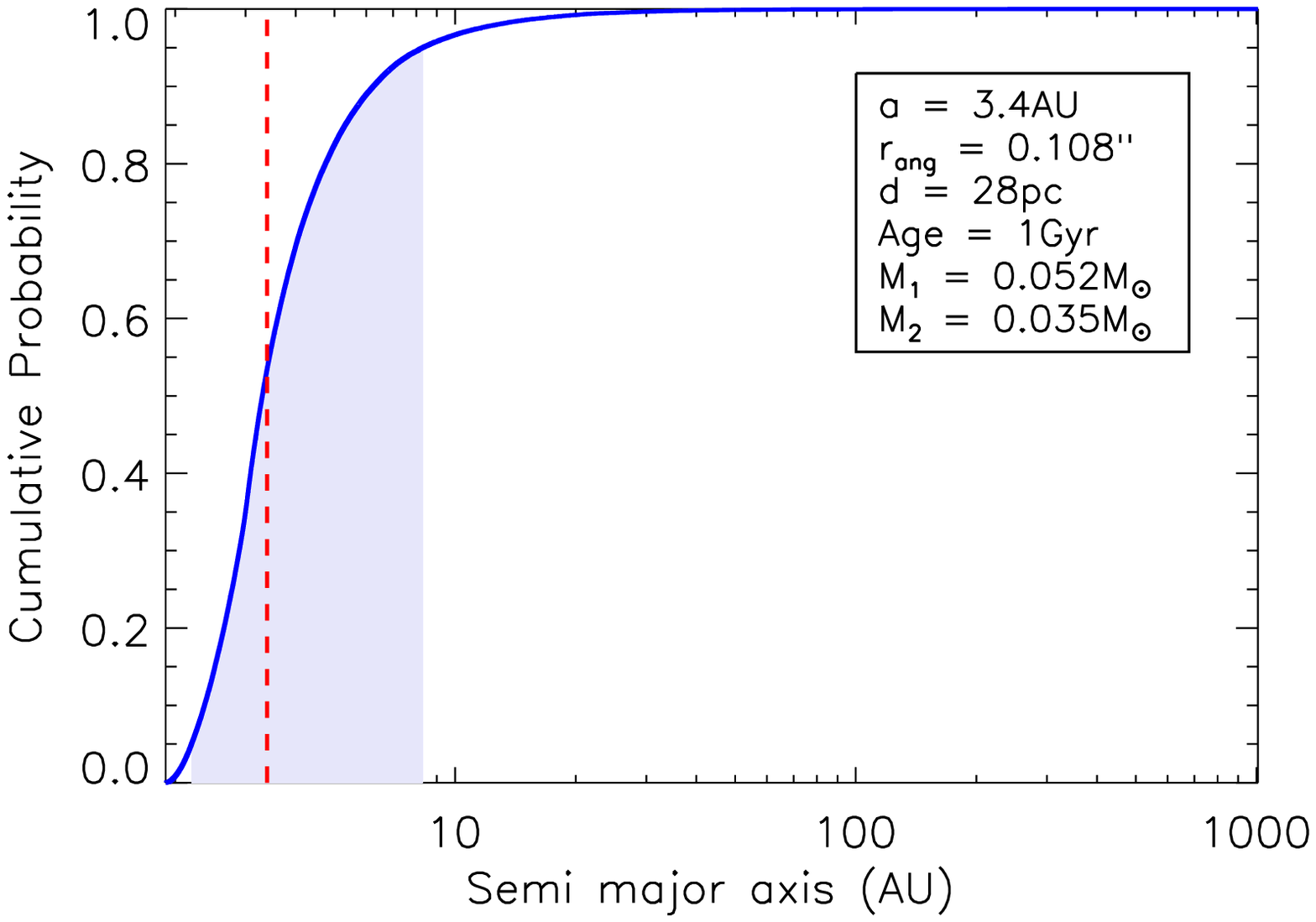}\includegraphics[width=.5\textwidth]{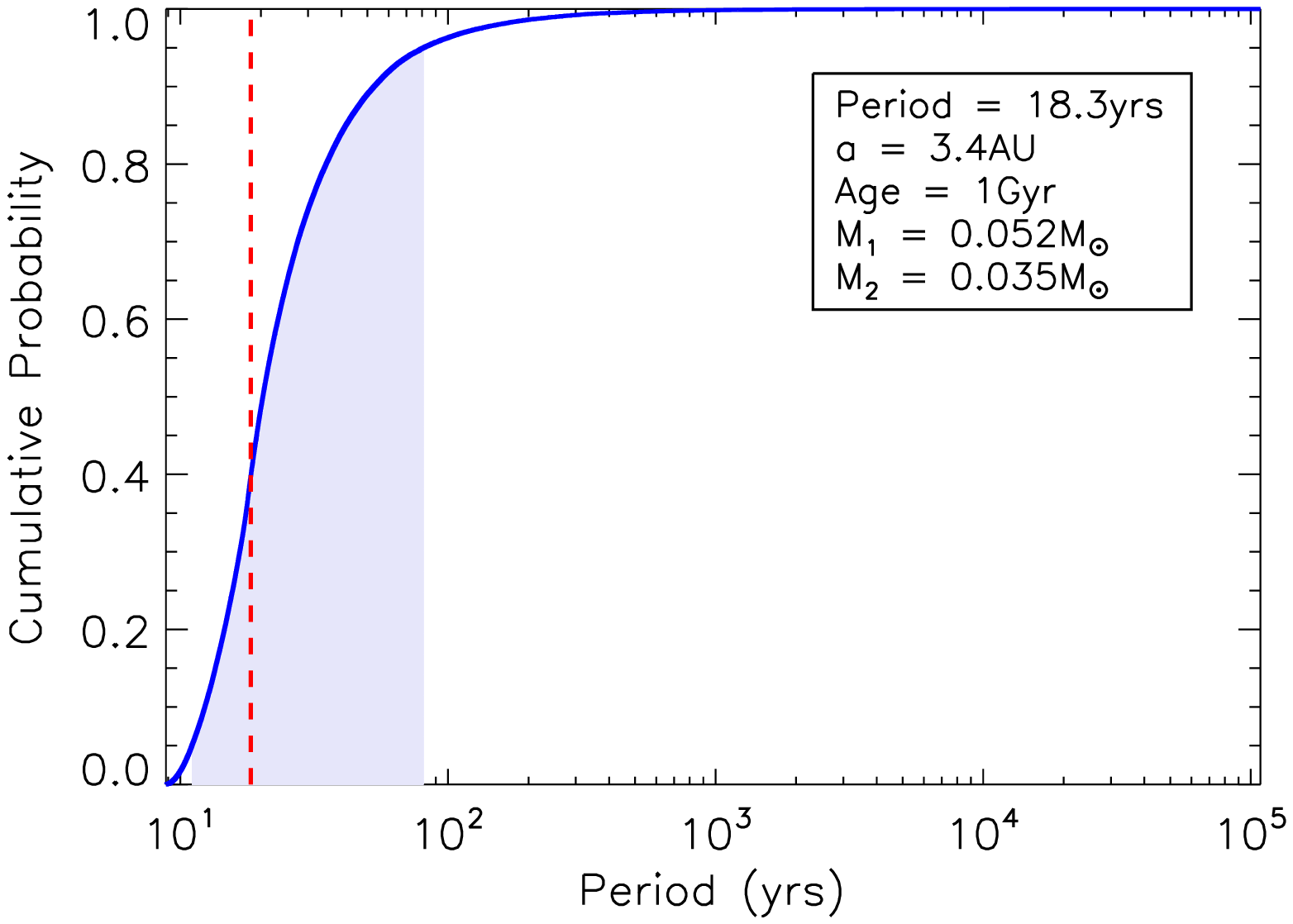}}\\
\caption{Cumulative probability distribution of possible semi-major axes and periods for two of the resolved sources from a Monte Carlo simulation for an age of 1Gyr, using one single observation for projected separation from the LGS-AO images in each case. The simulation parameters are shown on the box in the upper right corner. The most likely semi-major axis and period is represented by the dotted red line. The shaded regions indicate the central 68\% ($\pm1\sigma$ equivalent) of data points.}
\label{fig:montecarlo}
\end{figure}

The cumulative probability distributions for the semi-major axes of 2MASS J1341$-$3052 and SDSS J1511+0607 are shown in Figure~\ref{fig:montecarlo}. The most likely semi-major axes are demarcated by a dashed red line and the central 68\% ($\pm1\sigma$ equivalent) of the data are shaded in lavender. We estimated the periods for these orbits in years assuming $P^2 = a^3/M_{tot}$, with $M_{tot} = M_1+M_2$ in Solar masses estimated from the models of~\citet{2003A&A...402..701B} for ages of 0.5, 1, 5 and 10 Gyr (Table~\ref{tab:masses}). For 2MASS J1341$-$3052, we obtain most likely semi-major axis and period of $8.6^{+5.2}_{-1.8}$ AU and 63$-$85 years for decreasing ages, while for SDSS J1511+0607 the most likely semi-major axis and periods are $3.4^{+1.8}_{-0.8}$ AU and 15$-$21 years.




\subsection{On the frequency of short period VLM binaries}\label{sec:discussion}

Starting from a sample of LGS-AO observations of 43 brown dwarfs, we resolved 3 of 17 spectral binary candidates and none of the other targets. 2MASS J1733$-$1654 has a particularly asymmetrical PSF shape which could indicate the presence of a marginally resolved companion NE of the primary, but this is inconclusive from our images. The fraction of resolved systems from the spectral binary sample is $3/17 = 18^{+13}_{-6}\%$~(binomial uncertainties), which is consistent with the observed binary fractions reported in the literature from imaging programs~($10-20\%$;~\citealt{2003AJ....126.1526B, 2003IAUS..211..249C, 2006AJ....132..663B, 2007ApJ...668..492A, 2007ApJ...659..655B, 2012ApJ...757..141K}). Among the spectral binaries, 2MASS J0518$-$2828, WISEP J0720$-$0846, SDSS J0805+4812, 2MASS J1106+2754 and 2MASS J1209$-$1004 are known binaries unresolved in our images. This indicates a minimum binary fraction for spectral binaries in this sample of $(3+5)/17 = 47^{+12}_{-11}\%$. This is considerably higher than typical VLM binary search samples and indicates that the spectral binary sample is positively biased towards binaries. The fact that over half of the known binaries in this sample are not resolved suggests that imaging programs are similarly missing binaries, and that the true binary fraction may be significantly higher than what is currently reported. Note that the fraction reported here remains a lower limit; an unknown number of the 9 unresolved and unconfirmed spectral binaries may be true binaries with separations $\lesssim1.5-9.2$ AU.

Of the 12 confirmed spectral binaries to date (Table~\ref{tab:confSB}), about half have been unresolved in reported LGS-AO imaging, which is roughly consistent with the 3 resolved and 5 unresolved binaries in our sample. While follow-up of unresolved systems is more time-consuming and resource-intensive (radial velocity and astrometric monitoring), we speculate that the number of unresolved but confirmed spectral binaries will increase as follow-up is completed.
A high incidence of unresolved but confirmed spectral binaries implies a high incidence of currently unresolved binaries in general. For example, if all of the unresolved spectral binaries in our sample actually are binaries, this would indicate a ratio of unresolved-to-resolved systems of 4.7:1. Given that the resolved binary fraction is roughly $15\%$, this rate of unresolved pairs would imply an overall binary fraction of over $60\%$. It is more likely that the current pool of spectral binary candidates contains some number of contaminants, such as blue L dwarfs (BG14) and variable brown dwarfs~\citep{2013AJ....145...71K,2012ApJ...750..105R} which will need to be identified through more detailed spectral and photometric variability analysis. Nevertheless, it is important to note that all 17 spectral binary systems studied here have resolved or upper limit separations near or below the peak of the resolved binary separation of VLM dwarfs~\citep{2007ApJ...668..492A}. This is strong evidence that a large number of VLM binaries are being missed in current imaging surveys.

\section{Summary}\label{sec:summary}

We have observed 43 late-M, L and T dwarfs with Keck/NIRC2 LGS-AO, including 17 spectral binary candidates with high resolution Keck/NIRC2 LGS-AO imaging. Three sources were resolved: 2MASS J1341$-$3052, SDSS J1511+0607 and SDSS J2052$-$1609, while five other confirmed binaries were unresolved. Only one of our spectral binary candidates, 2MASS J1733$-$1654, has a candidate faint companion on the images, requiring confirmation. We used relative photometry to update the estimated component spectral types of our resolved systems. For SDSS J2052$-$1609, we combined our observations with those of~\citet{2011A&A...525A.123S} and archival data to make a preliminary determination of orbital parameters, finding a period of $33^{+4}_{-2}$ years and system mass of 0.0823$^{+0.0037}_{-0.0017}$M$_{\odot}$ consistent with a relatively young system (0.4$-$1.4 Gyr). For 2MASS J1341$-$3052 and SDSS J1511+0607, we estimated their most likely semi-major axes and periods based on their observed angular separations, distances and estimated masses. 2MASS J1341$-$3052 has a projected separation of 8.1$\pm$0.5 AU and a period in the range of $63-85$ years depending on the ages. For SDSS J1511+0607, we estimate a projected separation of 2.9$\pm$0.3 AU and a period in the $17-25$ years range.

For the remaining 14 unresolved spectral binaries we calculated separation limits based on their estimated component magnitude differences from template fitting and empirical sensitivity curves. Five of these unresolved systems are confirmed binaries with measured angular separations smaller than our upper limits, and therefore consistent. The other 9 unresolved systems have upper limits in angular separation of $0\farcs04-0\farcs28$, corresponding to projected separations limits of $1.5-9.2$AU.


The binary fraction of the spectral binary candidates in this sample is $47^{+12}_{-11}\%$, significantly higher than those from prior imaging surveys ($10-20\%$;~\citealt{2003AJ....126.1526B,2003ApJ...586..512B,2003AJ....126.2421C}) and the overall binary fraction ($20-25\%$;~\citealt{2006AJ....132..663B,2007ApJ...659..655B,2008A&A...492..545J}). While this sample is clearly biased towards binary systems, the high percentage of unresolved systems suggests that there may exist a large population of very tight brown dwarf binaries that cannot be confirmed with high resolution imaging. Confirmation of the 9 remaining unresolved spectral binaries depends on high resolution radial velocity or astrometric variability measurements that are currently ongoing. If these unresolved sources turn out to be binaries, a great advantage of their short projected separation is the high likelihood for full orbit and dynamical mass determinations. In any case, an unbiased, volume-limited sample of spectral binaries with complete follow-up is required in order to find the true underlying binary fraction.

\subsection*{Acknowledgements}
We would like to thank our referee Sandy Leggett for her helpful comments. The authors thank observing assistants Jason McIlroy, Heather Hershley, Terry Stickel, and Gary Puniwai and support astronomers Hien Tran, Luca Rizzi, Marc Kassis, and Al Conrad for their assistance during our observations. DBG acknowledges funding support from the IPAC graduate fellowship. DBG and AJB acknowledge support from the National Aeronautics and Space Administration under Grant No. NNX15AI75G. CRG was partially supported by a NASA Keck PI Data Award, administered by the NASA Exoplanet Science Institute. This research has made use of the Keck Observatory Archive (KOA), which is operated by the W. M. Keck Observatory and the NASA Exoplanet Science Institute (NExScI), under contract with the National Aeronautics and Space Administration. This publication makes use of data from the SpeX Prism Spectral Libraries, maintained by Adam Burgasser at \url{http://www.browndwarfs.org/spexprism}; the Dwarf Archives Compendium, maintained by Chris Gelino, Davy Kirkpatrick, Mike Cushing, David Kinder and Adam Burgasser at \url{http://DwarfArchives.org}; and the VLM Binaries Archive maintained by Nick Siegler at \url{http://vlmbinaries.org}. The authors wish to recognize and acknowledge the very significant cultural role and reverence that the summit of Mauna Kea has always had within the indigenous Hawaiian community. We are most fortunate to have the opportunity to conduct observations from this mountain.\\
\indent \emph{Facilities:} Keck NIRC2.


\begin{landscape}
\begin{deluxetable}{lccccccccr}
\tabletypesize{\footnotesize}
\tablecaption{Observation Log\label{tab:obs}}
\tablewidth{0pt}
\tablehead{
  \colhead{Name} & 
  \colhead{SpT} &  
  \colhead{2MASS $H$} & 
  \colhead{Date} & 
  \colhead{Reference Star} & 
  \colhead{Filter} &
  \colhead{$t_{exp}$(s)} &
  \colhead{Airmass} &
  \colhead{Strehl Ratio} &
  \colhead{S/N}}
\startdata
\cutinhead{Spectral Binaries from B10 and BG14}
SDSS J011912.22+240331.6    & T2 & $16.46\pm0.03$ & 2009 Aug 15 & 1140-0016097 & $H$ & 720  & 1.02 & 0.25 & 717\\
					        & & & 2013 Sep 22 & 1140-0016097 & $H$ & 120  & 1.01 & 0.01 & 103\\
						& & & 2013 Sep 23 & 1140-0016097 & $H$ & 120 & 1.05 & 0.23 & 897\\
SDSS J024749.90$-$163112.6 & T2 & $16.31\pm0.03$ & 2009 Aug 15 & 0734-0037544 & $H$ & 720  & 1.32 & 0.11 & 764\\
2MASS J03440892+0111251   & L0.5 & $13.91\pm0.04$ & 2013 Sep 22 & 0911-0037820 & $H$ & 175 & 1.07 & 0.15 & 3043\\
SDSS J035104.37+481046.8    & T1 & $15.57\pm0.14$ & 2009 Aug 15 & 1381-0118655 & $H$ & 720 & 1.38 & 0.14 & 894\\
2MASS J05185995$-$2828372	& T1 & $14.83\pm0.07$ & 2013 Sep 23 & 0615-0055796 & $H$ & 120  & 1.63 & 0.12 & 1718\\
WISE J07200320$-$0846513	& M9.5 & $9.92\pm0.02$ & 2014 Jan 19 & 0812-0137371 & $H$ & 60 & 1.28 & 0.02 & 1638\\
SDSS J080531.84+481233.0  	& L9 & $13.92\pm0.04$ & 2010 Mar 24 & 1382-0223846 & $K_s$ & 540 & 1.21 & 0.12 & 2624\\
					  	& & & 2013 Sep 23 & 1382-0223846 & $H$ & 120 & 1.61 & 0.09 & 3029\\
2MASS J11061197+2754225  	& T2.5 & $14.15\pm0.05$ & 2010 Mar 24 & 1179-0233699 & $H$ & 720 & 1.03 & 0.07 & 2559\\
2MASS J12095613$-$1004008	 & T3 & $15.33\pm0.09$ & 2014 Jan 13 & 0799-0230529 & $H$ & 120 & 1.15 & 0.04 & 1146\\
2MASS J13411160$-$3052505	 & L3 & $13.72\pm0.03$ & 2014 Jan 13 & 0591-0304901 & $J$ & 120 & 1.61 & 0.02 & 807\\
						& & & 2014 Jan 13 & 0591-0304901 & $H$ & 120 & 1.65 & 0.04 & 1508\\
						& & & 2014 Jan 19 & 0591-0304901 & $H$ & 120 & 1.59 & 0.04 & 1845\\
						& & & 2014 Jan 13 & 0591-0304901 & $K_s$ & 180 & 1.59 & 0.05 & 1861\\
SDSS J143553.25+112948.6    & T2 & $16.52\pm0.04$ & 2009 Aug 15 & 1014-0229971 & $H$ & 360 & 1.25 & 0.05 & 373\\
                                                & & & 2009 Aug 15 & 1014-0229971 & $K_s$ & 360 & 1.30 & 0.08 & 268\\
SDSS J151114.66+060742.9    & T0 & $14.96\pm0.08$ & 2009 Aug 15 & 0961-0243717 & $J$ & 360 & 1.36 & 0.01 & 560\\
  						& & & 2009 Aug 15 & 0961-0243717 & $H$ & 360 & 1.27 & 0.09 & 1078\\
            					& & & 2009 Aug 15 & 0961-0243717 & $K_s$ & 720 & 1.32 & 0.13 & 1101\\
        					& & & 2010 May 19 & 0961-0243717 & $H$ & 120 & 1.03 & 0.09 & 94\\
SDSS J151643.01+305344.4  	& T0.5 & $15.87\pm0.16$ & 2010 Mar 24 & 0930-0297471 & $H$ & 720 & 1.04 & 0.06 & 1007\\
SDSS J154727.23+033636.3  	& L2 & $15.07\pm0.06$ & 2009 Aug 15 & 0936-0258682 & $H$ & 720 & 1.40 & 0.01 & 805\\
2MASS J1711457+223204     	& L6.5 & $15.80\pm0.11$ & 2010 May 13 & 1125-0317350 & $J$ & 720 & 1.22 & 0.05 & 381\\
                                                & & & 2010 May 13 & 1125-0317350 & $H$ & 720 & 1.27 & 0.07 & 844\\
2MASS J1733423$-$165449    	& L0.5 & $12.81\pm0.06$ & 2009 Aug 15 & 0730-0518366 & $J$ & 720 & 1.39 & 0.08 & 2324\\
                                                & & & 2009 Aug 15 & 0730-0518366 & $H$ & 720 & 1.36 & 0.10 & 2748\\
SDSS J205235.31$-$160929.8 & T1 & $15.45\pm0.03$ & 2009 Aug 15 & 0738-0802833 & $J$ & 360 & 1.30 & 0.06 & 543\\
                                                & & & 2009 Aug 15 & 0738-0802833 & $H$ & 360 & 1.27 & 0.00 & 1003\\
                                                & & & 2009 Aug 15 & 0738-0802833 & $K_s$ & 360 & 1.28 & 0.17 & 729\\
\cutinhead{Visually Selected Spectral Binary Candidates}
2MASS J0019457+521317       & M9 & $12.07\pm0.02$ & 2009 Aug 15  & 1422-0011510 & $J$ & 360 & 1.30 & 0.09 & 2502\\
                                                & & & 2009 Aug 15 & 1422-0011510 & $H$ & 720  & 1.28 & 0.25 & 5028\\
2MASS J00320509+0219017   & L1.5 & $13.39\pm0.02$ & 2009 Aug 15 & 0923-0006944  & $H$ & 720 & 1.10 & 0.11 & 2811\\
SDSS J003259.36+141036.6    & L8 & $15.65\pm0.14$ & 2009 Aug 15 & 1041-0005438  & $H$ & 720 & 1.02 & 0.10 & 1112\\
                                                & & & 2009 Aug 15 & 1041-0005438 & $K_s$ & 360  & 1.02 & 0.11 & 721\\
2MASS J02361794+0048548  & L6.5 & $15.27\pm0.07$ & 2013 Sep 22 & 0908-0027044 & $H$ & 180  & 1.06 & 0.14 & 1959\\
SDSS J075840.33+324723.4  	& T2 & $14.11\pm0.04$ & 2010 Mar  24 & 1227-0198441  & $H$ & 720 & 1.06 & 0.05 & 2333\\
SDSS J093109.56+032732.5  	& L7.5 & $16.27\pm0.24$ & 2010 Mar 24 & 0934-0195871 & $H$ & 720 & 1.18 & 0.03 & 611\\
2MASS J09490860$-$1545485 & T2 & $15.26\pm0.11$ & 2010 Mar 24 & 0742-0216967 & $H$ & 720 & 1.24 & 0.05 & 1399\\
SDSS J103321.92+400549.5  	& L6 & $16.05\pm0.04$ & 2010 Mar 24 & 1300-0206652 & $H$ & 720 & 1.07 & 0.06 & 828\\
SDSS J112118.57+433246.5  & L7.5 & $16.56\pm0.04$ & 2010 Mar 24 & 3015-00408-1 & $H$ &  720 & 1.16 & 0.08 & 674\\
2MASS J11582073+0435022   & sdL7 & $14.68\pm0.06$ & 2010 May 13 & 0945-0201532 & $H$ & 720 & 1.04 & 0.15 & 1776\\
                                                & & & 2010 May 13 & 0945-0201532 & $K_s$ & 720 & 1.04 & 0.23 & 1882\\
SDSS J120602.51+281328.7    & T3 & $15.83\pm0.03$ & 2010 Mar 24 & 1182-0220446 & $H$ & 720 & 1.11 & 0.10 & 197\\
2MASS J14283132+5923354   & L4 & $13.88\pm0.04$ & 2010 May 19 & 1493-0219430  & $H$ & 720 & 1.29 & 0.10 & 3219\\
2MASS J1707333+430130 	& L0.5 & $13.18\pm0.03$ & 2009 Aug 15 & 1330-0334426 & $H$ & 720 & 1.41 & 0.08 & 2425\\
2MASS J1721039+334415       & L3 & $12.95\pm0.03$ & 2009 Aug 15 & 1237-0272128 & $H$ & 720 & 1.21 & 0.11 & 953\\
                                                & & & 2010 Mar 24 & 1237-0272128 & $H$ & 720 & 1.03 & 0.07 & 2721\\
2MASS J21265916+7617440	& T0 & $13.59\pm0.04$ & 2013 Sep 23 & 1662-0097897  & $H$ & 120 & 2.16 & 0.08 & 79\\
SDSS J214956.55+060334    	& M9 & $12.63\pm0.03$ & 2009 Aug 15 & 0960-0568745 & $H$ & 720 & 1.08 & 0.09 & 2670\\
\cutinhead{Additional Targets}
WISEP J004701.06+680352.1   & L7.5 & $13.97\pm0.04$ & 2013 Sep 23 & 1580-0021858 & $H$ & 90 & 1.56 & 0.28 & 2410\\
2MASS J03001631+2130205   & L6p & $14.73\pm0.07$ & 2013 Sep 22 & 1115-0038422 & $H$ & 120 & 1.06 & 0.13 & 1583\\
2MASS J03020122+1358142   & L3 & $15.43\pm0.09$ & 2013 Sep 23 & 1039-0030470 & $H$ & 120 & 1.01 & 0.05 & 944\\
HYT0429+1535				& \nodata & \nodata & 2013 Sep 22 & 1031-0059287 & $CH_4$ & 120 & 1.01& 0.16 & 131\\
						& & & 2013 Sep 22 & 1031-0059287 & $H$ & 120  & 1.00 & 0.17 & 88\\
2MASSI J0443058$-$320209	& L5 & $14.35\pm0.06$ & 2013 Sep 22 & 0579-0075735 & $H$ & 120  & 1.67 & 0.07 & 2206\\
WISE J052857.69+090104.4	& M9.5p & $15.44\pm0.12$ & 2014 Jan 19 & 0990-0058827 & $H$ & 135  & 1.40 & 0.07 & 99\\
SDSS J115013.17+052012.3	& L6 & $15.46\pm0.14$ & 2014 Jan 13 & 0279-01016-1  & $H$  & 60 & 1.03 & 0.17 & 993\\
ULAS J132605.18+120009.9    & T6p & $17.93\pm0.09$ & 2010 May 19 & 1019-0249297 & $H$ & 720 & 1.02 & 0.24 & 371\\
2MASS J14140586+0107102   & L4.8: & $15.73\pm0.19$ & 2010 May 13 & 0317-00292-1 & $H$ & 720 & 1.07 & 0.20 & 1168\\
2MASS J20025073$-$0521524 & L6 & $14.28\pm0.05$ & 2009 Aug 15 &  0846-0581639 & $H$ & 360 & 1.35 & 0.12 & 1709\\
\enddata
\end{deluxetable}
\end{landscape}

\begin{deluxetable}{lll}
\tabletypesize{\footnotesize}
\tabletypesize{\scriptsize}
\tablecolumns{3} 
\tablewidth{0pt}
 \tablecaption{Updated index selection criteria for B10 and BG14.\label{tab:indices}}
 \tablehead{
    \colhead{$x$} & 
    \colhead{$y$} & 
    \colhead{Limits}}
 \startdata
 \cutinhead{\citet{2010ApJ...710.1142B}}
 H$_2$O-\emph{J} & H$_2$O-\emph{K} & $0.325<x<0.65$ and $y>0.615x+0.300$\\
 CH$_4$-\emph{H} & CH$_4$-\emph{K} & $0.6<x<1.0$ and $y>1.063x-0.288$\\
 CH$_4$-\emph{H} & $K/J$ & $0.65<x<1.00$ and $y>0.471x-0.096$\\
 H$_2$O-\emph{H} & $H$-dip & $0.44<x<0.68$ and $y<0.49$\\
 SpT & H$_2$O-\emph{J}/H$_2$O-\emph{H} & L8.5$<x<$T3.5, $y<0.925$ and $y<-0.037x+2.106$\\
 SpT & H$_2$O-\emph{J}/CH$_4$-\emph{K} & L8$<x<$T4.5 and $y<0.041x-0.517$\\
 \cutinhead{\citet{2014ApJ...794..143B}}
 SpT & CH$_4$-\emph{H} & $M7.5<x<L8$ and $y<-4.3\times10^{-4}x^{2} + 0.0253x + 0.6824$\\
 H$_2$O-\emph{J} & CH$_4$-\emph{H} & $0.60<x<0.92$ and $y<-0.094x+1.096$.\\
 H$_2$O-\emph{J} & \emph{H}-bump & $0.65<x<0.90$ and $y>0.16x+0.806$.\\
 CH$_4$-\emph{J} & CH$_4$-\emph{H} & $0.6<x<1.04$, $y<1.04$ and $y<-0.562x+1.417$.\\
 CH$_4$-\emph{J} & \emph{H}-bump & $0.60<x<0.74$, $y>0.91$ and $y>1.00x+0.24$.\\
 CH$_4$-\emph{H} & \emph{J}-slope & $0.94<x<1.03$, $y>1.03$ and $y>1.250x-0.207$.\\
 CH$_4$-\emph{H} & \emph{J}-curve & $0.95<x<1.03$ and $y>1.245x^2-1.565x+2.312$.\\
 CH$_4$-\emph{H} & \emph{H}-bump & $0.94<x<1.04$, $y>0.92$ and $y<1.36x^2-4.26x+3.877$.\\
 \emph{J}-slope & \emph{H}-dip & $1.03<x<1.13$ and $y<0.20x+0.27$.\\
 \emph{J}-slope & \emph{H}-bump & $1.025<x<1.130$, $y>-2.75x+3.84$ and $y>0.91$.\\
 \emph{K}-slope & H$_2$O-\emph{Y} & $0.93<x<0.96$ and $y>12.036x^2-20.000x+9.037$.\\
 \emph{J}-curve & \emph{H}-bump & $2.00<x<2.45$, $y>0.92$ and $y>0.269x^2-1.326x+2.527$.\\
\enddata
\end{deluxetable}


\begin{landscape}
\begin{deluxetable}{lccccccrcccc}
\tabletypesize{\footnotesize}
\tabletypesize{\scriptsize}
\tablecolumns{12} 
\tablewidth{0pt}
 \tablecaption{Projected separation constraints for spectral binaries and all other targets.\label{tab:seplim}}
 \tablehead{
  & & \multicolumn{2}{l}{Spectral Type}
 & & & & & & & \multicolumn{2}{c}{Separation}\\
\cline{2-4}
\cline{11-12}
 \colhead{Source} &
 \colhead{Optical\tablenotemark{*}} &
 \colhead{Primary}& 
 \colhead{Secondary}& 
  \colhead{2MASS $\Delta J$} & 
  \colhead{2MASS $\Delta H$} & 
  \colhead{2MASS $\Delta K$} & 
  \colhead{Confidence} &
  \colhead{Distance (pc)} & 
  \colhead{Ref.} &
  \colhead{Angular ($mas$)} &
  \colhead{Projected (AU)}}
\startdata
\cutinhead{Spectral Binaries}
SDSS J011912.22+240331.6 & T2.0 & T0.3$\pm$0.7 & T3.7$\pm$0.5 & -0.28$\pm$0.13 & 0.30$\pm$0.11 & 0.82$\pm$0.20 & $>99\%$ & 43$\pm$3 & 1 & $<43$ & $<1.9$\\
SDSS J024749.90$-$163112.6 & T2.0 & L8.4$\pm$0.6 & T5.7$\pm$0.5 & 0.57$\pm$0.30 & 1.62$\pm$0.33 & 2.10$\pm$0.30 & $100\%$ & 40$\pm$3 & 1 & $<96$ & $<3.8$\\
2MASS J03440892+0111251 & L0.5 & L0.3$\pm$0.4 & T3.8$\pm$1.5 & 2.77$\pm$0.38 & 3.10$\pm$0.65 & 3.50$\pm$0.70 & $98\%$ & 41$\pm$4 & 1 & $<226$ & $<9.2$\\
SDSS J035104.37+481046.8 & T1.0 & L6.3$\pm$0.7 & T5.4$\pm$0.9 & 0.45$\pm$0.47 & 1.62$\pm$0.51 & 2.45$\pm$0.52 & $>99\%$ & 37$\pm$4 & 1 & $<102$ & $<$3.8\\
2MASS J05185995$-$2828372 & \nodata & L6.4$\pm$0.4 & T5.6$\pm$0.5 & 0.54$\pm$0.25 & 1.75$\pm$0.27 & 2.48$\pm$0.28 & $100\%$ & 23$\pm$1\tablenotemark{\pi} & 4 & $<98$ & $<2.2$\\
WISE J07200320$-$0846513 & \nodata & M8.9$\pm$0.0 & T5.2$\pm$0.7 & 3.50$\pm$0.24 & 4.15$\pm$0.36 & 4.57$\pm$0.41 & $100\%$ & 6$\pm$1\tablenotemark{\pi} & 2 & $<500$ & $<3.0$\\	
SDSS J080531.84+481233.0 & L9.0 & L5.3$\pm$0.1 & T5.8$\pm$0.4 & 1.92$\pm$0.17 & 2.85$\pm$0.18 & 3.49$\pm$0.30 & $100\%$ & 23$\pm$1\tablenotemark{\pi} & 4 & $<164$ & $<3.8$\\	
2MASS J11061197+2754225 & T2.5 & L8.9$\pm$0.5 & T4.2$\pm$0.4 & -0.24$\pm$0.14 & 0.49$\pm$0.17 & 1.03$\pm$0.16 & $100\%$ & 21$\pm$1\tablenotemark{\pi} & 3 & $<74$ & $<1.5$\\
2MASS J12095613$-$1004008 & \nodata & T1.1$\pm$0.0 & T6.0$\pm$0.4 & 0.95$\pm$0.11 & 1.77$\pm$0.15 & 2.10$\pm$0.28 & $>99\%$ & 22$\pm$1\tablenotemark{\pi} & 4 & $<100$ & $<2.1$\\	
2MASS J13411160$-$30525049\tablenotemark{\dagger} & L3.0 & L2.3$\pm$0.6 & T6.0$\pm$1.0 & 2.68$\pm$0.08 & 4.03$\pm$0.12 & 4.23$\pm$0.07 & 96\% & 29$\pm$3 & 1 & 279$\pm$17 & $8.9\pm0.4$\\	
SDSS J143553.25+112948.6 & T2.0 & L8.9$\pm$0.7 & T5.6$\pm$0.5 & 0.28$\pm$0.24 & 1.38$\pm$0.28 & 2.11$\pm$0.37 & $>99\%$ & 44$\pm$4 & 1 & $<90$ & $<4.0$\\	
SDSS J151114.66+060742.9\tablenotemark{\dagger} & T0.0 & L5.2$\pm$0.9 & T4.9$\pm$0.4 & 0.25$\pm$0.13 & 1.41$\pm$0.13 & 2.38$\pm$0.30 & $>99\%$ & 28$\pm$5\tablenotemark{\pi} & 5 & 108$\pm$11 & 2.9$\pm$0.3\\	
SDSS J151643.01+305344.4 & T0.5	& L7.6$\pm$0.8 & T2.3$\pm$0.3 & -0.36$\pm$0.24 & 0.21$\pm$0.21 & 0.80$\pm$0.27 & 99\% & 39$\pm$4 & 1 & $<70$ & $<2.7$\\		
SDSS J154727.23+033636.3 & L2.0 & L1.8$\pm$0.2 & T6.6$\pm$1.0 & 3.32$\pm$0.53 & 4.38$\pm$0.60 & 4.98$\pm$0.65 & 90\% & 53$\pm$6 & 1 & Unconstrained & Unconstrained\\	
2MASS J1711457+223204 & L6.5 & L5.5$\pm$0.5 & T5.3$\pm$1.0 & 1.21$\pm$0.41 & 2.29$\pm$0.56 & 3.08$\pm$0.64 & $>99\%$ & 30$\pm$4\tablenotemark{\pi} & 4 & $<156$ & $<4.7$\\
2MASS J1733423$-$165449 & L0.5 & L0.1$\pm$0.2 & T4.0$\pm$1.3 & 2.76$\pm$0.30 & 3.14$\pm$0.54 & 3.59$\pm$0.60 & $98\%$ & 24$\pm$2 & 1 & $<219$ &$<5.3$\\	
SDSS J205235.31$-$160929.8\tablenotemark{\dagger} & T1.0 & L5.9$\pm$1.6 & T2.1$\pm$0.5 & 0.03$\pm$0.25 & 0.44$\pm$0.30 & 1.12$\pm$0.40 & $>99\%$ & 30$\pm$1\tablenotemark{\pi} & 4 & 103$\pm$2 & 3.2$\pm$0.5\\
\cutinhead{Visual Spectral Binaries}
2MASSI J0019457+521317 & M9.0 & M8.5$\pm$0.2 & T6.9$\pm$1.1 & 4.82$\pm$0.54 & 5.53$\pm$0.65 & 5.91$\pm$0.71 & 50\% & 20$\pm$2 & 1 & Unconstrained & Unconstrained\\
2MASS J00320509+0219017 & L1.5	& L0.1$\pm$0.3 & T6.4$\pm$1.4 & 4.06$\pm$0.68 & 4.87$\pm$0.83 & 5.36$\pm$0.85 & 49\% & 33$\pm$4 & 1 & Unconstrained & Unconstrained\\
SDSS J003259.36+141036.6 & L8.0 & L6.2$\pm$0.7 & T2.4$\pm$1.9 & 0.55$\pm$0.70 & 1.03$\pm$0.95 & 1.54$\pm$0.97 & 50\% & 33$\pm$6\tablenotemark{\pi} & 4 & $<96$ & $<3.2$\\
SDSSp J023617.93+004855.0 & L6.5 & L5.1$\pm$0.5 & T1.9$\pm$1.2 & 1.04$\pm$0.49 & 1.33$\pm$0.59 & 1.81$\pm$0.58 & 7\% & 39$\pm$4 & 1 & $<95$ & $<3.7$\\
SDSS J075840.33+324723.4 & T2.0 & T2.3$\pm$0.0 & T2.2$\pm$0.0 & -0.28$\pm$0.0 & -0.07$\pm$0.0 & 0.193$\pm$0.0& $92\%$ & 16$\pm$2 & 1 & $<65$ & $<1.0$ \\ 
SDSS J093109.56+032732.5 & L7.5 & L7.2$\pm$0.3 & T6.6$\pm$1.8 & 2.66$\pm$0.49 & 3.45$\pm$0.78 & 3.87$\pm$0.91 & 90\% & 37$\pm$4 & 1 & $<288$ & $<10.7$\\ 
2MASS J09490860$-$1545485 & T2.0 & T1.1$\pm$0.2 & T3.5$\pm$2.0 & 0.54$\pm$1.01 & 0.83$\pm$1.28 & 0.83$\pm$1.26 & 95\% & 18$\pm$2\tablenotemark{\pi} & 5 & $<101$ & $<1.8$\\ 
SDSS J103321.92+400549.5 & L6.0 & L4.9$\pm$0.6 & T4.4$\pm$2.2 & 1.96$\pm$0.72 & 2.52$\pm$1.02 & 2.99$\pm$1.07 & 84\% & 54$\pm$6 & 1 & $<206$ & $<11.2$\\ 
SDSS J112118.57+433246.5 & L7.5 & L6.8$\pm$0.5 & T5.0$\pm$1.5 & 1.52$\pm$0.64 & 2.21$\pm$0.80 & 2.74$\pm$0.84 & $95\%$ & 52$\pm$6 & 1 & $<164$ & $<8.4$\\ 
2MASS J11582077+0435014 & sdL7 & L6.4$\pm$0.0 & T2.6$\pm$0.8 & 1.42$\pm$0.12 & 1.57$\pm$0.22 & 1.98$\pm$0.26 & $100\%$ & 28$\pm$2 & 1 & $<93$ & $<2.5$\\ 
SDSS J120602.51+281328.7 & T3.0 & T2.0$\pm$0.4 & T5.0$\pm$0.8 & 0.46$\pm$0.35 & 1.05$\pm$0.49 & 1.26$\pm$0.50 & 74\% & 29$\pm$3 & 1 & $<87$ & $<2.5$\\ 
2MASS J14283132+5923354 & L4.0 & L4.4$\pm$0.7 & T6.4$\pm$2.0 & 3.30$\pm$0.72 & 4.24$\pm$1.00 & 4.77$\pm$1.06 & 68\% & 21$\pm$3 & 1 & Unconstrained & Unconstrained\\
2MASS J17073334+4301304 & L0.5 & M8.7$\pm$0.1 & T6.9$\pm$0.7 & 4.41$\pm$0.47 & 5.28$\pm$0.51 & 5.75$\pm$0.53 & 60\% & 35$\pm$3 & 1 & Unconstrained & Unconstrained\\
2MASS J1721039+334415 & L3.0 & L2.5$\pm$0.0 & T3.8$\pm$2.0 & 2.91$\pm$0.62 & 3.10$\pm$0.99 & 3.32$\pm$1.09 & 99\% & 19$\pm$2 & 1 & $<270$ & $<5.1$\\ 
2MASS J21265916+7617440 & T0.0 & L8.5$\pm$1.0 & T4.5$\pm$2.0 & 0.42$\pm$0.82 & 1.21$\pm$1.01 & 1.75$\pm$1.03 & 63\% & 12$\pm$2 & 1 & $<106$ & $<1.3$\\
SDSS J214956.55+060334 & M9.0 & M8.2$\pm$0.0 & T6.6$\pm$1.2 & 4.83$\pm$0.59 & 5.53$\pm$0.70 & 5.91$\pm$0.74 & 58\% & 29$\pm$3 & 1 &  Unconstrained & Unconstrained\\
\enddata
\tablenotetext{*}{Unresolved or combined optical spectral type.}
\tablenotetext{\dagger}{Resolved binary with \emph{measured} delta magnitudes.}
\tablenotetext{\pi}{Parallactic distance. Otherwise, spectrophotometric distance, assuming relative magnitudes from template fitting.}
\tablerefs{(1) This paper; (2)~\citet{2015AJ....149..104B}; (3)~\citet{2013A&A...560A..52M}; (4)~\citet{2012ApJS..201...19D}; (5)~\citet{2012ApJ...752...56F}.}
\end{deluxetable}
\end{landscape}

\begin{deluxetable}{lccc}
\tabletypesize{\footnotesize}
\tabletypesize{\scriptsize}
\tablecolumns{4} 
\tablewidth{0pt}
 \tablecaption{Properties of three resolved binary systems.\label{tab:binprops}}
 \tablehead{
  \colhead{Parameter} &
  \colhead{2MASS J1341$-$3052} &
  \colhead{SDSS J1511+0607}& 
  \colhead{SDSS J2052$-$1609}}
\startdata
Primary SpT & L2.5$\pm$1.0 & L5.5$\pm$1.0 & L6.0$\pm$2.0\\
Secondary SpT & T6.0$\pm$1.0 & T5.0$\pm$0.5 & T2.0$\pm$0.5\\
$\Delta J$ & 2.68$\pm$0.08 &  0.25$\pm$0.13 & 0.03$\pm$0.25\\
$\Delta H$ & 4.03$\pm$0.12 & 1.41$\pm$0.13 & 0.44$\pm$0.30\\
$\Delta K_s$ & 4.23$\pm$0.07 & 2.38$\pm$0.30 & 1.12$\pm$0.40\\
Distance (pc) & 29$\pm$3\tablenotemark{1} & 28$\pm$5\tablenotemark{2} & 30$\pm$1\tablenotemark{3}\\
Ang. Sep. ($mas$) & 279$\pm$17 & 108$\pm$11 & 103$\pm$2\\
Proj. Sep. (AU) & 8.1$\pm$0.5 & 2.9$\pm$0.3 & 3.0$\pm$0.1\\
PA ($^{\circ}$) & 317.9$\pm$0.6 & 335.0$\pm$4.3 & 68.4$\pm$1.1\\
Epoch (JD) & 2456671.15 & 2455058.77 & 2455058.94\\
\enddata
\tablenotetext{1}{This paper.}
\tablenotetext{2}{\citet{2012ApJ...752...56F}}
\tablenotetext{3}{\citet{2012ApJS..201...19D}}
\end{deluxetable}


\begin{deluxetable}{lclccl}
\tablecaption{Resolved Separation Measurements for SDSS~J2052$-$1609AB\label{tab:astrometry2052}}
\tabletypesize{\small}
\tablewidth{0pt}
\tablehead{
\colhead{UT Date} &
\colhead{JD} &
\colhead{Instrument} &
\colhead{$\Delta\alpha$} &
\colhead{$\Delta\delta$} &
\colhead{Ref}  \\
& & &
\colhead{(mas)} &
\colhead{(mas)} \\
}
\startdata
2005 Oct 11 & 2453654  & Keck/NIRC2 & 25.4$\pm$1.2 & 114.3$\pm$1.4 & 1,2 \\
2007 Apr 23 & 2454213 & Keck/NIRC2 &  54.6$\pm$2.1 & 88.8$\pm$1.5 & 1,2 \\
2008 Jun 24 & 2454642 & HST/NICMOS &  79.2$\pm$0.9 & 65.6$\pm$0.8 & 3 \\
2009 Jun 19 & 2455002 & VLT/NACO & 93.1$\pm$1.0 & 38.9$\pm$0.9 & 3\\
2009 Aug 15 & 2455058 & Keck/NIRC2 & 95.7$\pm$0.7 & 38.1$\pm$0.7 & 1 \\
2010 May 01 & 2455317 & Keck/NIRC2 &  103.4$\pm$1.7 & 20.6$\pm$1.5  & 1,4 \\
\enddata
\tablerefs{(1) This paper; (2) NIRC2 PI M.\ Liu; (3) \citet{2011A&A...525A.123S}; (4) NIRC2 PI B.\ Biller.}
\end{deluxetable}

\begin{deluxetable}{lll}
\tablecaption{Orbital Analysis of SDSS~J2052$-$1609AB Relative Astrometry \label{tab:orbit2052}}
\tabletypesize{\small}
\tablewidth{0pt}
\tablehead{
\colhead{Parameter} &
\colhead{Best} &
\colhead{Median} \\}
\startdata
Best $\chi^2$ (DOF)  & 12.05 (6) &  \nodata  \\
$P$\tablenotemark{a} (yr) & 32   & 33$^{+4}_{-2}$ \\
$a$ (AU) & 4.4 &  4.5$^{+0.5}_{-0.2}$ \\
$e$\tablenotemark{a} & 0.005 &  0.014$^{+0.023}_{-0.010}$ \\
$i$ ($\degr$) & 45 & 45$^{+5}_{-3}$ \\
$\omega$ ($\degr$) & 98 & 100$^{+15}_{-13}$ \\
$\Omega$ ($\degr$) & 327 & 327$^{+3}_{-4}$ \\
$M_0$ ($\degr$) & 318 & 313$^{+15}_{-8}$ \\
$d$\tablenotemark{a} (pc)  & 30.5 & 30.7$^{+0.2}_{-0.4}$ \\
$M_{tot}$ (M$_{\odot}$) & 0.081 & 0.0823$^{+0.0037}_{-0.0017}$ \\
\enddata
\tablenotetext{a}{Parameter was constrained to a limited value range in MCMC analysis.}
\end{deluxetable}

\begin{deluxetable}{lccccc}
\tabletypesize{\footnotesize}
\tabletypesize{\scriptsize}
\tablewidth{0pt}
 \tablecaption{Estimated masses and orbit results from Monte Carlo simulation.\label{tab:masses}}
 \tablehead{
 \colhead{System} & 
 \colhead{Age (Gyr)} &
 \colhead{Primary Mass (M$_{\bigodot}$)} &
 \colhead{Secondary Mass (M$_{\bigodot}$)} &
 \colhead{Semi-major Axis (AU)} &
 \colhead{Period (years)}}
 \startdata
{\bf 2MASS J1341$-$3052AB} & 0.5 & 0.052$\pm$0.005 & 0.022$\pm$0.005 & $8.6^{+5.2}_{-1.8}$ & $85^{+104}_{-20}$\\ 
SpT = L2.5$\pm$1.0 and T6.0$\pm$1.0 & 1.0 & 0.065$\pm$0.004 & 0.030$\pm$0.007 & $8.6^{+5.3}_{-1.8}$ & $71^{+91}_{-18}$\\
Primary T$_{eff}$ = 1904$\pm$165K & 5.0 & 0.075$\pm$0.002 & 0.054$\pm$0.007 & $8.6^{+5.3}_{-1.8}$ & $64^{+79}_{-15}$\\
Secondary T$_{eff}$ =1027$\pm$144K & 10 & 0.075$\pm$0.001 & 0.061$\pm$0.006 & $8.6^{+5.2}_{-1.8}$ & $63^{+76}_{-14}$\\
\hline
{\bf SDSS J1511+0607AB} & 0.5 & 0.041$\pm$0.004 & 0.026$\pm$0.003 & $3.4^{+1.8}_{-0.8}$ & $21^{+25}_{-5}$\\
SpT = L5.0$\pm$1.0 and T5.0$\pm$0.5 & 1.0 & 0.052$\pm$0.005 & 0.035$\pm$0.004 & $3.4^{+1.8}_{-0.9}$ & $18^{+22}_{-5}$\\
Primary T$_{eff}$ = 1617$\pm$139K & 5.0 & 0.070$\pm$0.002 & 0.059$\pm$0.004 & $3.4^{+1.7}_{-0.9}$ & $15^{+18}_{-3}$\\
Secondary T$_{eff}$ = 1115$\pm$107K & 10 & 0.072$\pm$0.001 & 0.065$\pm$0.003 & $3.4^{+1.8}_{-0.9}$ & $15^{+17}_{-4}$\\
\enddata
\end{deluxetable}

\begin{landscape}
\begin{deluxetable}{lcccccccccl}
\tabletypesize{\footnotesize}
\tabletypesize{\scriptsize}
\tablecolumns{11} 
\tablewidth{0pt}
 \tablecaption{Confirmed Spectral Binaries\label{tab:confSB}}
 \tablehead{
  & & \multicolumn{2}{c}{Spectral Type}\\
\cline{2-5}
 \colhead{Name} &
 \colhead{Combined} &
 \colhead{Combined} & 
 \colhead{Primary}& 
 \colhead{Secondary}& 
 \colhead{2MASS-$J$}& 
 \colhead{$J-K_s$}& 
 \colhead{$\Delta J$}& 
 \colhead{Separations} & 
 \colhead{Confirmation} &
 \colhead{Ref.}\\ 
  \colhead{}&
 \colhead{Optical} &
 \colhead{NIR} & 
 \colhead{} & 
 \colhead{} & 
 \colhead{} & 
 \colhead{} & 
 \colhead{} & 
 \colhead{(AU)} & 
 \colhead{method\tablenotemark{*}} & 
 \colhead{SB;Conf.}}
 \startdata 
SDSS J000649.16$-$085246.3 & M9 & \nodata  & M8.5$\pm$0.5 & T5$\pm$1 & 14.14$\pm$0.04 & 1.01$\pm$0.05 & 3.15$\pm$0.31 & 0.29$\pm$0.01 & RV  & 5; 5\\
2MASS J03202839$-$0446358  & M8: & L1 & M8.5$\pm$0.3 & T5$\pm$0.9 & 12.13$\pm$0.03 & 1.13$\pm$0.04 & 3.50$\pm$0.20 & 0.404$\pm$0.042 & RV & 8; 3\\
2MASS J05185995$-$2828372 & \nodata & \nodata & L8.6$\pm$0.3 & T6.4$\pm$1.0 & 15.98$\pm$0.10 & 1.82$\pm$0.12 & 0.13$\pm$0.19 & 1.8$\pm$0.5 & DI & 12; 10\\ 
WISEP J072003.20$-$084651.2 & M9.5 & \nodata & M8.9$\pm$0.0 & T5.2$\pm$0.7 & 10.63$\pm$0.02 & 1.16$\pm$0.03 & 3.30$\pm$0.20 & 0.84$\pm$0.17 & DI & 4; 4\\
SDSS J080531.84+481233.0 & L4 & L9.5 & L5.1$\pm$0.4 & T5.7$\pm$0.5 & 14.73$\pm$0.04 & 1.46$\pm$0.05 & 1.50$\pm$0.09 & 0.9-2.3 & AV & 9; 13\\
SDSS J092615.38+584720.9  & T4.5 & \nodata & T4.0$\pm$0.1 & T5.3$\pm$0.7 & 16.77$\pm$0.14 & $<1.57$ & 0.40$\pm$0.20 & 2.6$\pm$0.5 & DI & 16; 10, 11\\
2MASS J11061197+2754225 & \nodata & T2.5 & T0.4$\pm$0.3 & T4.0$\pm$0.8 & 14.82$\pm$0.04 & 1.02$\pm$0.07 & -0.37$\pm$0.06 & $<$2.67 & OL & 7, 14; 15\\
2MASS J12095613$-$1004008 & \nodata & T3	 & T1.2$\pm$0.3 & T5.9$\pm$0.6 & 15.91$\pm$0.08 & 0.85$\pm$0.16 & 1.50$\pm$0.20 & 4.8$\pm$0.2 & DI & 1; 18\\
2MASS J13153094$-$2649513 & L5 & \nodata & L4.9$\pm$0.5 & T6.1$\pm$2.1 & 15.07$\pm$0.05 & 1.63$\pm$0.07 & 3.03$\pm$0.03 & 6.6$\pm$0.9 & DI & 6; 6\\
2MASS J13411160$-$30525049 & L3 & \nodata & L2.3$\pm$0.6 & T6.0$\pm$1.0 & 14.61$\pm$0.03 & 1.53$\pm$0.04 & 3.28$\pm$0.53 & 7.8$\pm$0.5 & DI & 2; 1\\
SDSS J151114.66+060742.9  & \nodata & T0$\pm$2 & L5.2$\pm$0.9 & T4.9$\pm$0.5 & 16.02$\pm$0.08  & 1.47$\pm$0.13 & 0.54$\pm$0.32 & 2.9$\pm$0.3 & DI  & 7; 1\\
SDSS J205235.31$-$160929.8 & \nodata & T1$\pm$1 & L5.8$\pm$1.8 & T2.1$\pm$0.5 & 16.33$\pm$0.12 & 1.21$\pm$0.19 & 0.04$\pm$0.18 & 3.0$\pm$0.1 & DI & 7; 17\\
\enddata
\tablenotetext{*}{RV = Radial velocity variability, DI = Direct imaging, AV = Astrometric variability, OL = Overluminosity.}
\tablerefs{(1) This paper; (2)~\citet{2014ApJ...794..143B}; (3)~\citet{2008ApJ...678L.125B}; (4)~\citet{2015AJ....149..104B}; (5)~\citet{2012ApJ...757..110B}; (6)~\citet{2011ApJ...739...49B}; (7)~\citet{2010ApJ...710.1142B}; (8)~\citet{2008ApJ...681..579B};  (9)~\citet{2007AJ....134.1330B}; (10)~\citet{2006ApJS..166..585B}; (11)~\citet{2011ApJ...743..141C}; (12)~\citet{2004ApJ...604L..61C}; (13)~\citet{2012ApJS..201...19D}; (14)~\citet{2008ApJ...685.1183L}; (15)~\citet{2013A&A...560A..52M}; (16)~\citet{2008ApJ...676.1281M}; (17)~\citet{2011A&A...525A.123S}; (18)~\citet{2010ApJ...722..311L}.}
\end{deluxetable}
\end{landscape}


\clearpage


\begin{figure}[htp]
\centering
\subfloat[][\emph{2MASS J1341-3052 $J$}.]
{\includegraphics[trim=0 0 140 0, clip, width=.3\textwidth]{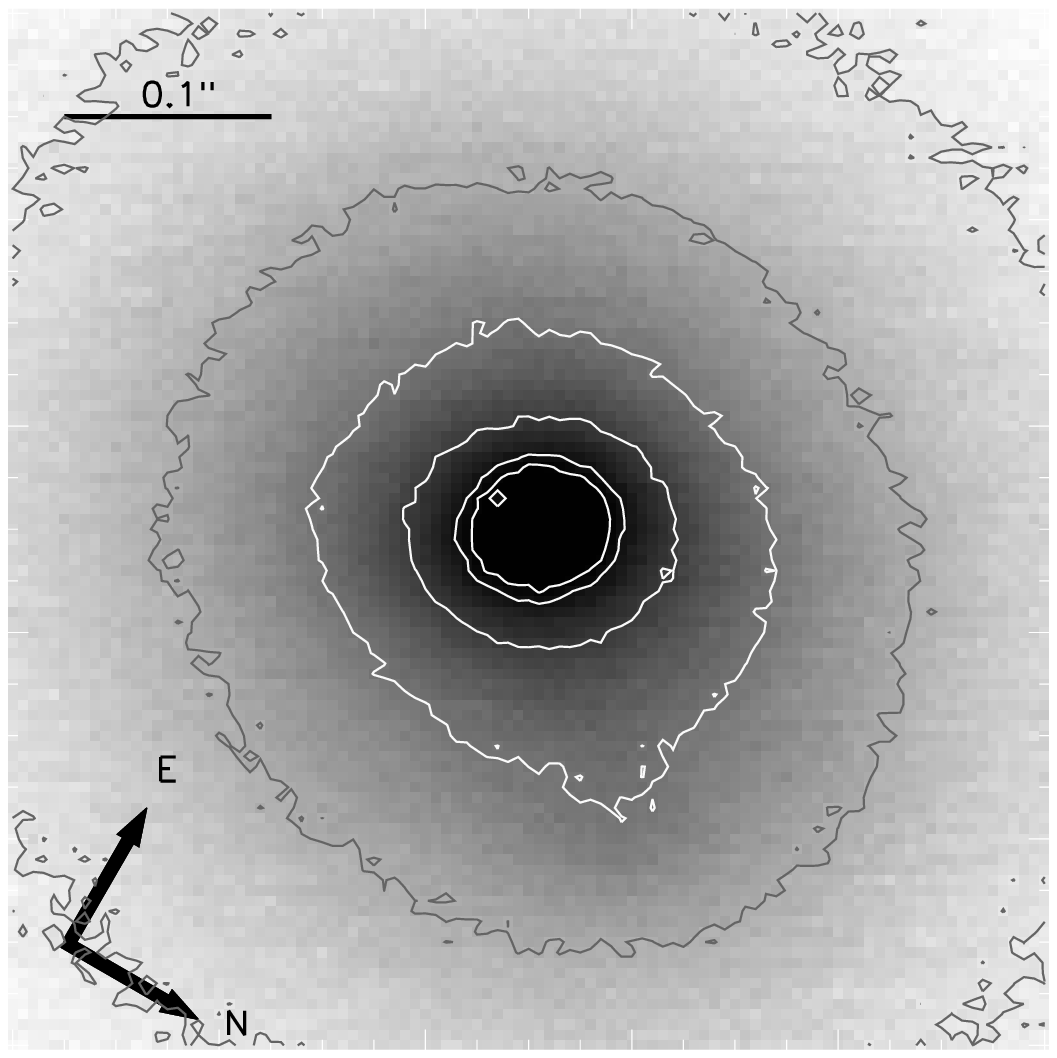}} \quad
\subfloat[][\emph{2MASS J1341-3052 $H$}.]
{\includegraphics[trim=0 0 140 0, clip, width=.3\textwidth]{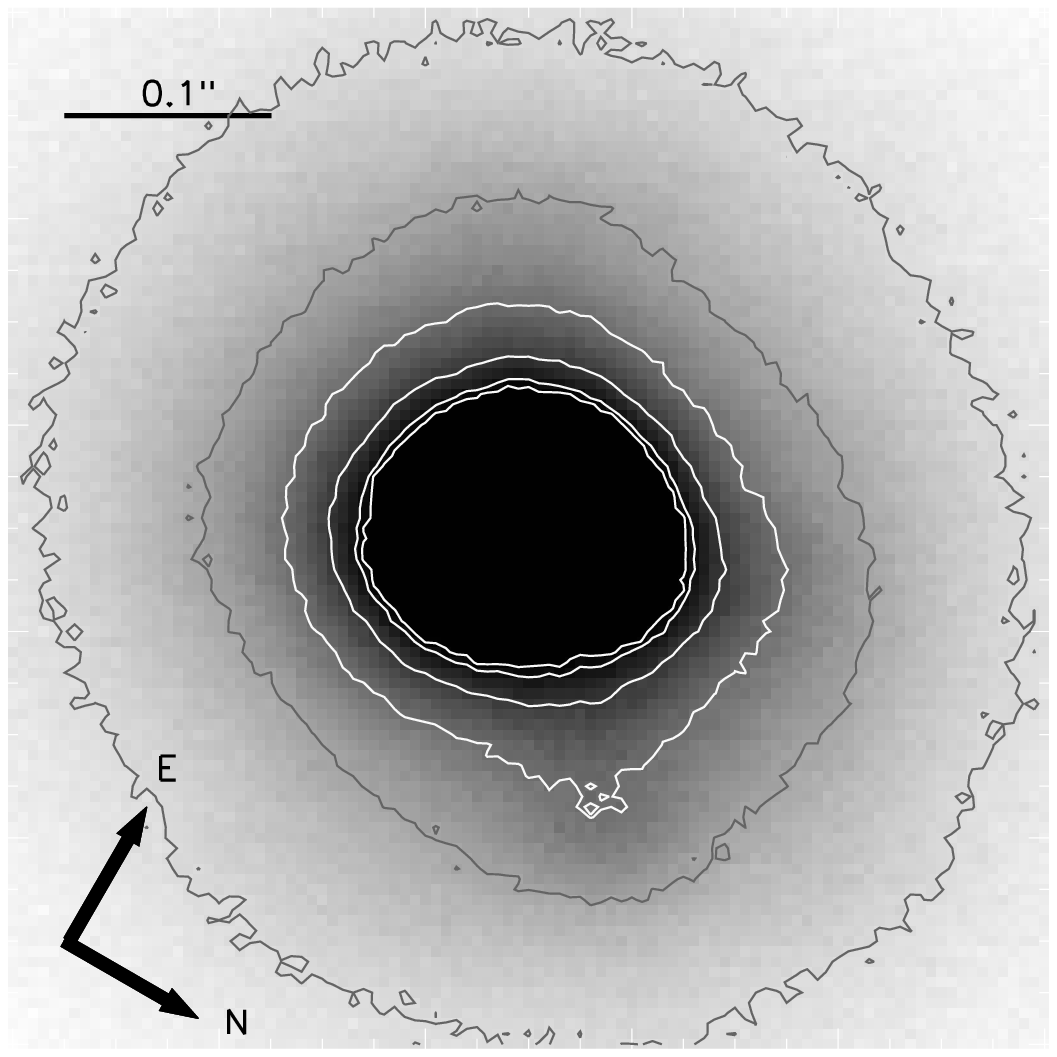}} \quad
\subfloat[][\emph{2MASS J1341-3052 $K_s$}.]
{\includegraphics[trim=0 0 140 0, clip, width=.3\textwidth]{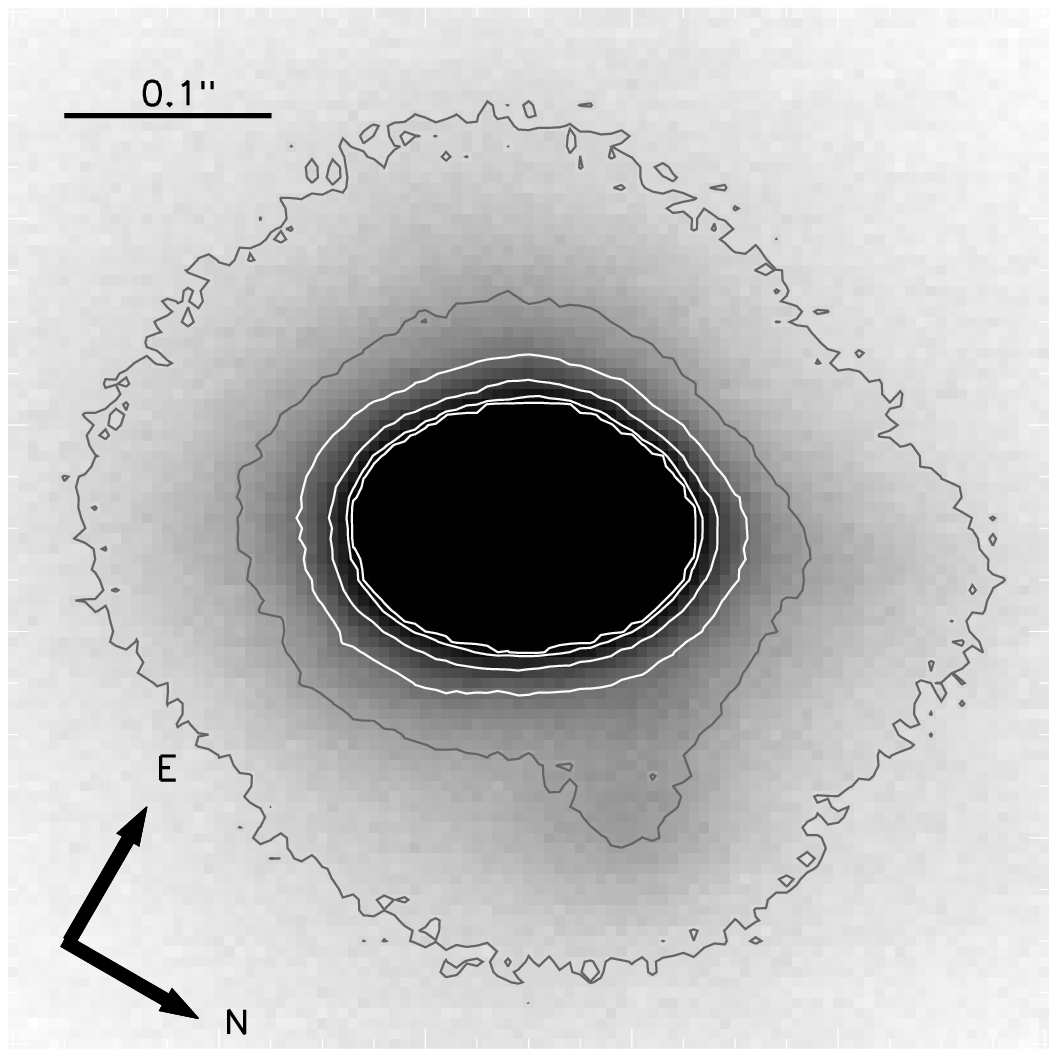}} \\
\subfloat[][\emph{SDSS J1511+0607 $J$}.]
{\includegraphics[trim=0 0 140 0, clip, width=.3\textwidth]{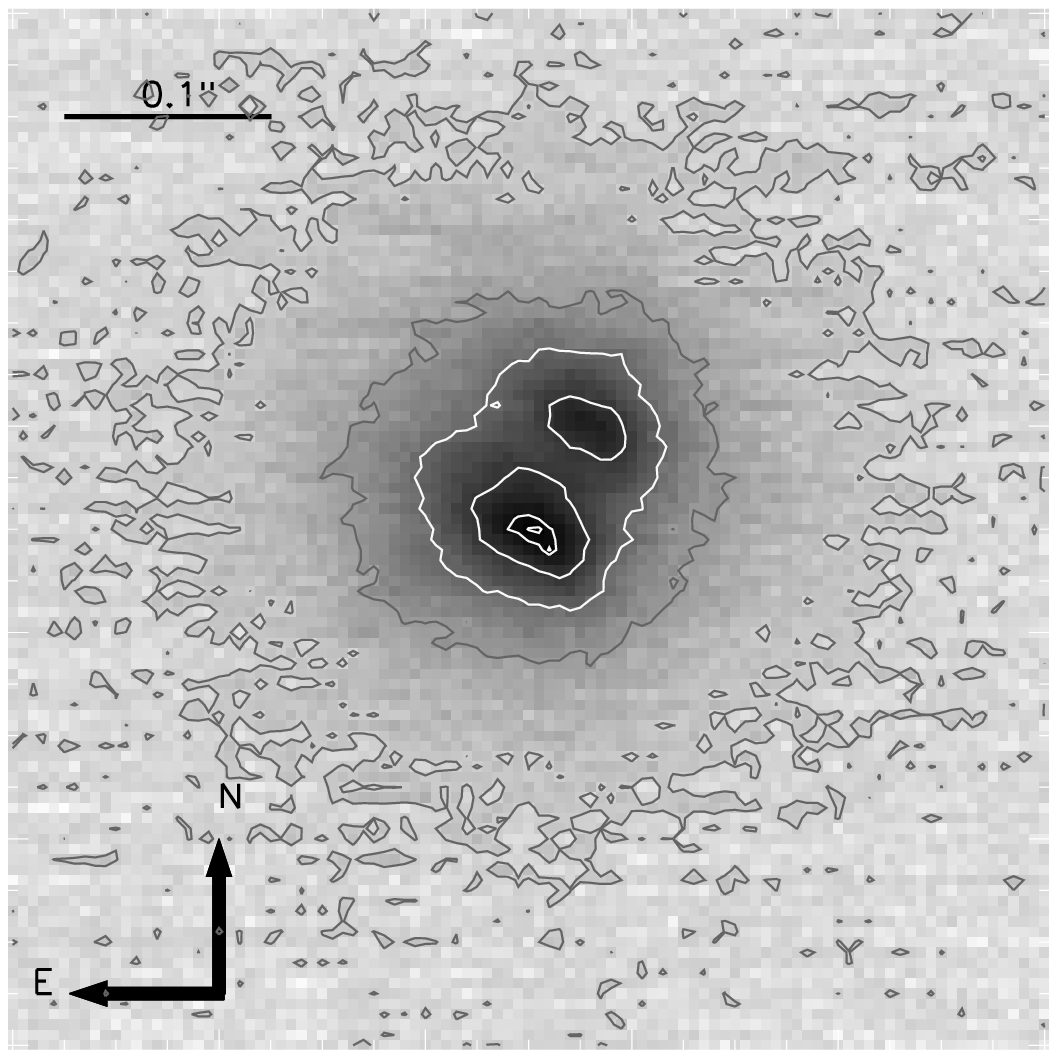}} \quad
\subfloat[][\emph{SDSS J1511+0607 $H$}.]
{\includegraphics[trim=0 0 140 0, clip, width=.3\textwidth]{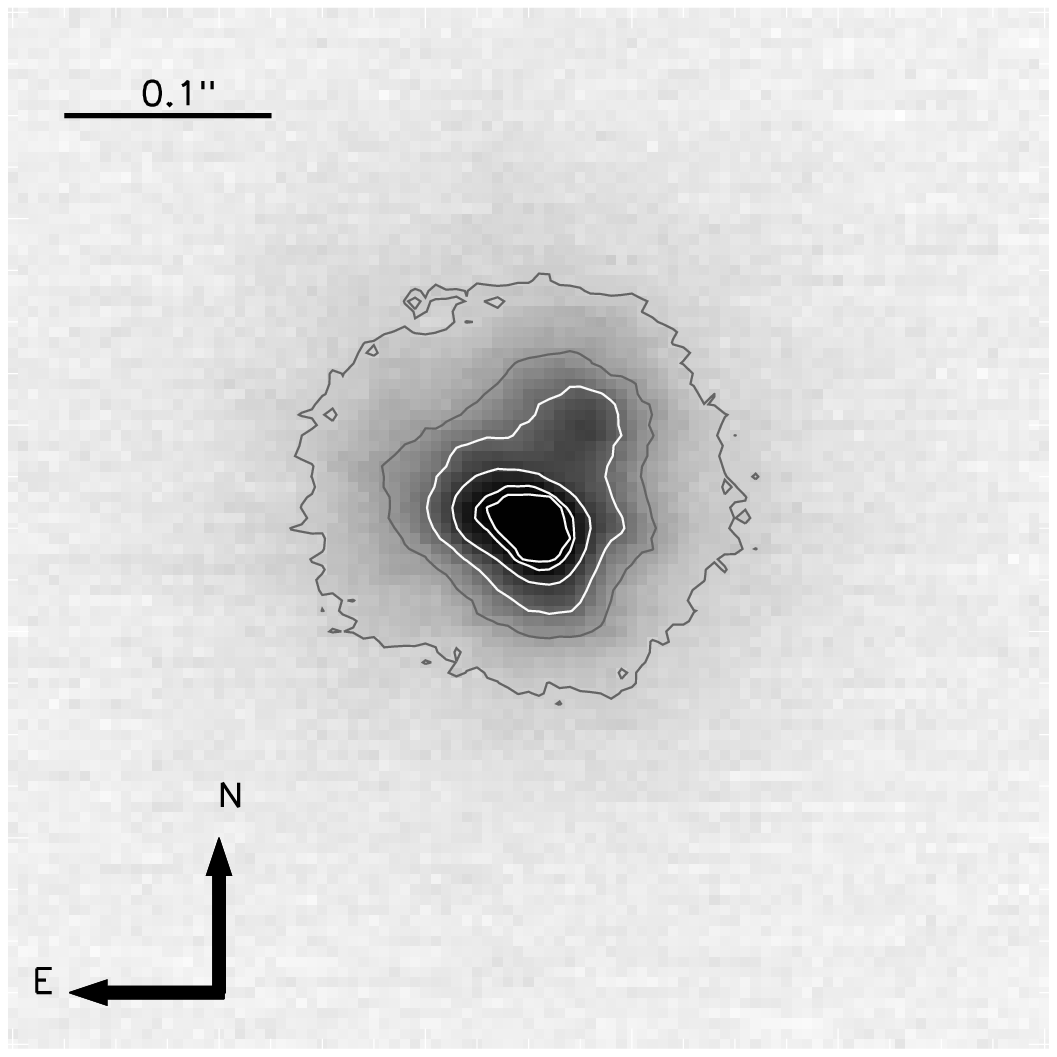}} \quad
\subfloat[][\emph{SDSS J1511+0607 $K_s$}.]
{\includegraphics[trim=0 0 140 0, clip, width=.3\textwidth]{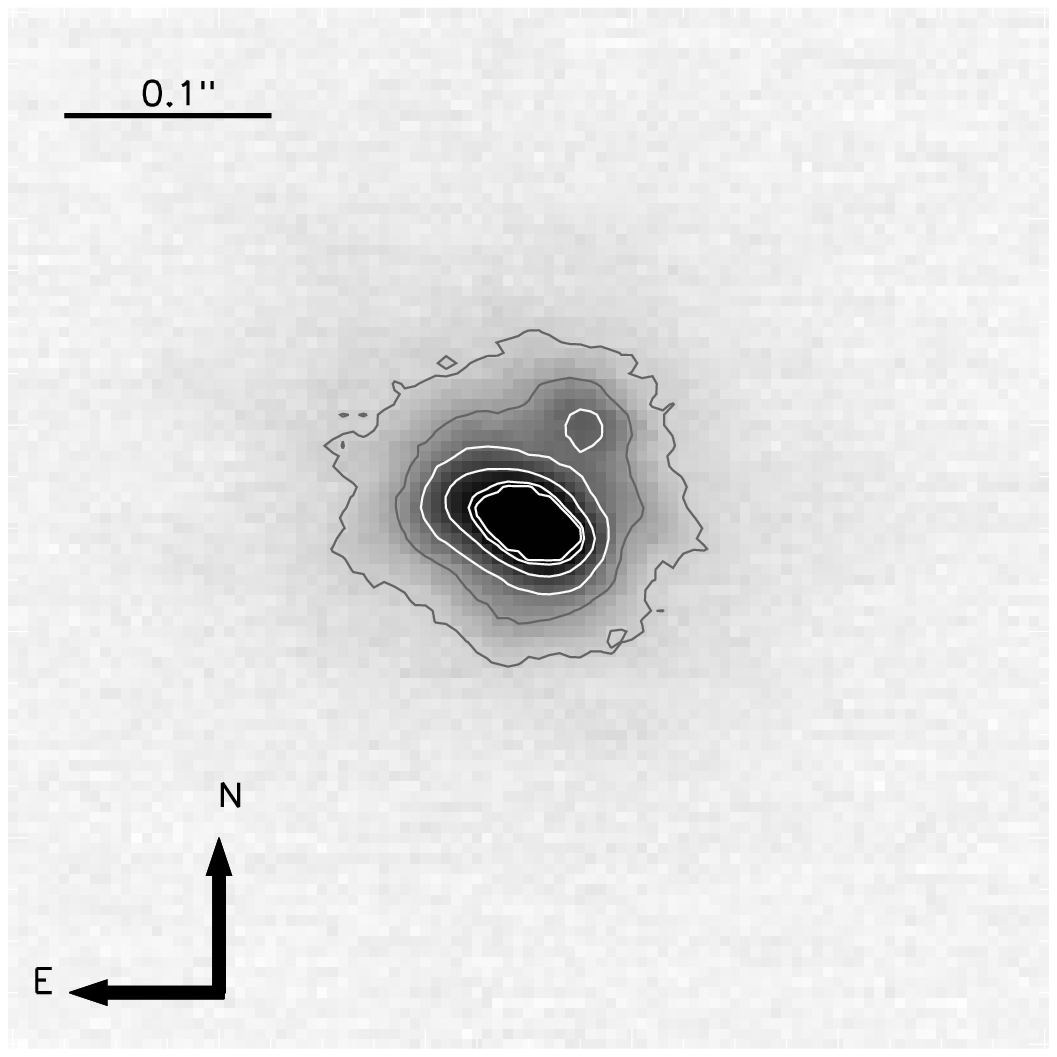}} \\
\subfloat[][\emph{SDSS J2052-1609 $J$}.]
{\includegraphics[trim=0 0 140 0, clip, width=.3\textwidth]{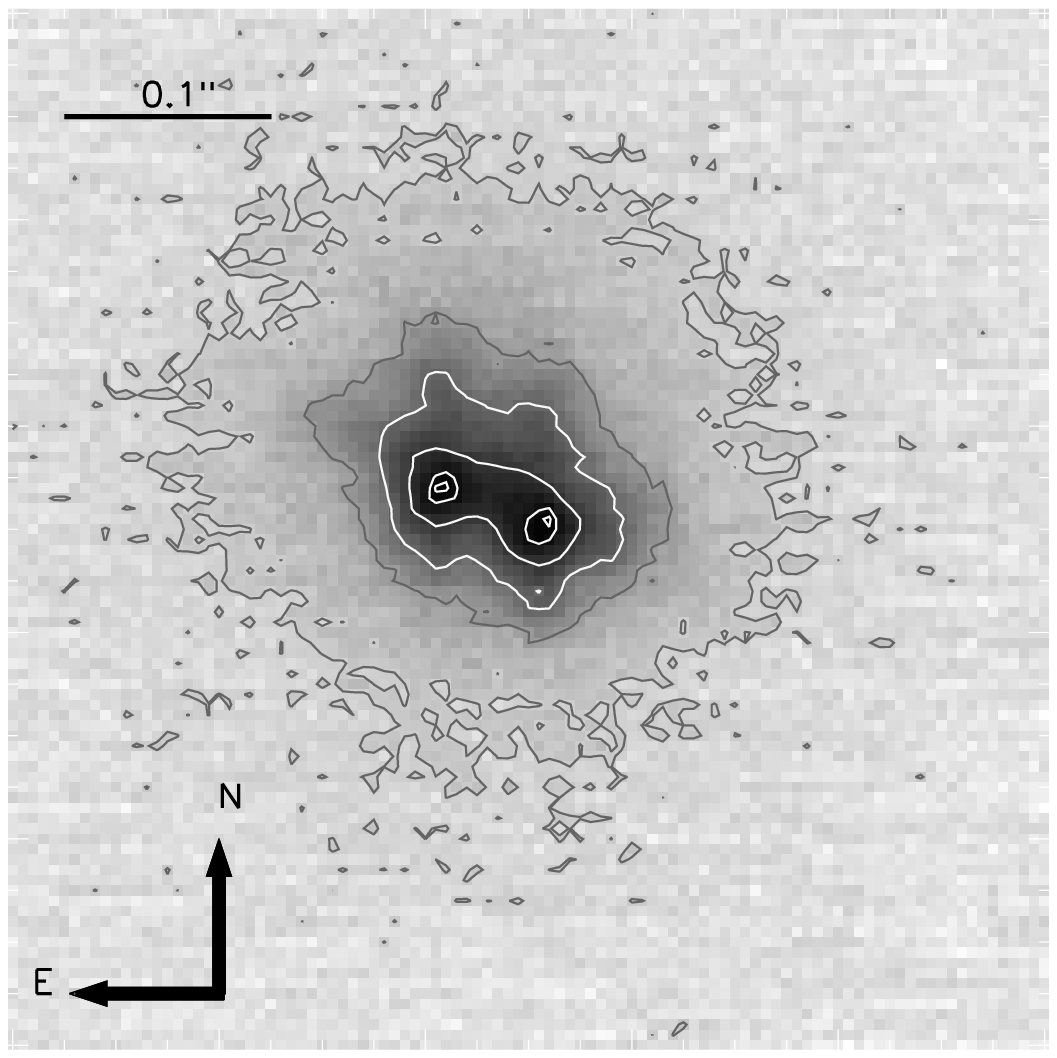}} \quad
\subfloat[][\emph{SDSS J2052-1609 $H$}.]
{\includegraphics[trim=0 0 140 0, clip, width=.3\textwidth]{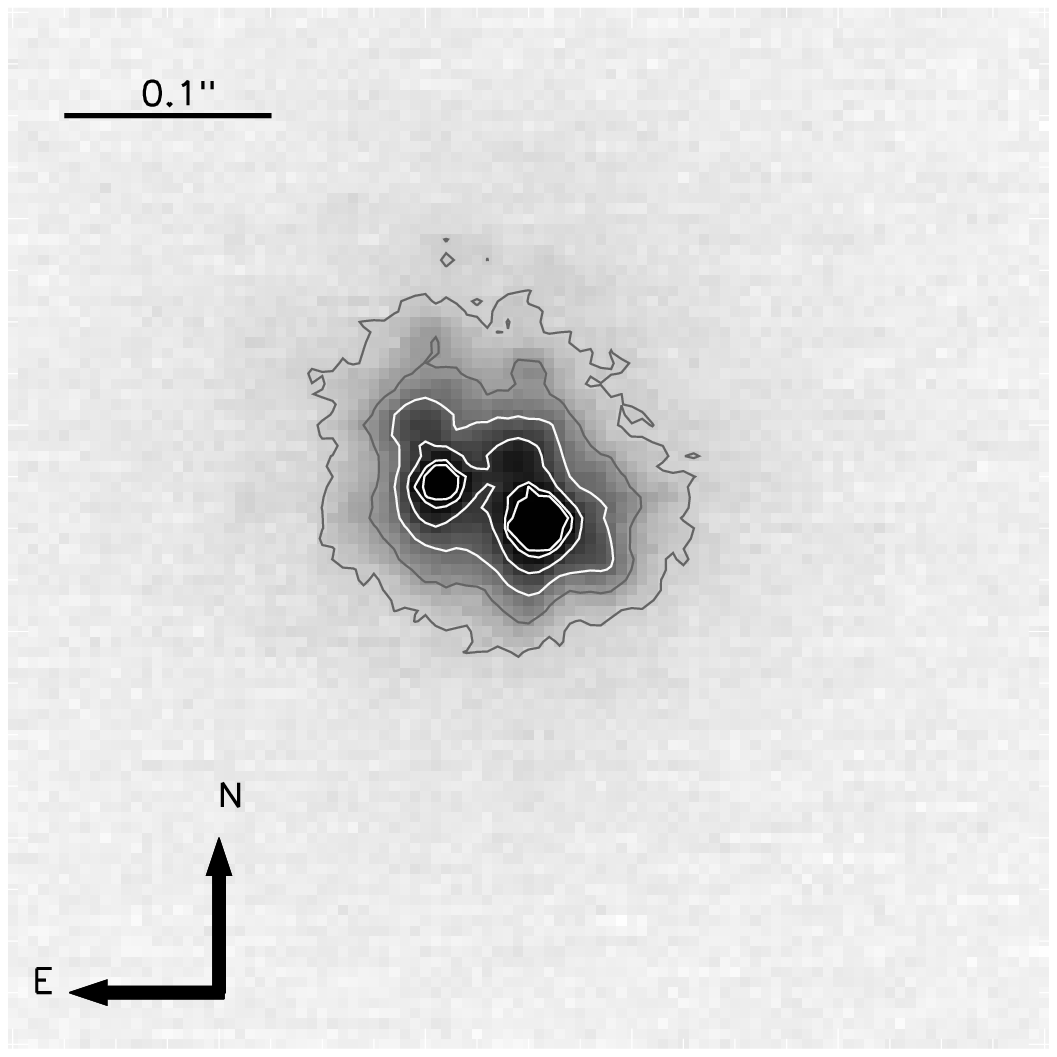}} \quad
\subfloat[][\emph{SDSS J2052-1609 $K_s$}.]
{\includegraphics[trim=0 0 140 0, clip, width=.3\textwidth]{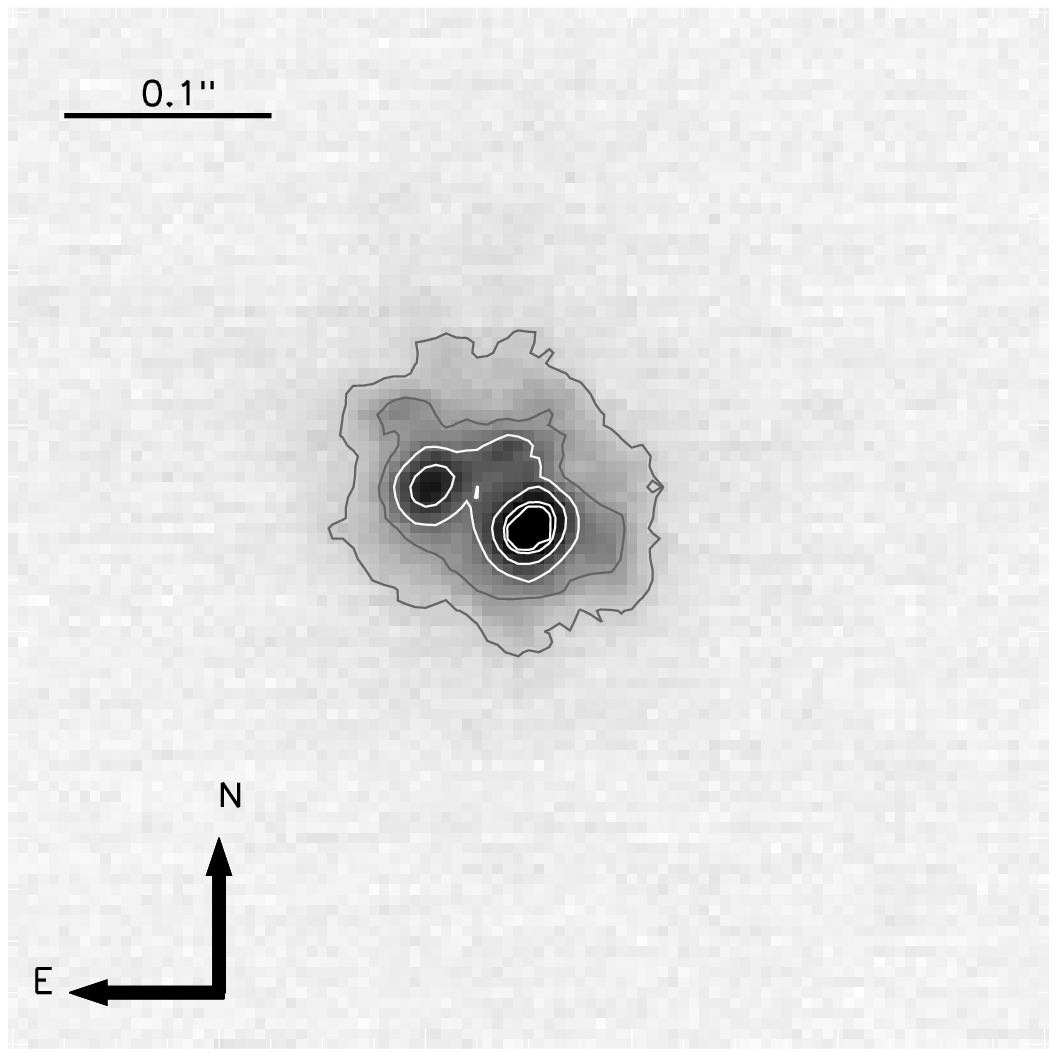}}
\caption{Keck NIRC2 LGS-AO images of the three binaries resolved in this sample in $JHK_s$ bands.}
\label{fig:binaries}
\end{figure}

\begin{figure}[htp]
\centering
\renewcommand{\thesubfigure}{\roman{subfigure}}
\subfloat[][\emph{2MASSI J0019+5213}.]
{\includegraphics[trim=0 0 140 0, clip, width=.3\textwidth]{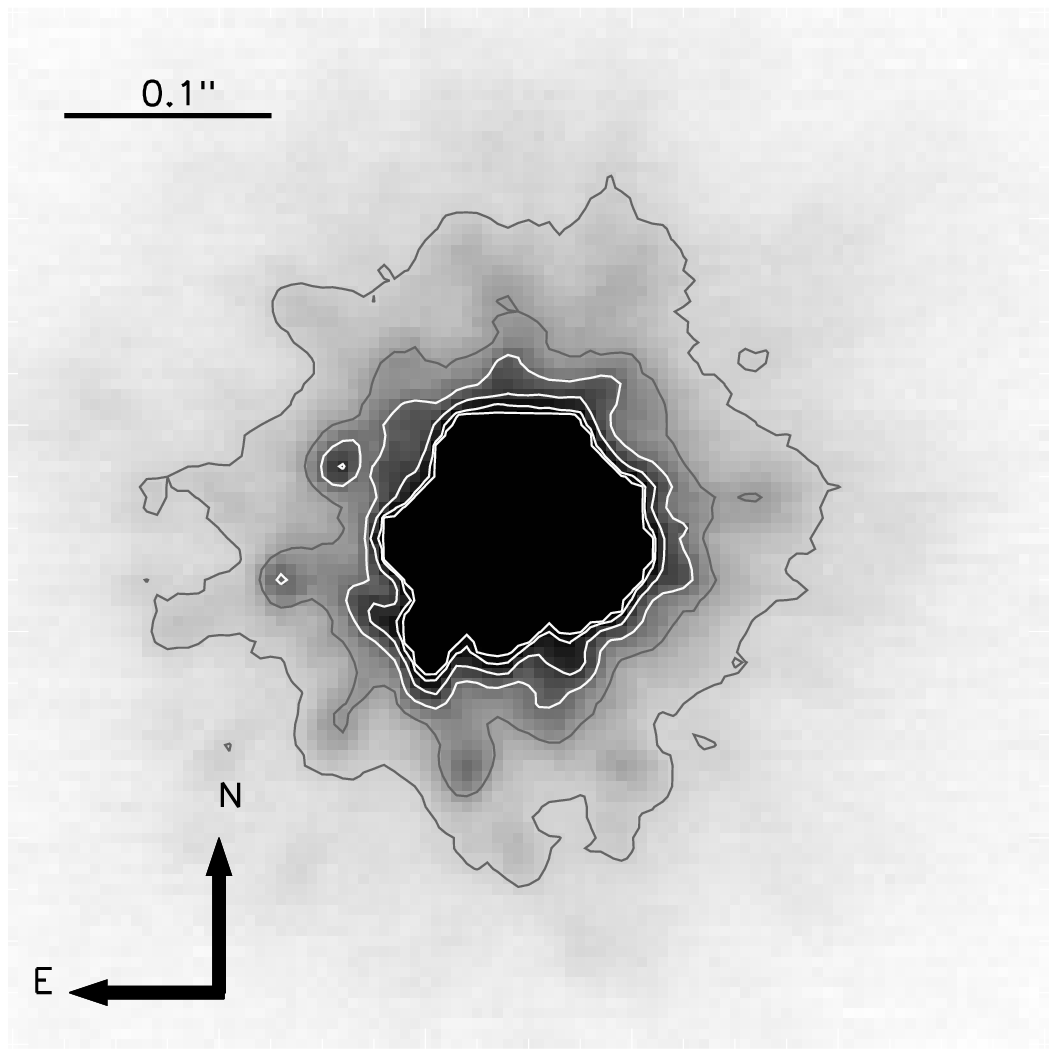}} \quad
\subfloat[][\emph{2MASS J0032+0219}.]
{\includegraphics[trim=0 0 140 0, clip, width=.3\textwidth]{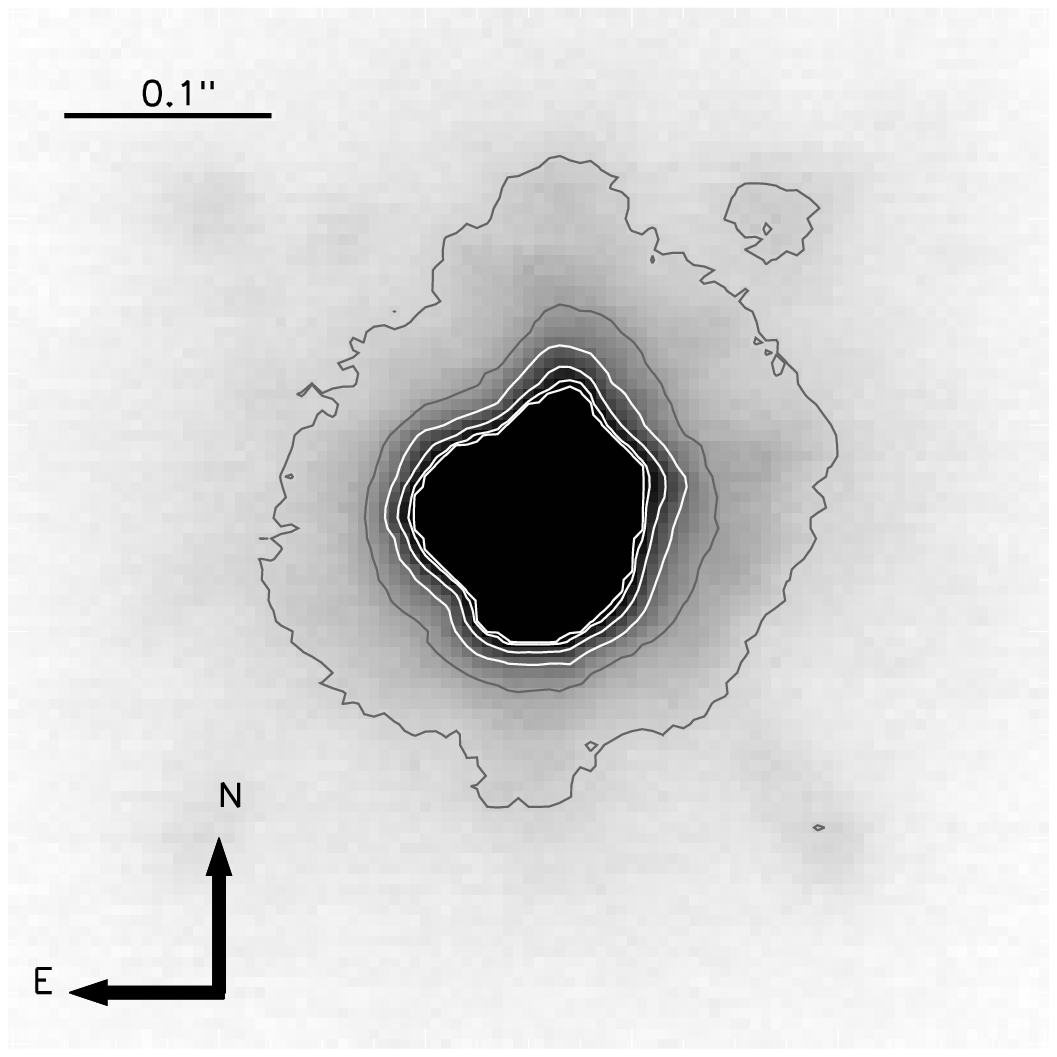}} \quad
\subfloat[][\emph{SDSS J0032+1410}.]
{\includegraphics[trim=0 0 140 0, clip, width=.3\textwidth]{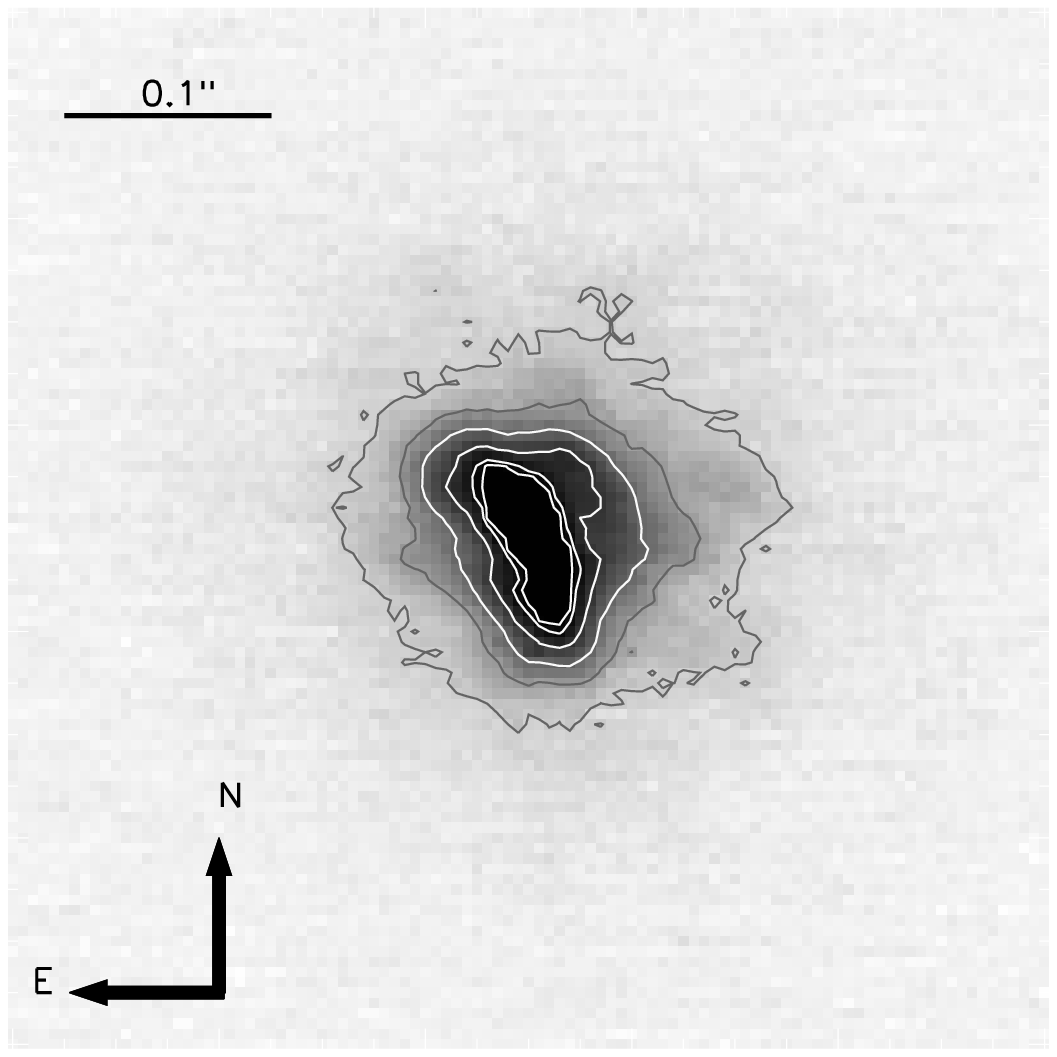}} \quad
\subfloat[][\emph{WISEP J0047+6803}.]
{\includegraphics[trim=0 0 140 0, clip, width=.3\textwidth]{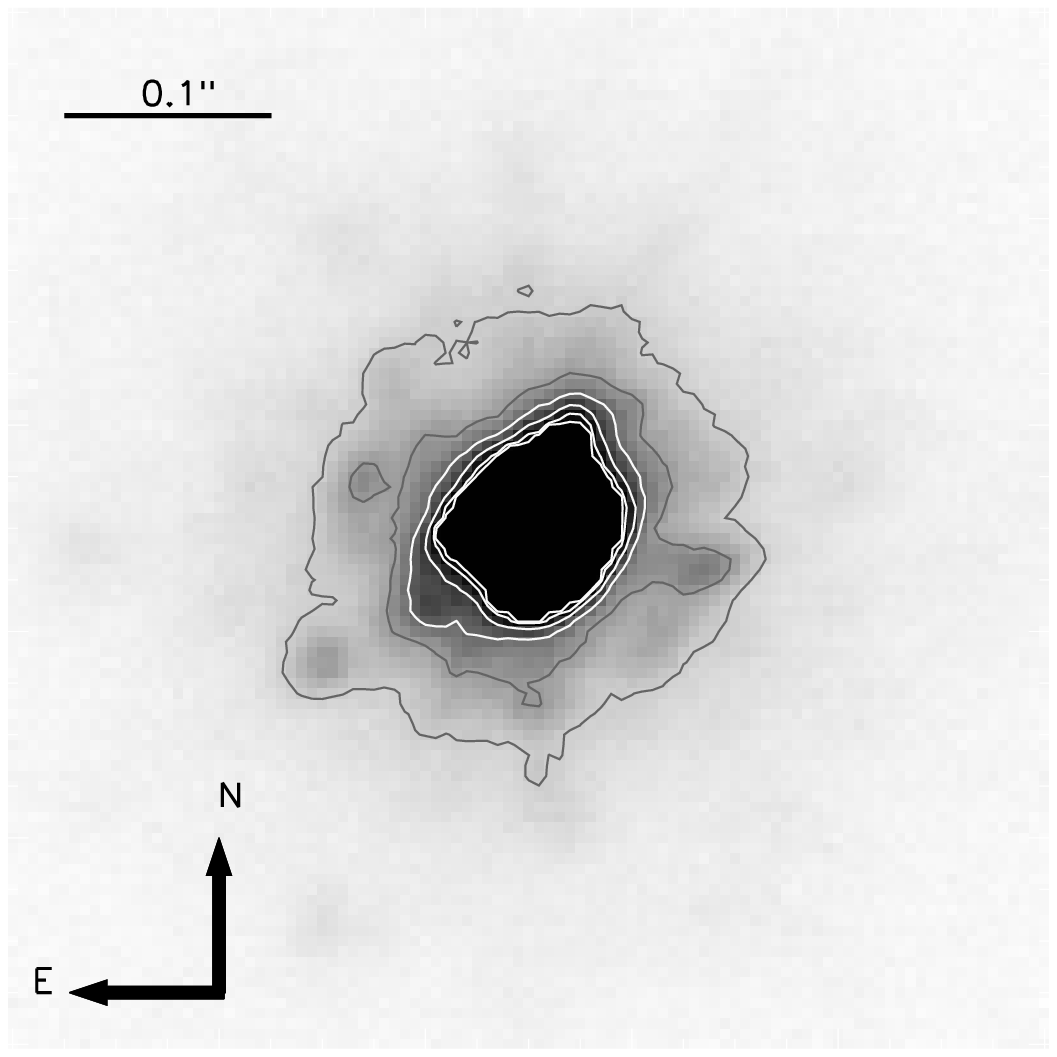}} \quad
\subfloat[][\emph{SDSS J0119+2403}.]
{\includegraphics[trim=0 0 140 0, clip, width=.3\textwidth]{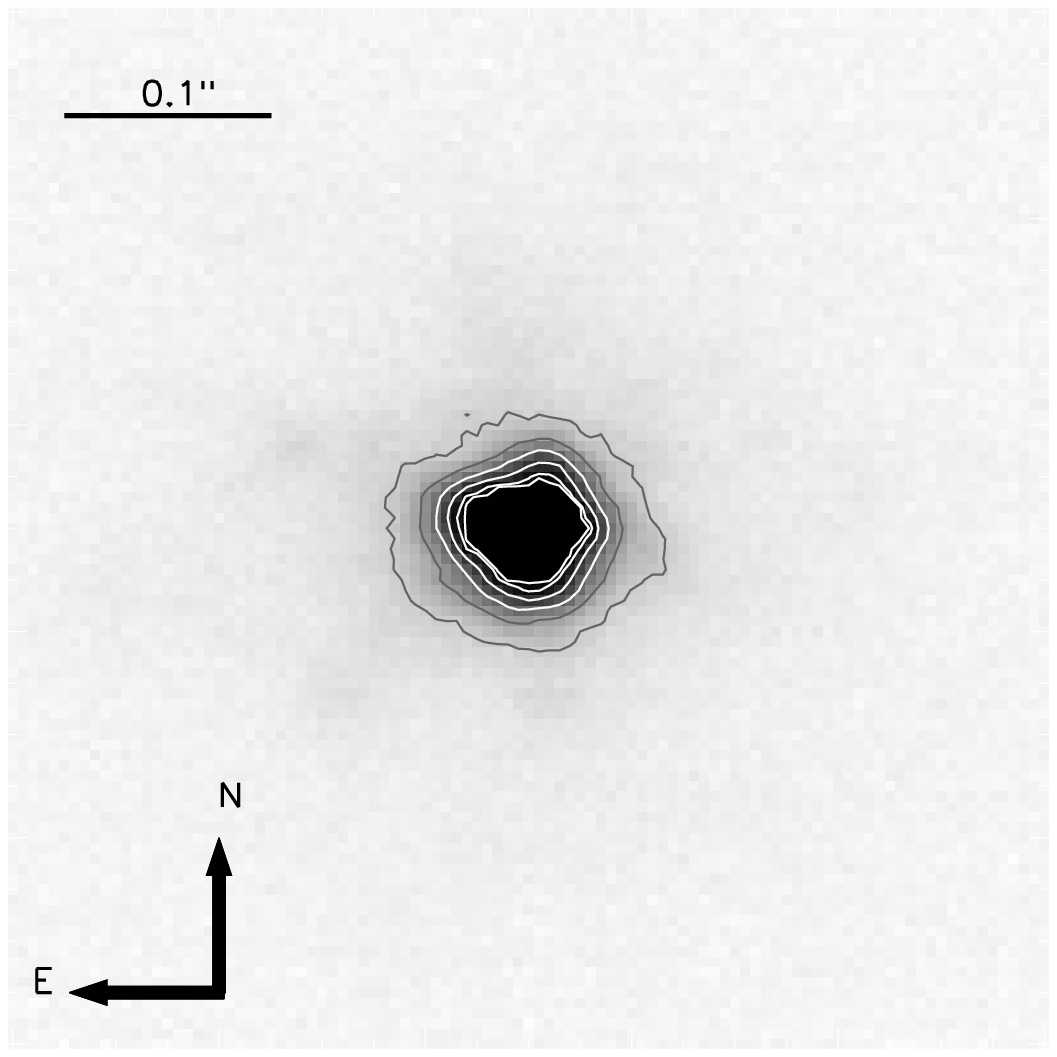}} \quad
\subfloat[][\emph{SDSSp J0236+0048}.]
{\includegraphics[trim=0 0 140 0, clip, width=.3\textwidth]{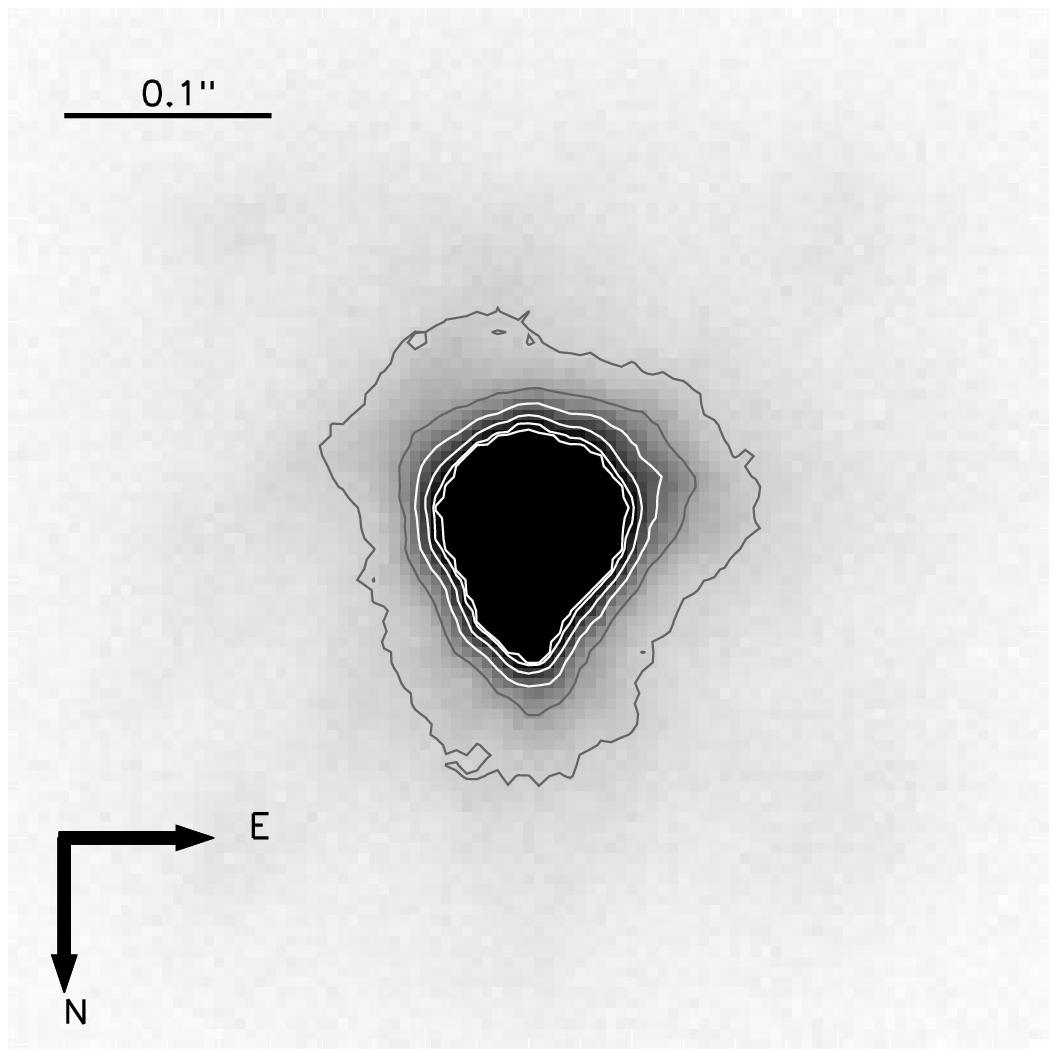}} \quad
\subfloat[][\emph{SDSS J0247-1631}.]
{\includegraphics[trim=0 0 140 0, clip, width=.3\textwidth]{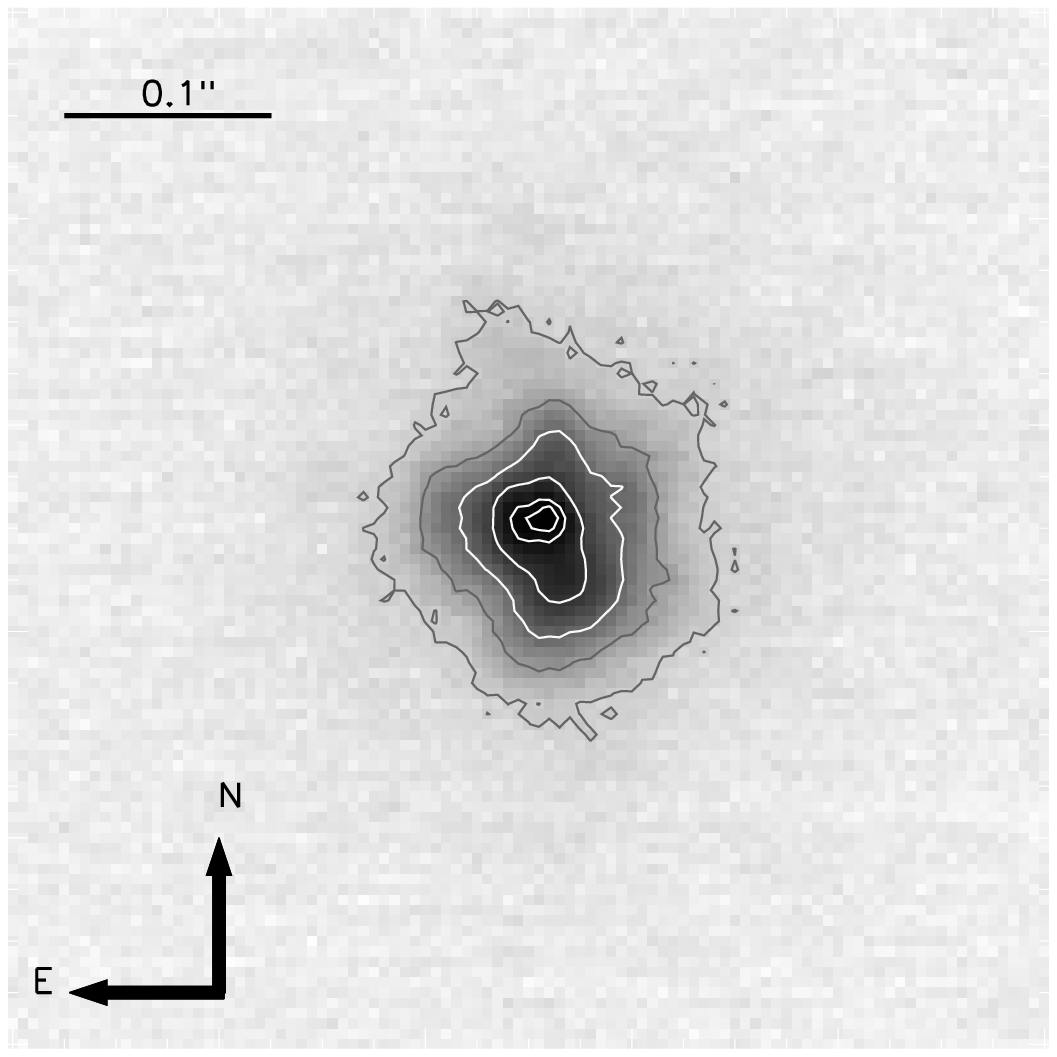}} \quad
\subfloat[][\emph{2MASS J0300+2130}.]
{\includegraphics[trim=0 0 140 0, clip, width=.3\textwidth]{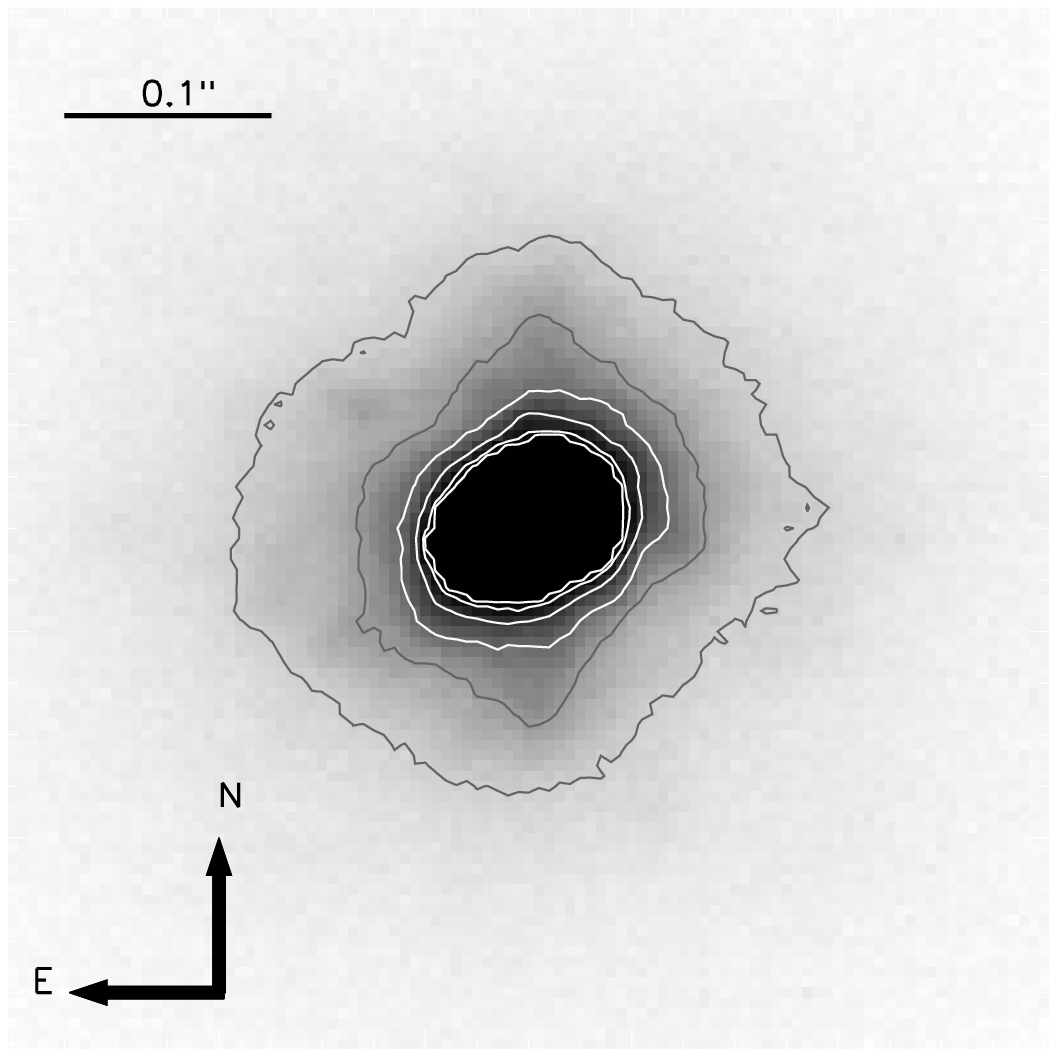}} \quad
\subfloat[][\emph{2MASS J0302+1358}.]
{\includegraphics[trim=0 0 140 0, clip, width=.3\textwidth]{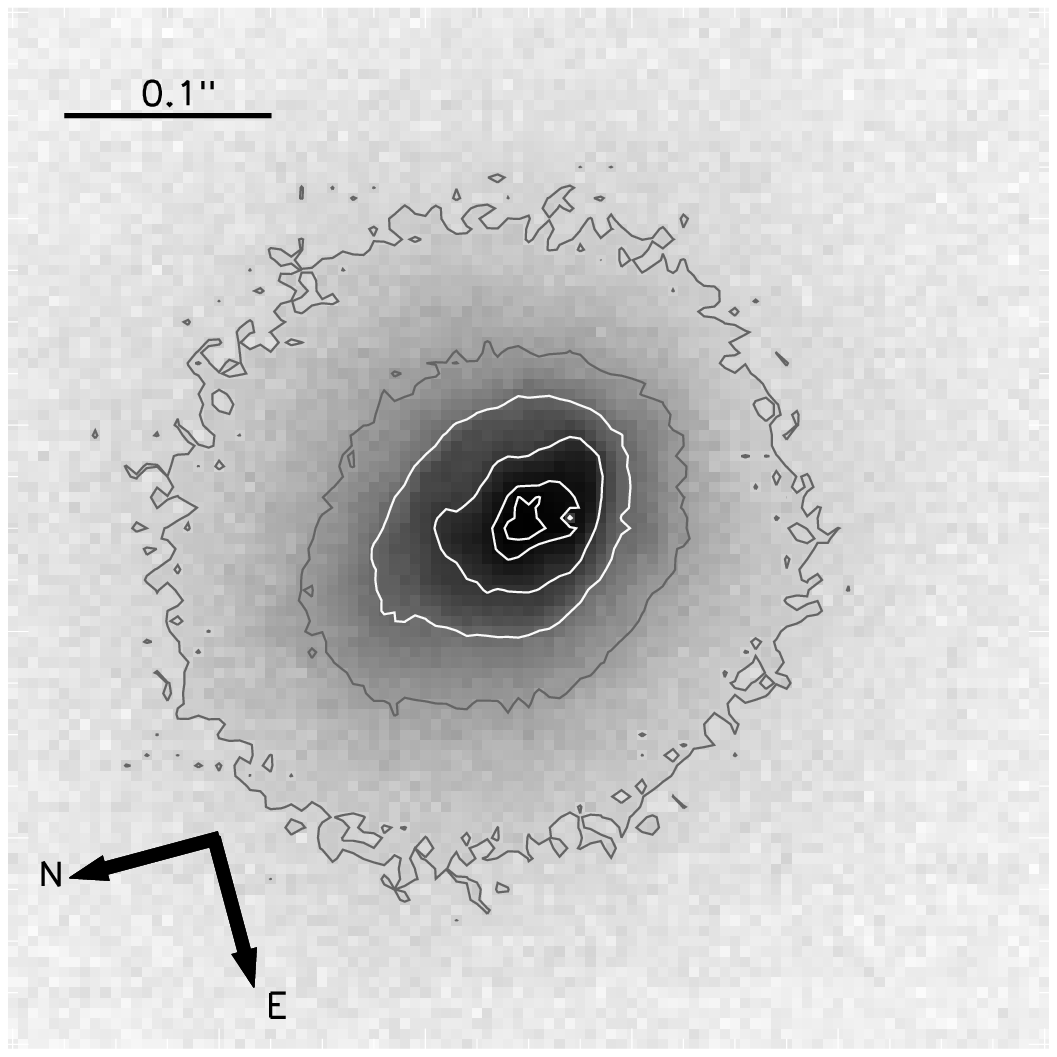}} \quad
\subfloat[][\emph{2MASS J0344+0110}.]
{\includegraphics[trim=0 0 140 0, clip, width=.3\textwidth]{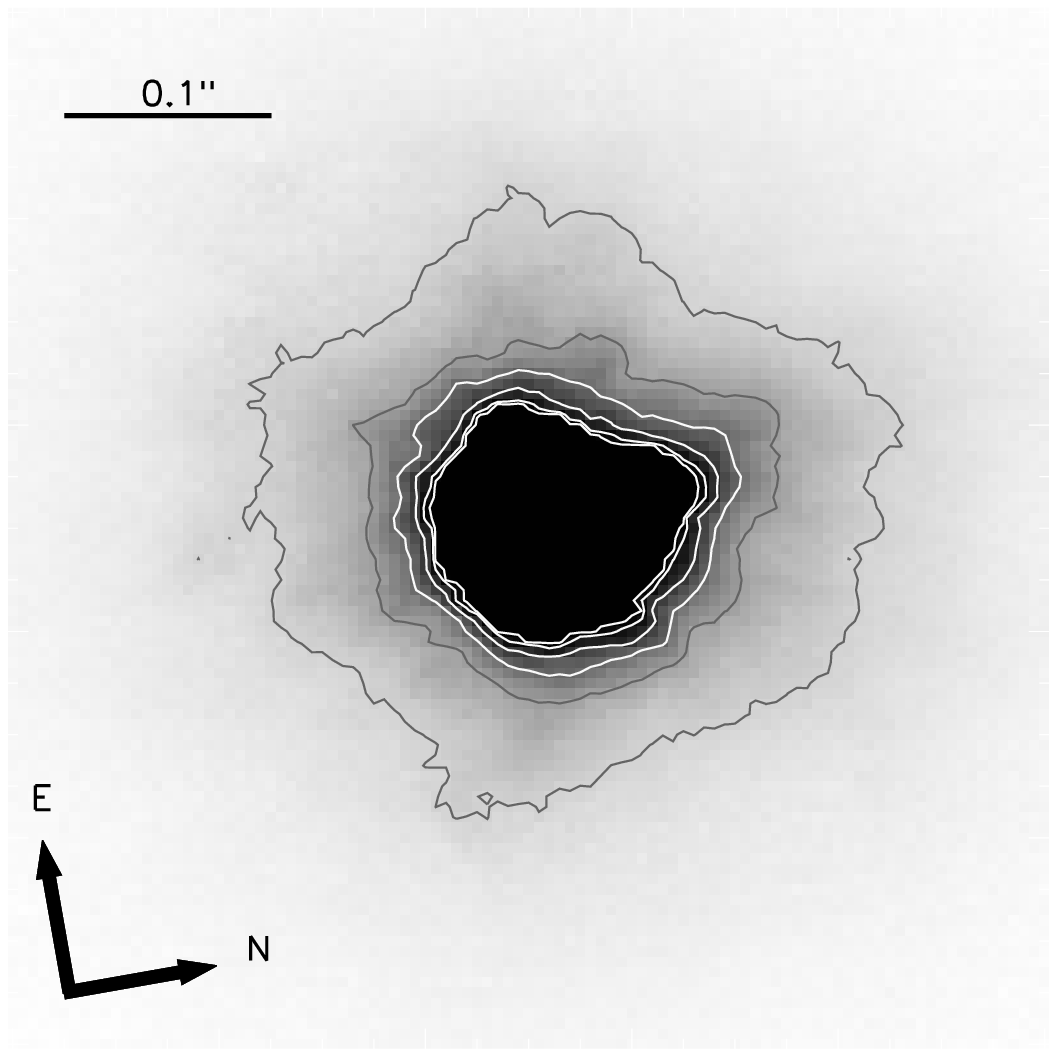}} \quad
\subfloat[][\emph{SDSS J0351+4810}.]
{\includegraphics[trim=0 0 140 0, clip, width=.3\textwidth]{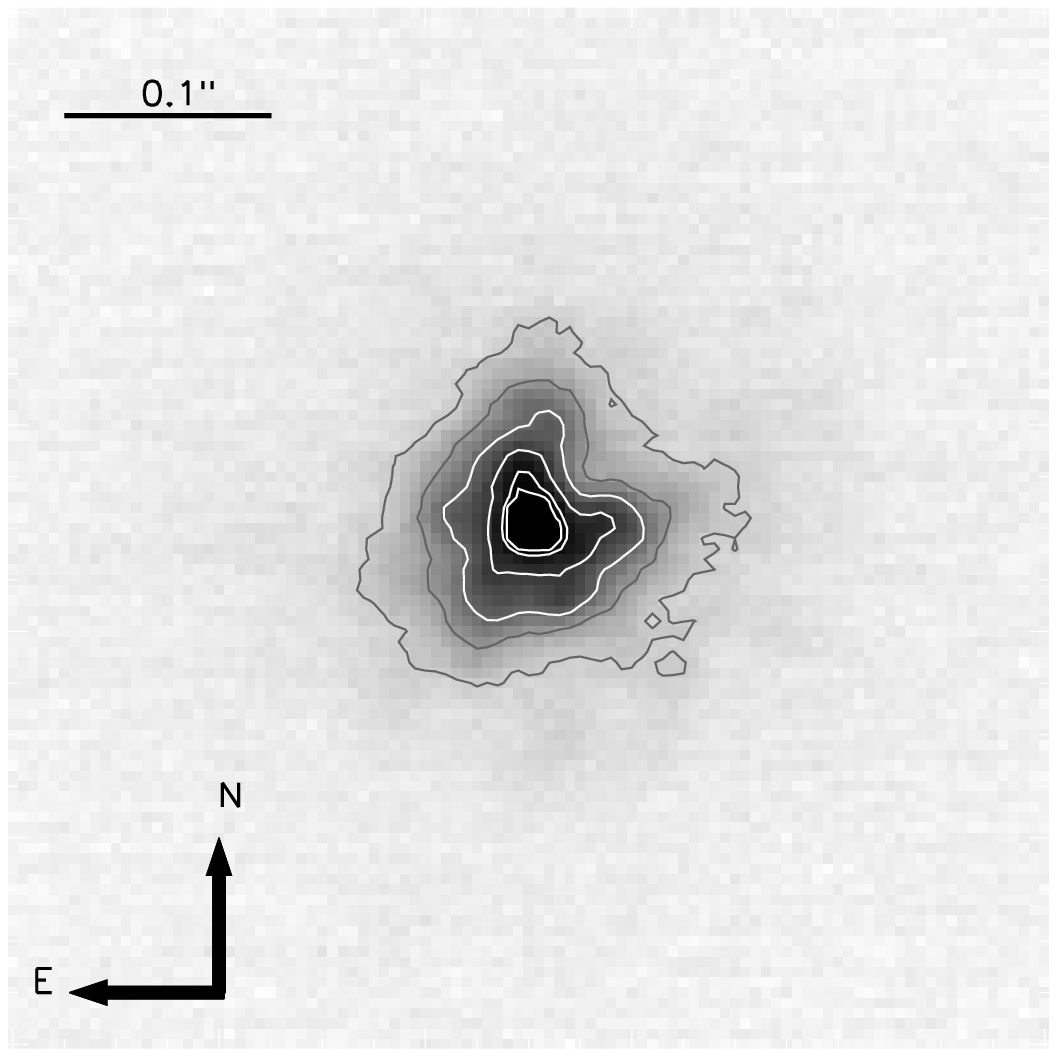}} \quad
\subfloat[][\emph{HTY J0429+1535}.]
{\includegraphics[trim=0 0 140 0, clip, width=.3\textwidth]{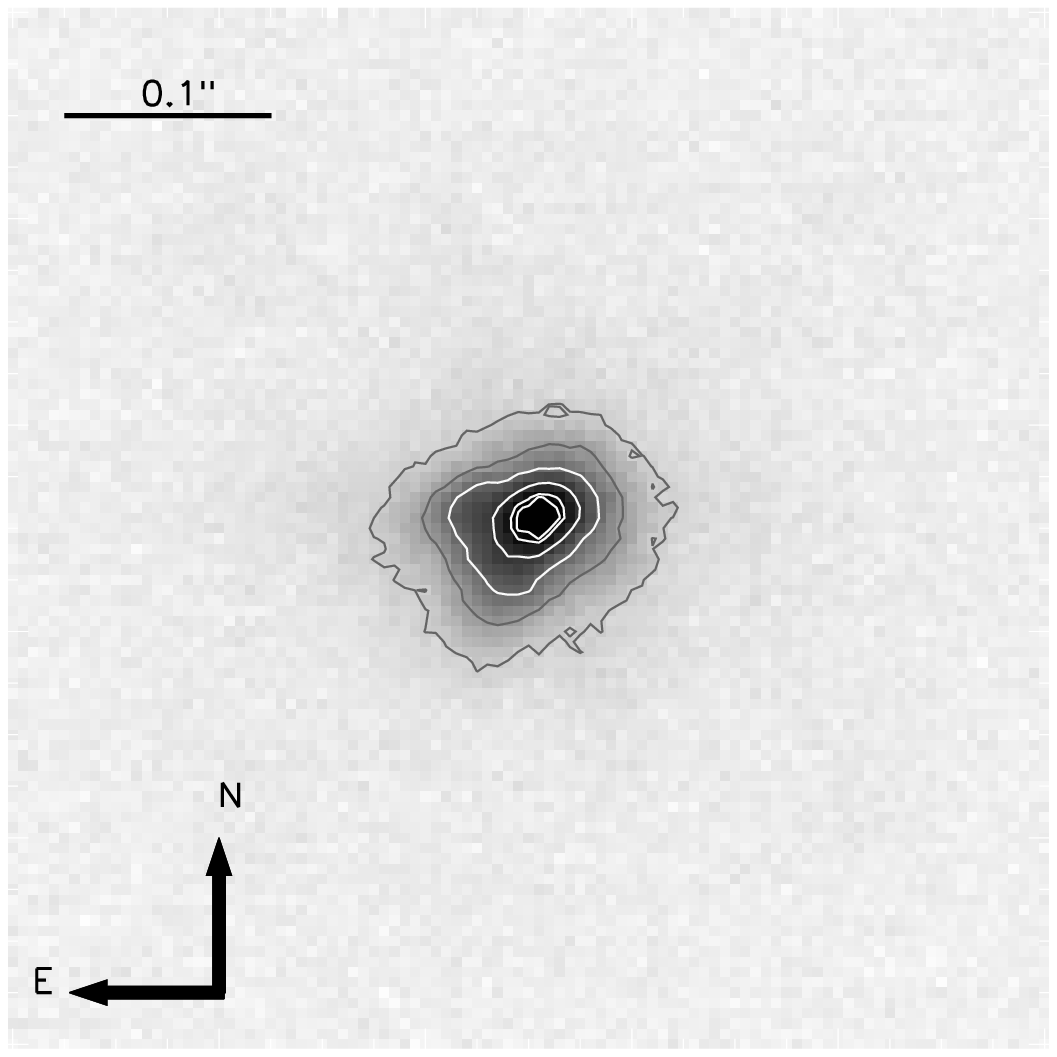}} \quad
\caption{Keck NIRC2 LGS-AO images in the $H$ band of all targets. Contours are drawn at 20, 40, 60, 80, 95 and 99\% of the image minimum.\label{fig:unresbin}}
\ContinuedFloat
\end{figure}

\begin{figure}[htp]
\renewcommand{\thesubfigure}{\roman{subfigure}}
\centering
\subfloat[][\emph{2MASS J0443-3202}.]
{\includegraphics[trim=0 0 140 0, clip, width=.3\textwidth]{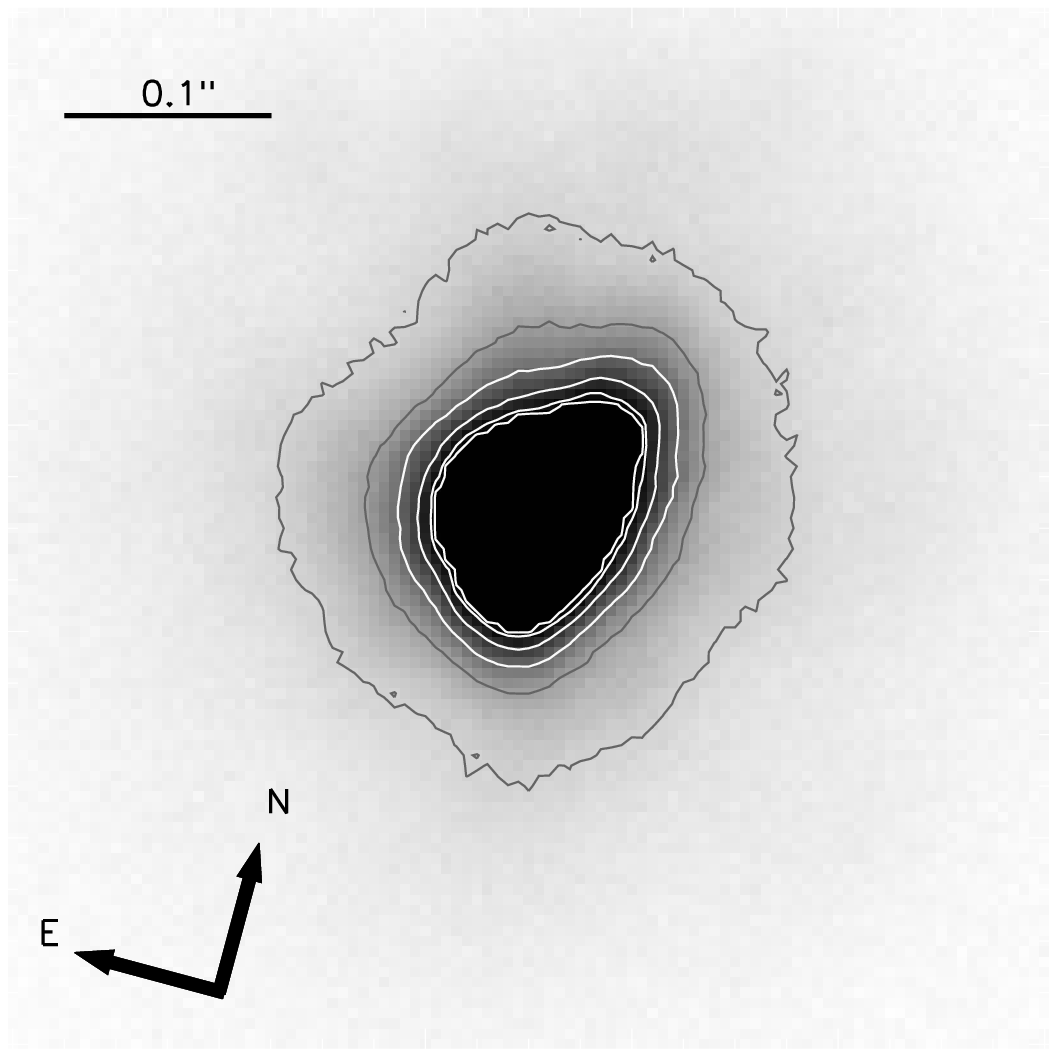}} \quad
\subfloat[][\emph{2MASS J0518$-$2828}.]
{\includegraphics[trim=0 0 140 0, clip, width=.3\textwidth]{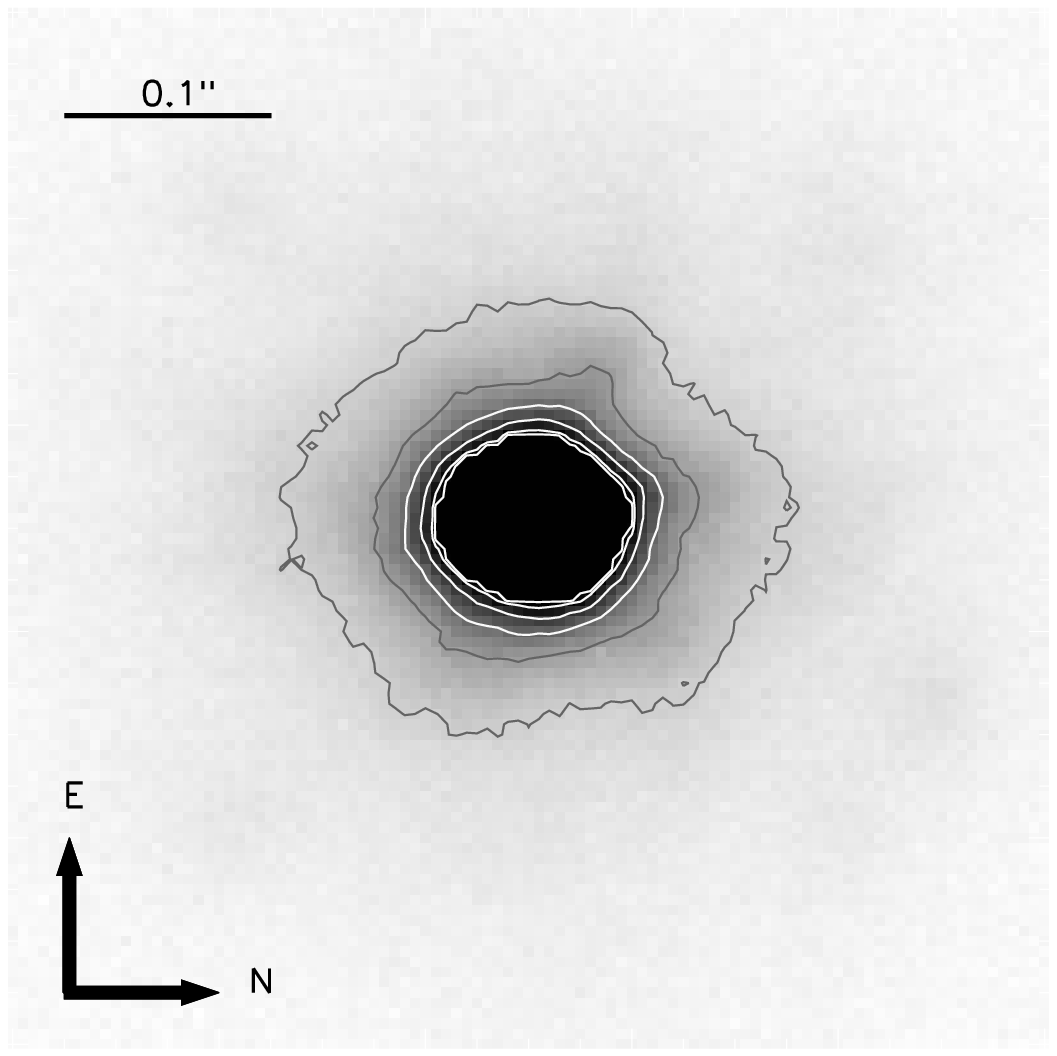}} \quad
\subfloat[][\emph{WISE J0528+0901}.]
{\includegraphics[trim=0 0 140 0, clip, width=.3\textwidth]{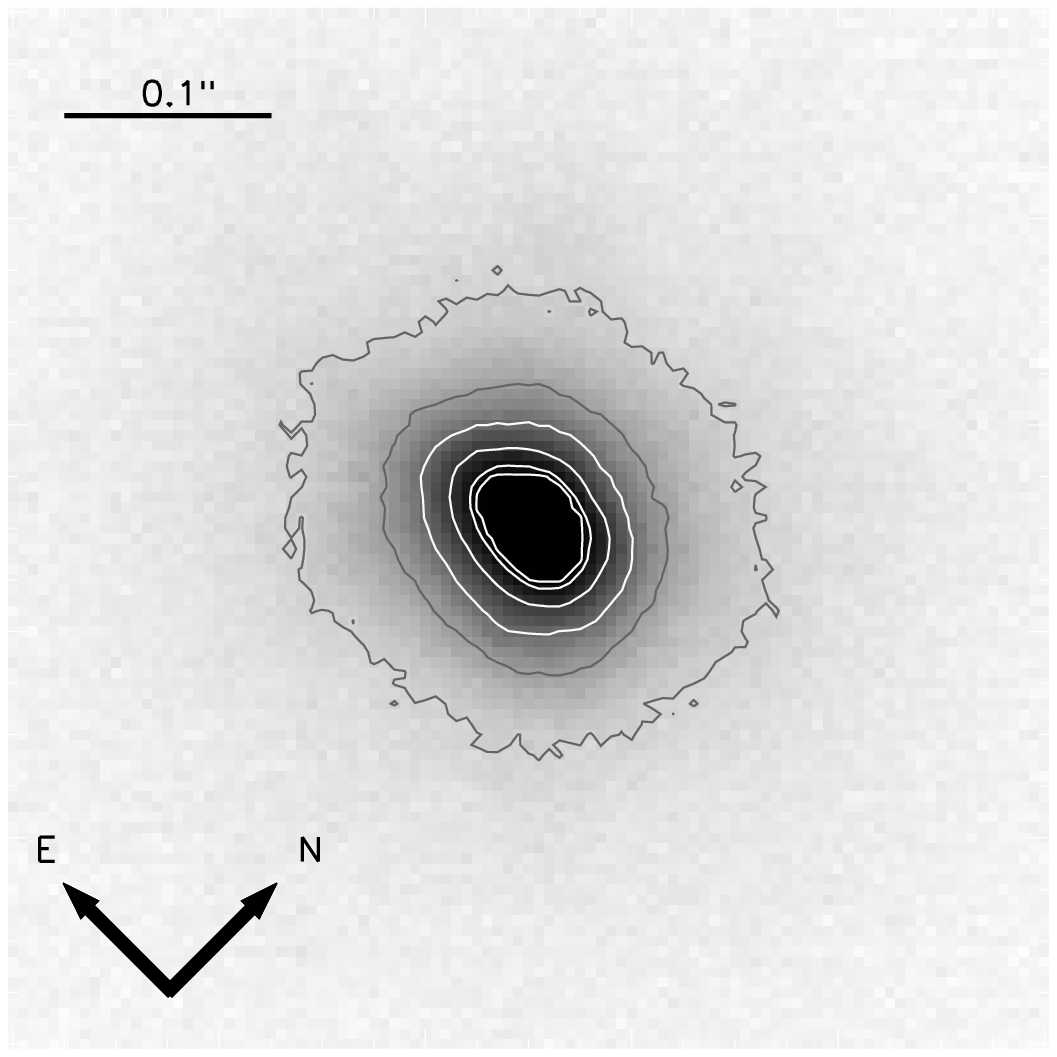}} \quad
\subfloat[][\emph{WISE J0720-0846}.]
{\includegraphics[trim=0 0 140 0, clip, width=.3\textwidth]{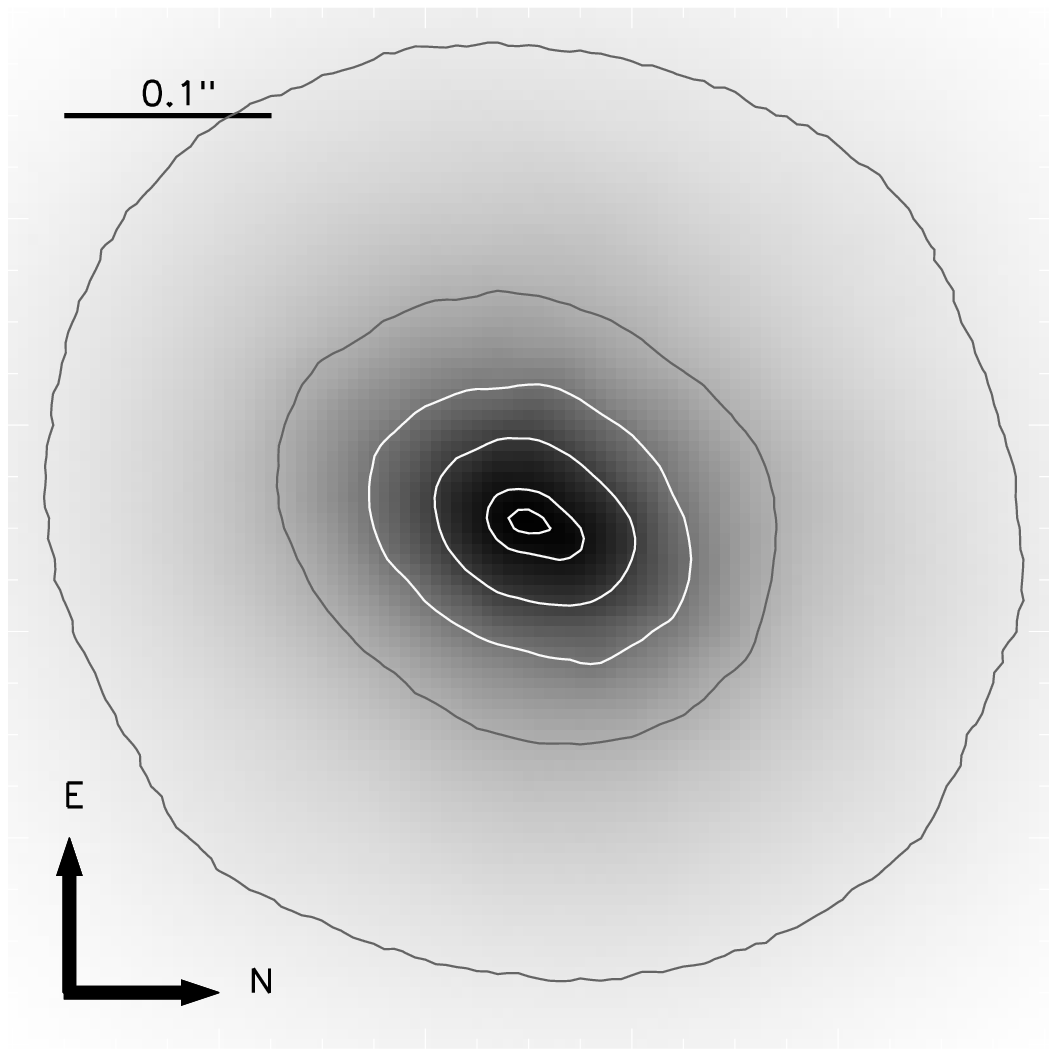}} \quad
\subfloat[][\emph{SDSS J0758+3247}.]
{\includegraphics[trim=0 0 140 0, clip, width=.3\textwidth]{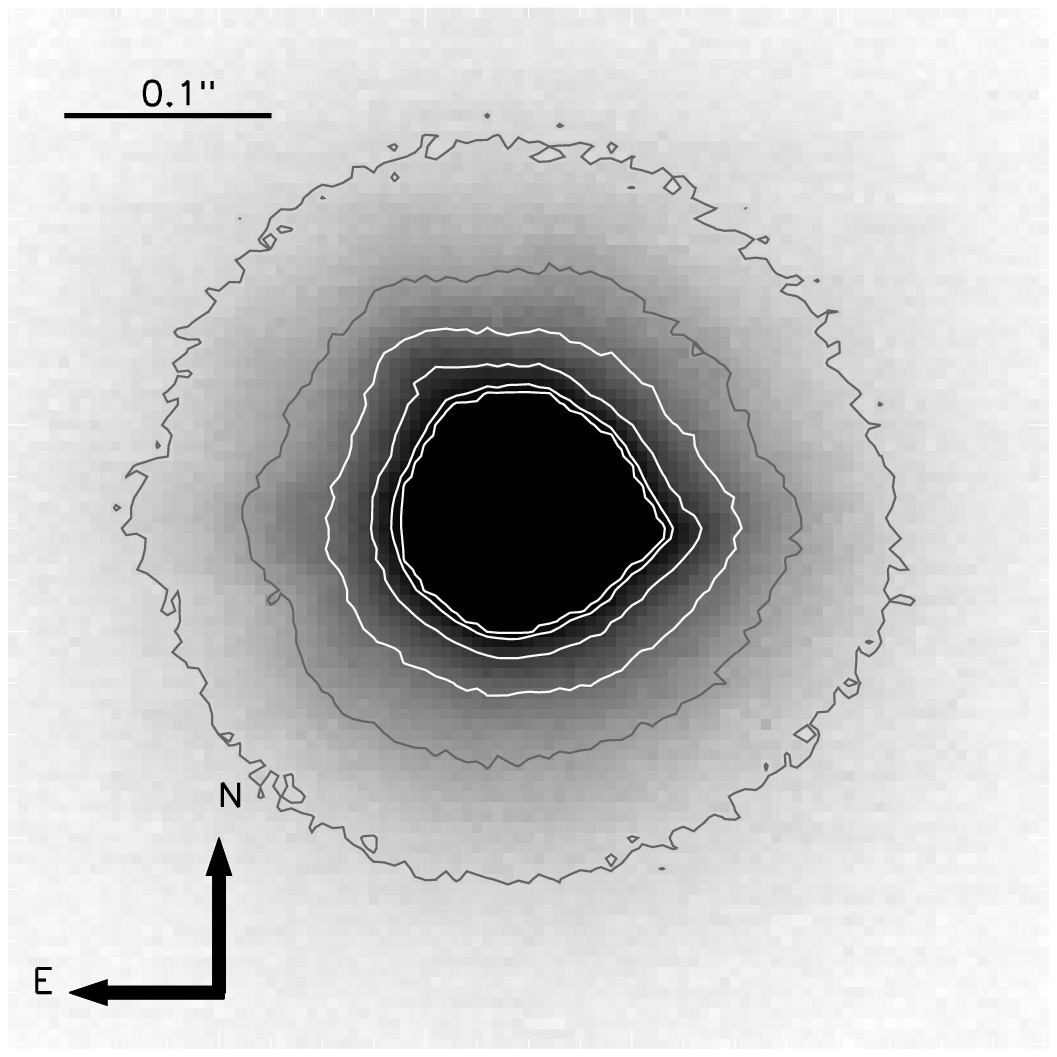}} \quad
\subfloat[][\emph{SDSS J0805+4812}.]
{\includegraphics[trim=0 0 140 0, clip, width=.3\textwidth]{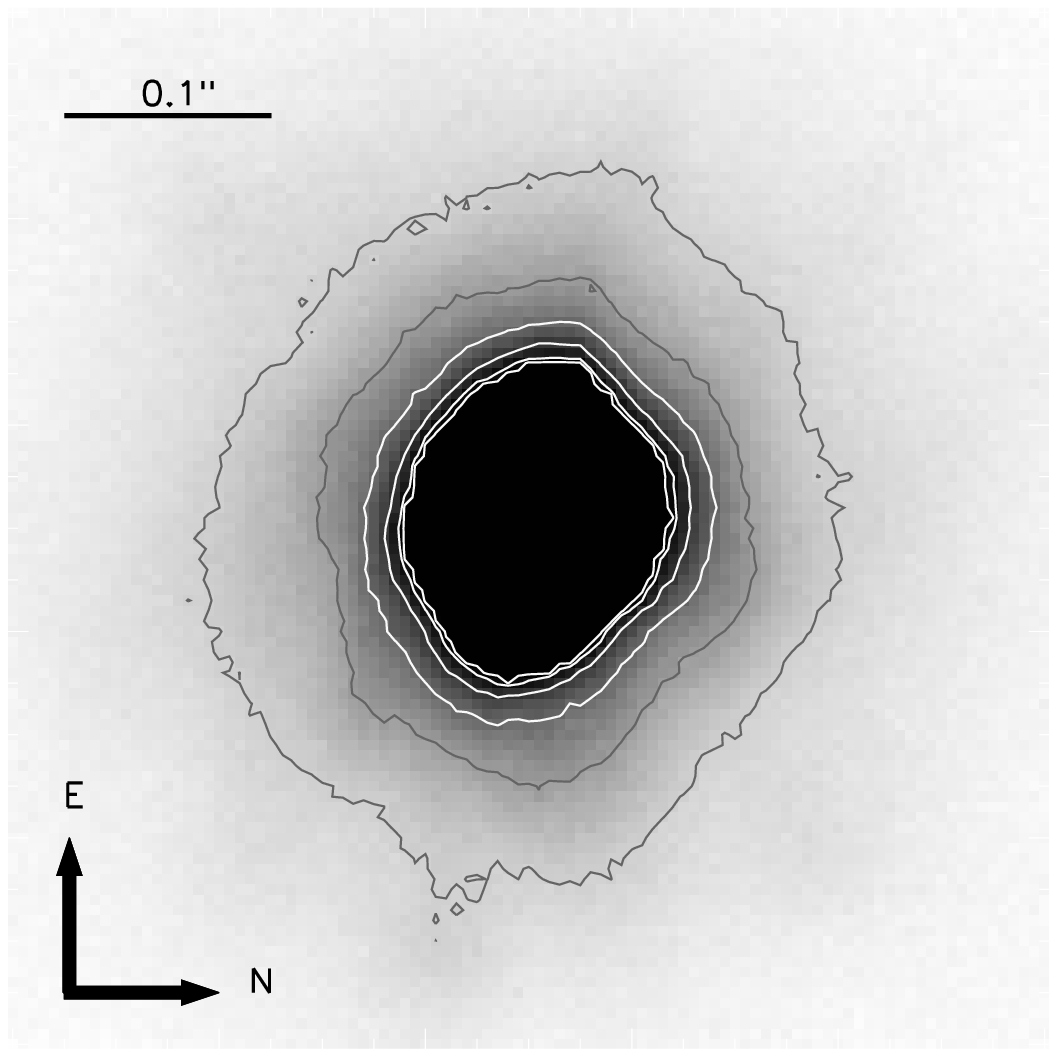}} \quad
\subfloat[][\emph{SDSS J0931+0327}.]
{\includegraphics[trim=0 0 140 0, clip, width=.3\textwidth]{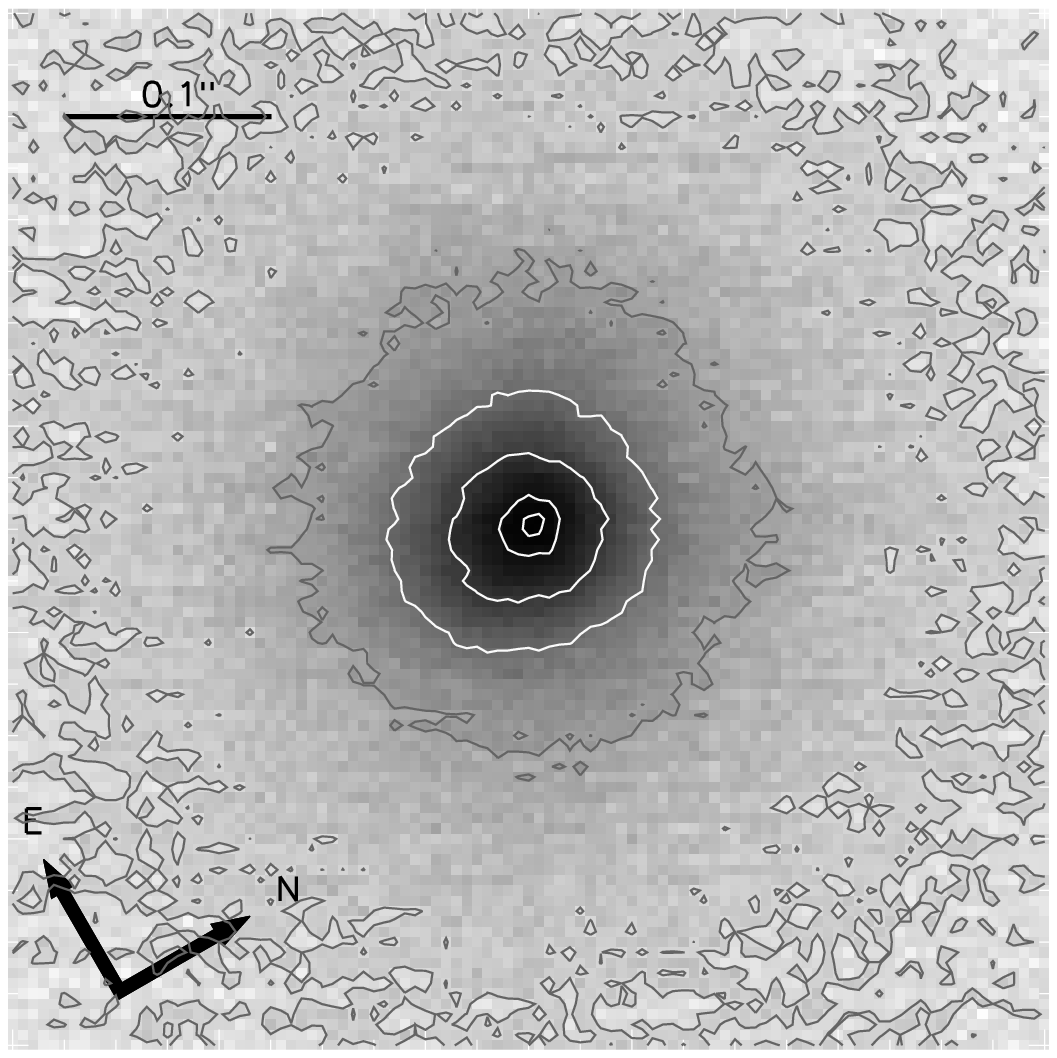}} \quad
\subfloat[][\emph{2MASS J0949$-$1545}.]
{\includegraphics[trim=0 0 140 0, clip, width=.3\textwidth]{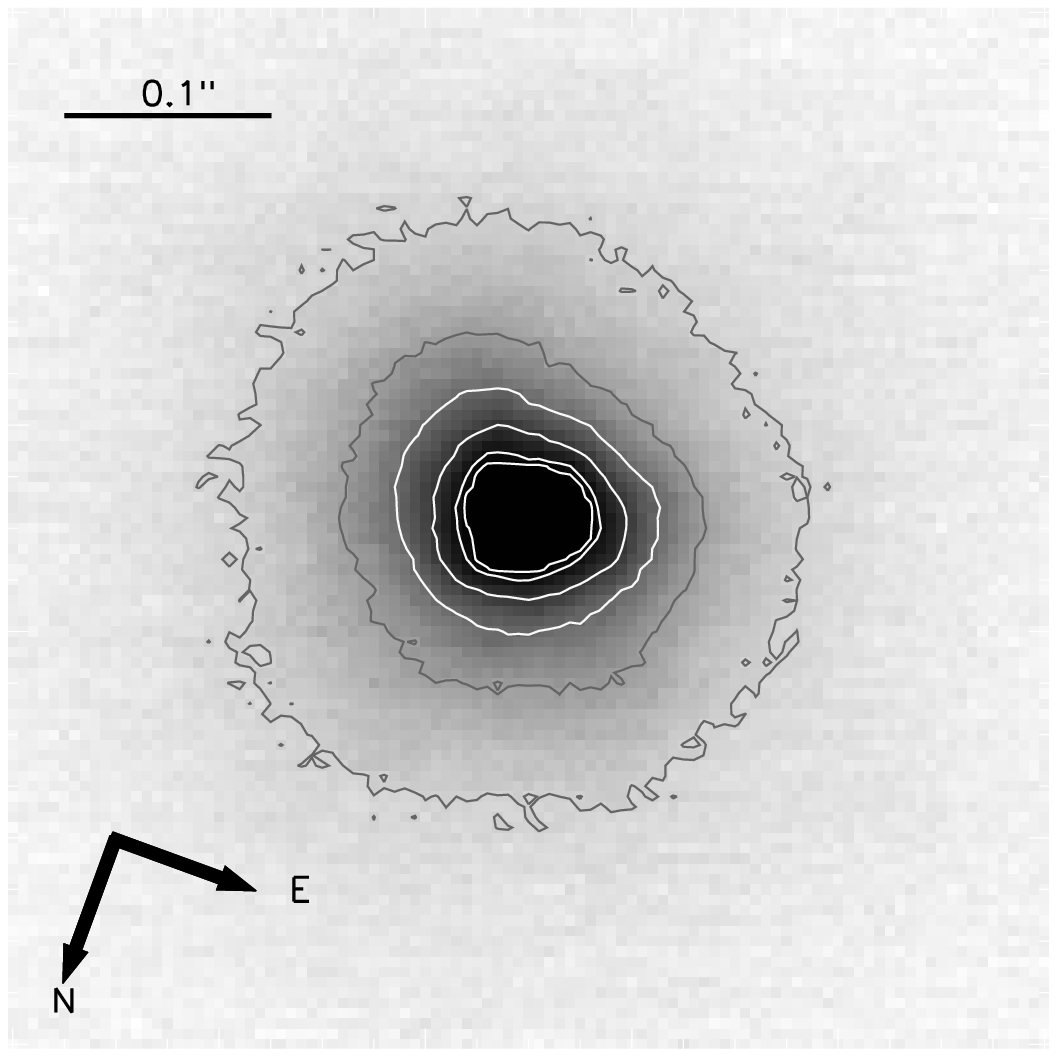}} \quad
\subfloat[][\emph{SDSS J1033+4005}.]
{\includegraphics[trim=0 0 140 0, clip, width=.3\textwidth]{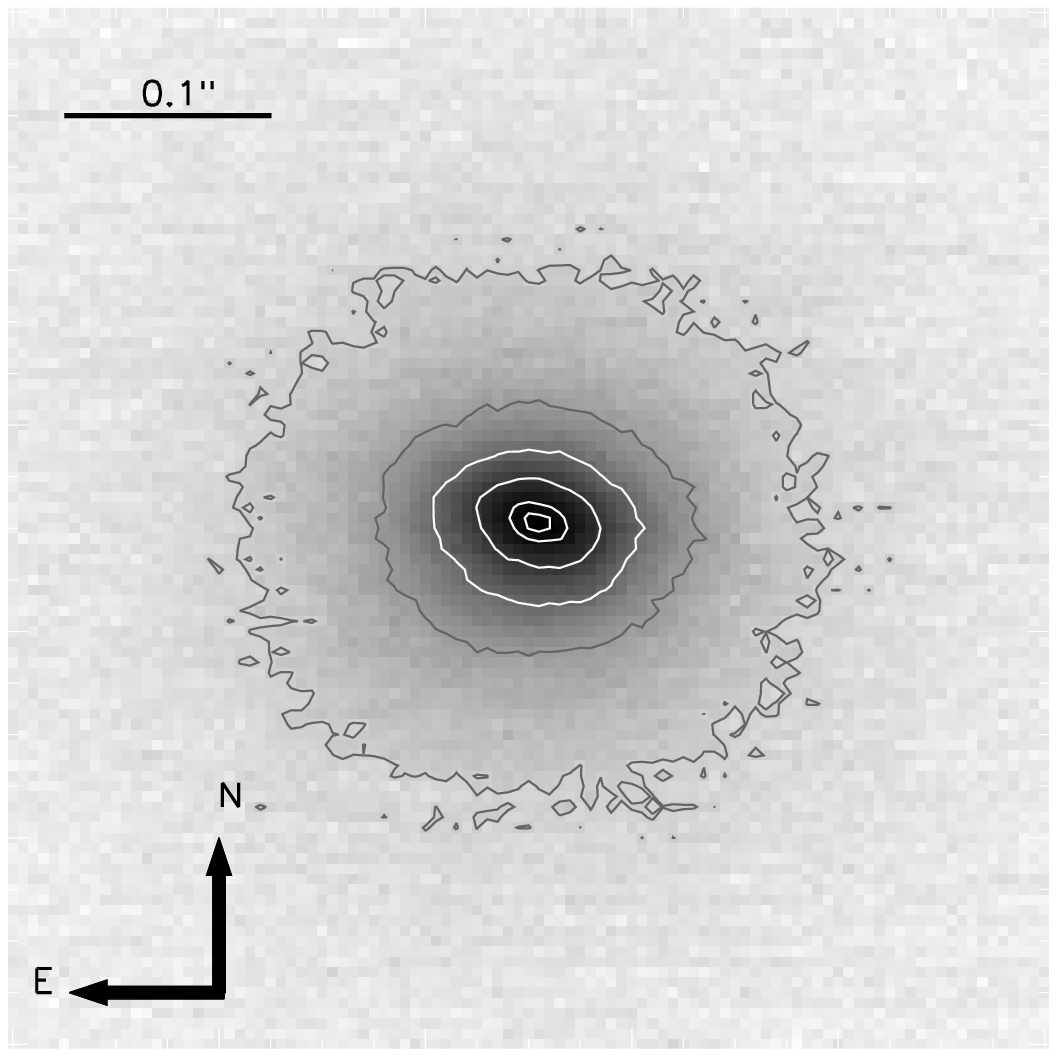}} \quad
\subfloat[][\emph{2MASS J1106+2754}.]
{\includegraphics[trim=0 0 140 0, clip, width=.3\textwidth]{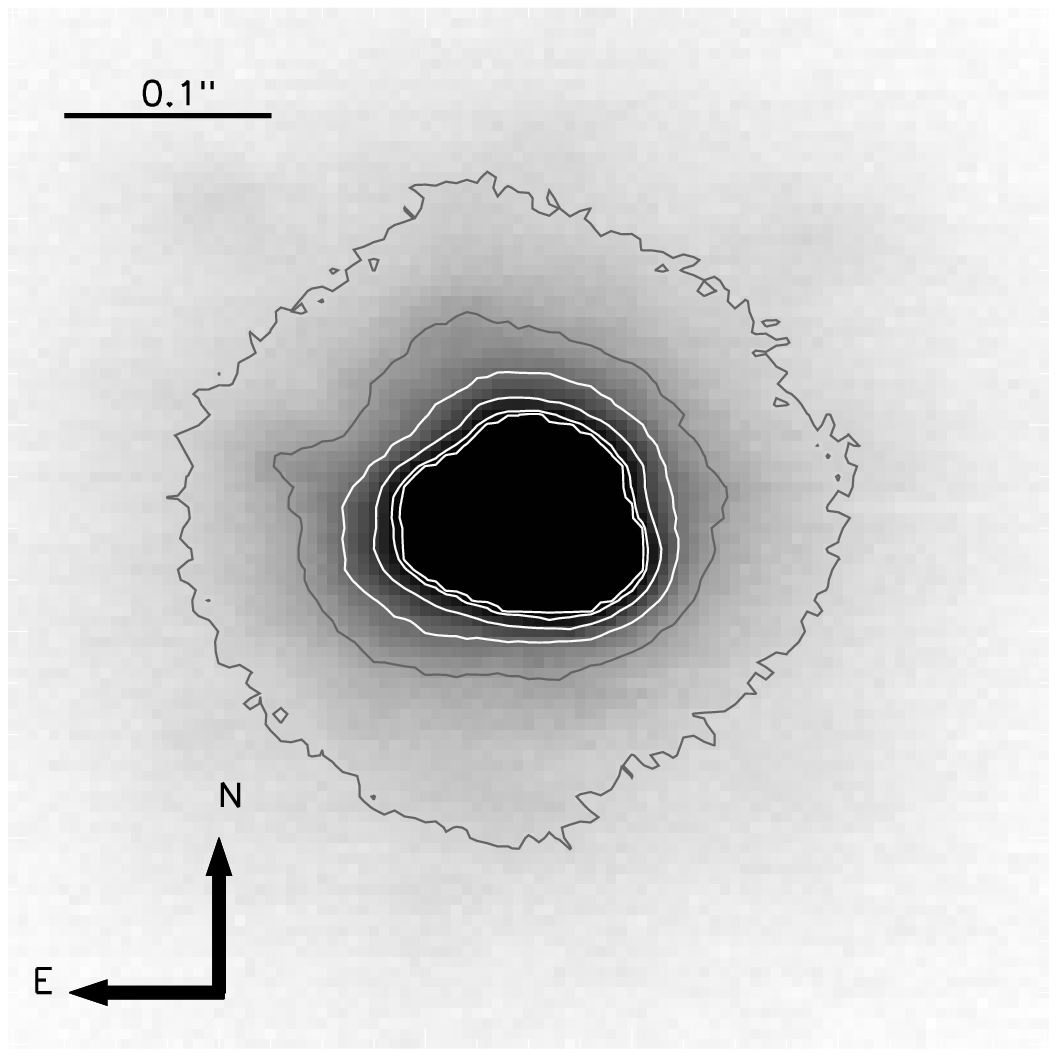}} \quad
\subfloat[][\emph{SDSS J1121+4332}.]
{\includegraphics[trim=0 0 140 0, clip, width=.3\textwidth]{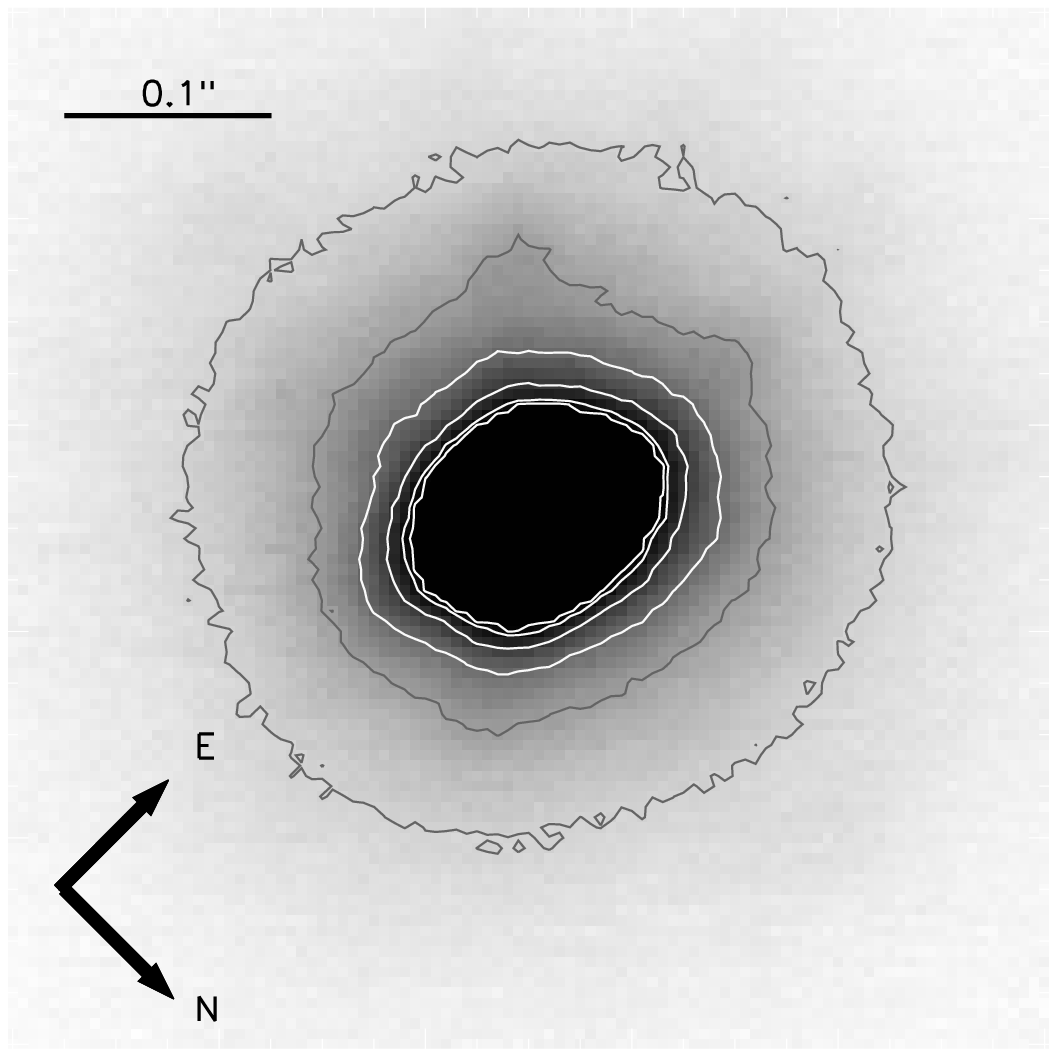}} \quad
\subfloat[][\emph{2MASS J1150+0520}.]
{\includegraphics[trim=0 0 140 0, clip, width=.3\textwidth]{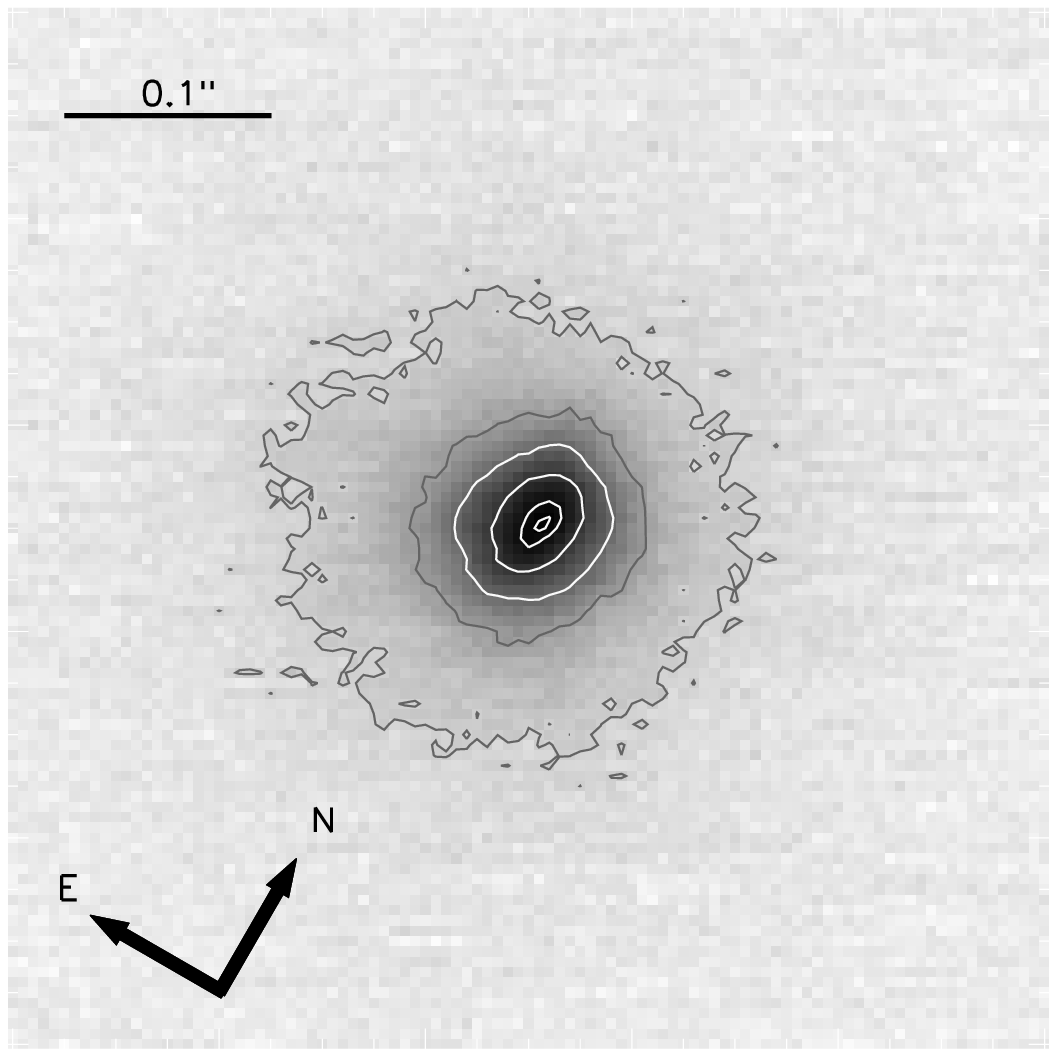}} \quad
\caption{\emph{(Continued.)}\label{fig:unresbin}}
\ContinuedFloat
\end{figure}

\begin{figure}[htp]
\renewcommand{\thesubfigure}{\roman{subfigure}}
\centering
\subfloat[][\emph{2MASS J1158+0435}.]
{\includegraphics[trim=0 0 140 0, clip, width=.3\textwidth]{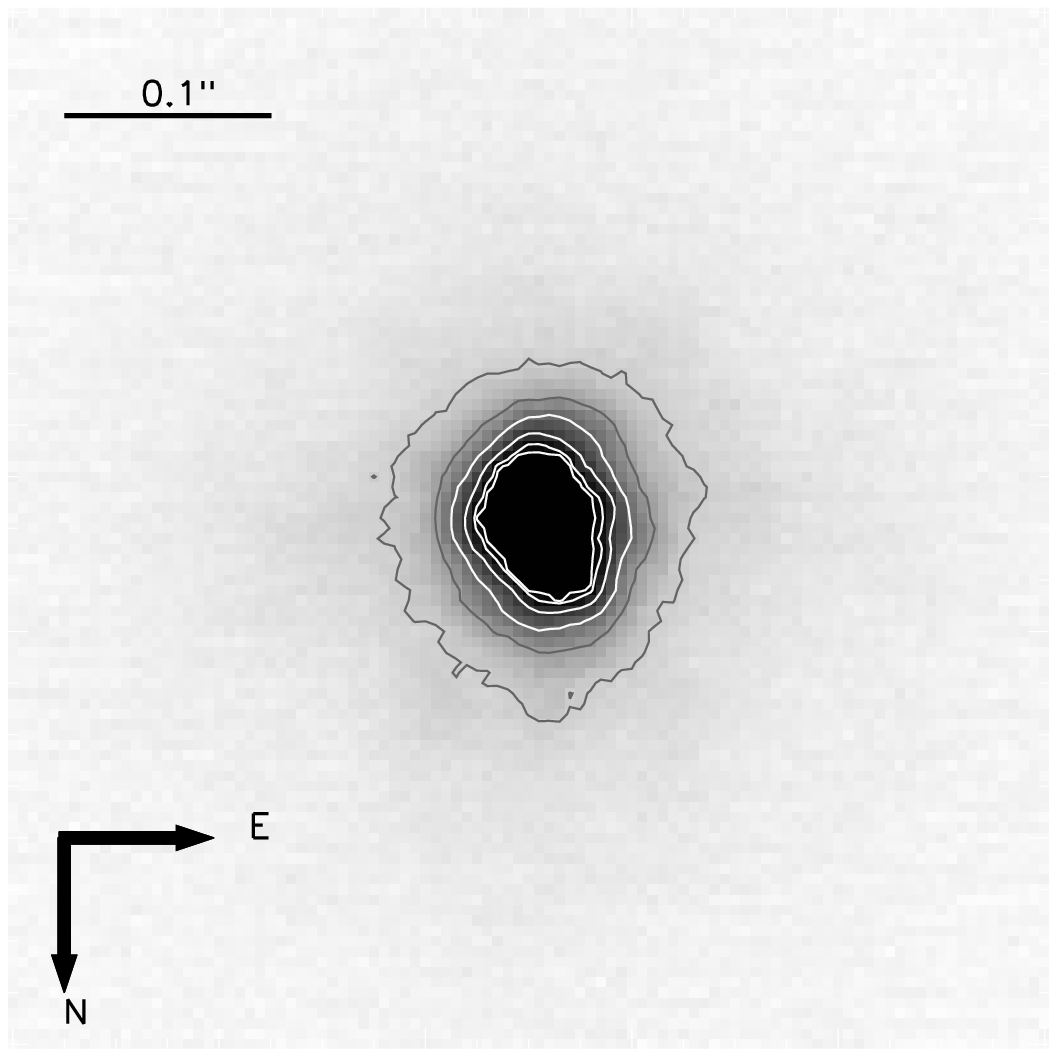}} \quad
\subfloat[][\emph{SDSS J1206+2813}.]
{\includegraphics[trim=0 0 140 0, clip, width=.3\textwidth]{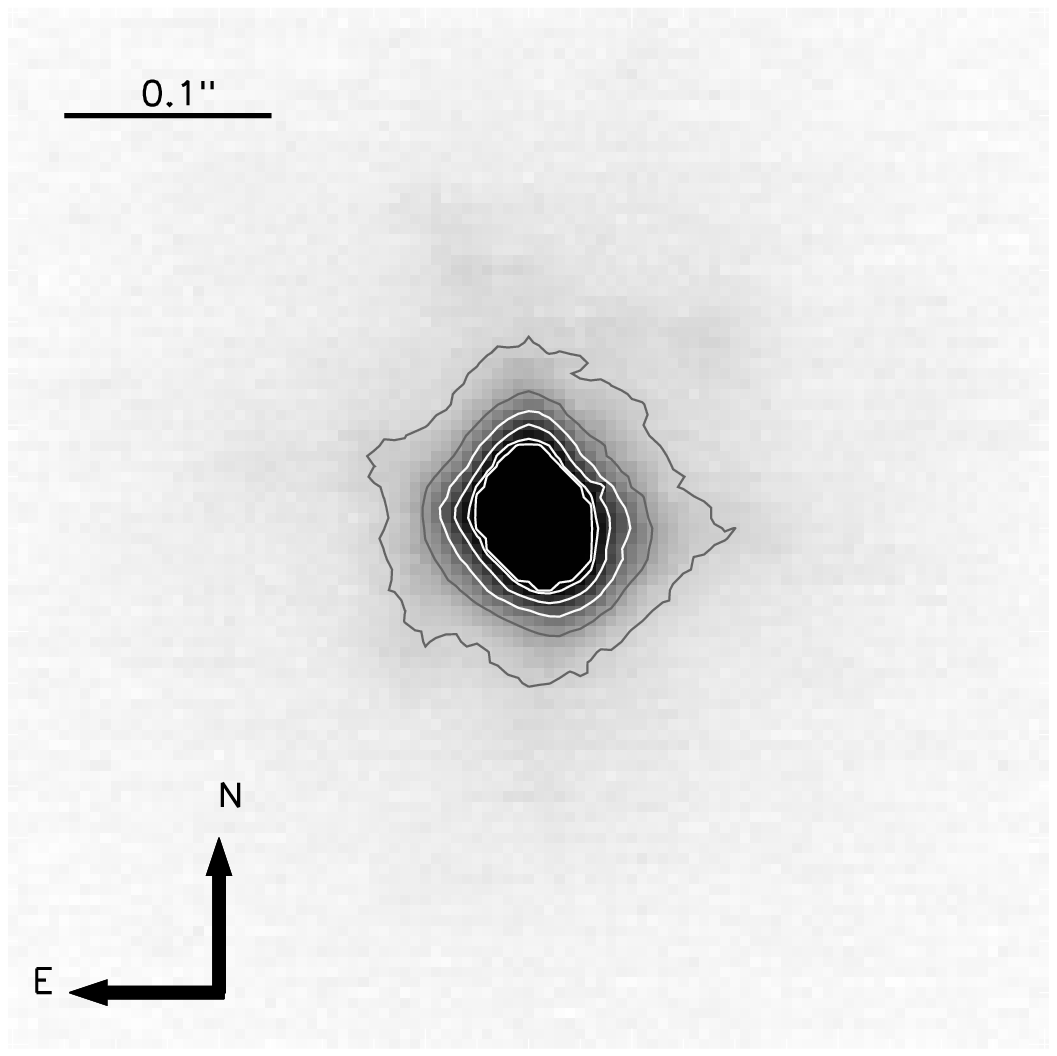}} \quad
\subfloat[][\emph{2MASS J1209-1004}.]
{\includegraphics[trim=0 0 140 0, clip, width=.3\textwidth]{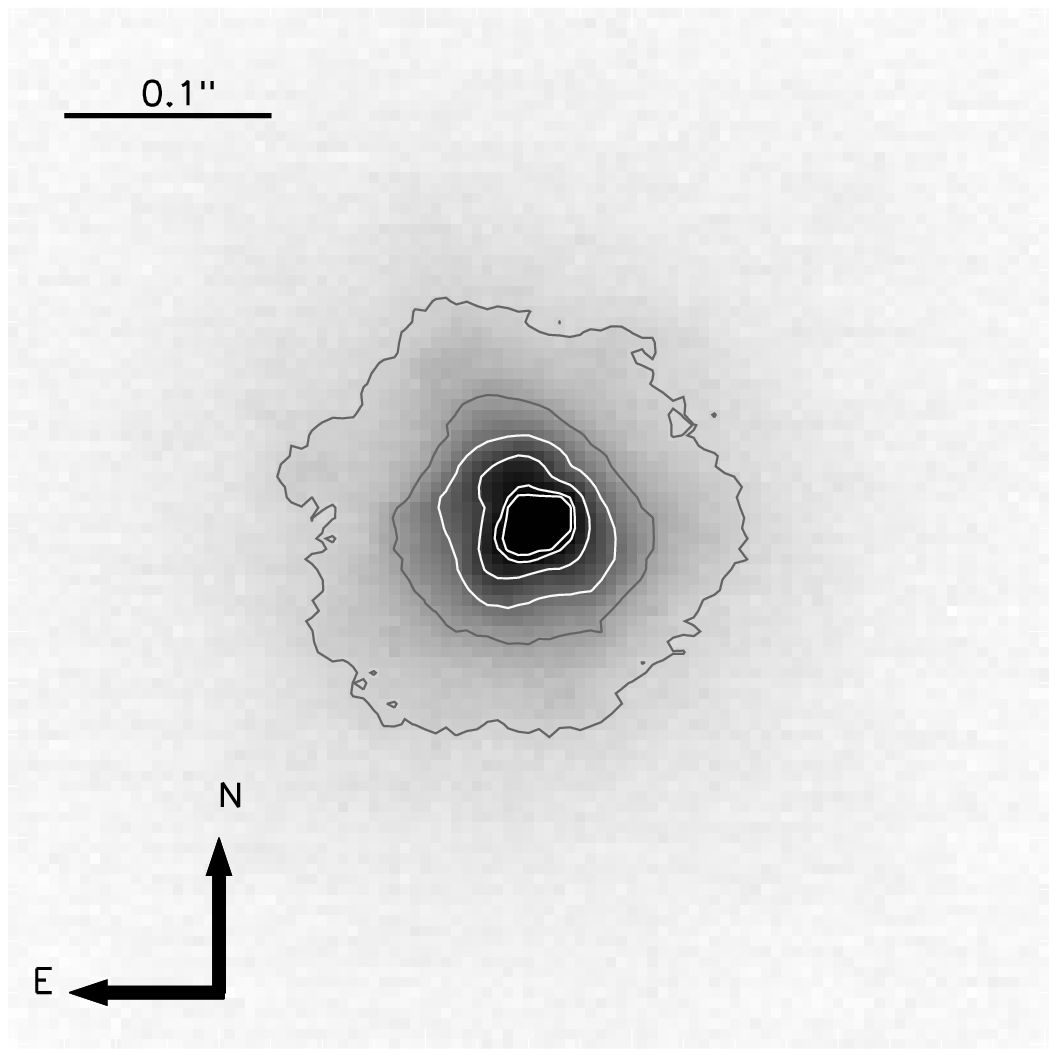}} \quad
\subfloat[][\emph{ULAS J1326+1200}.]
{\includegraphics[trim=0 0 140 0, clip, width=.3\textwidth]{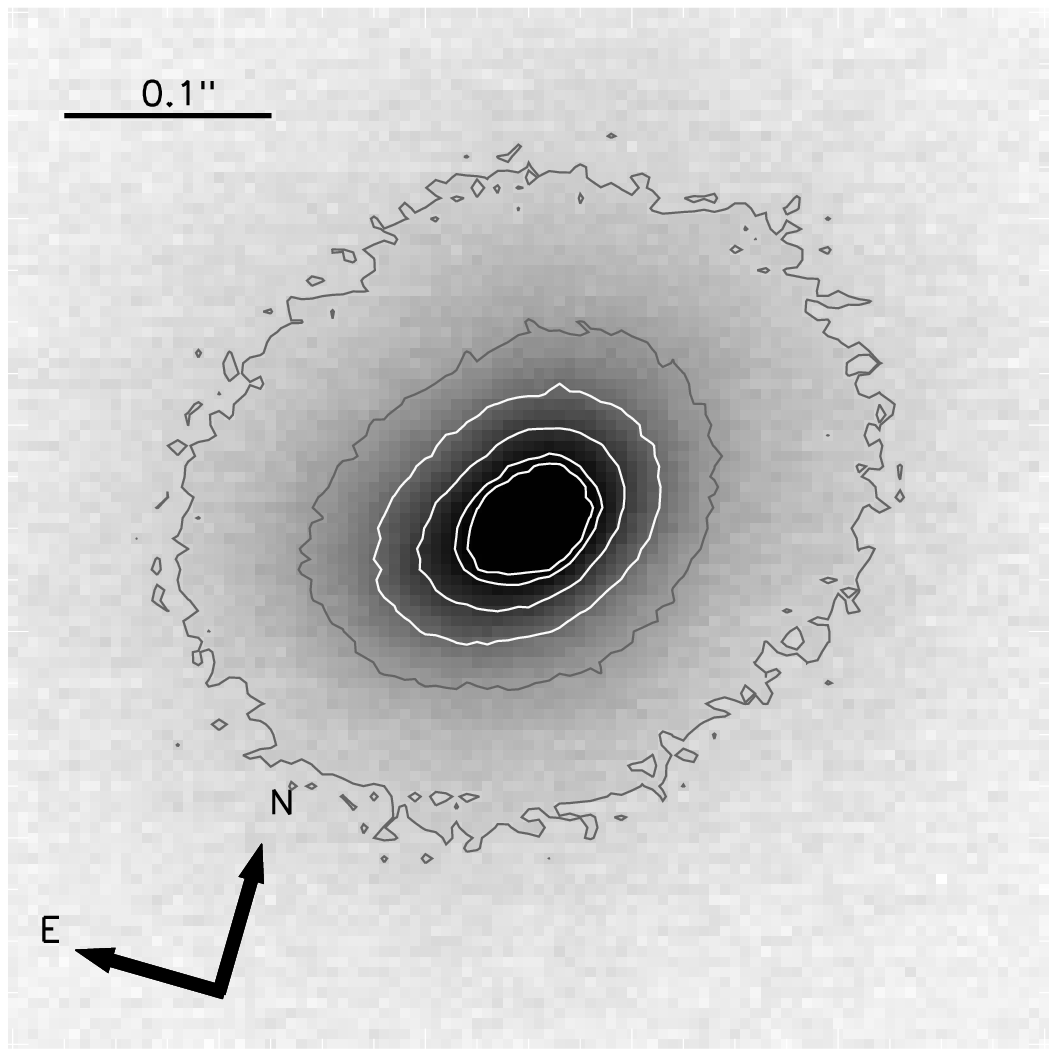}} \quad
\subfloat[][\emph{2MASS J1414+0107}.]
{\includegraphics[trim=0 0 140 0, clip, width=.3\textwidth]{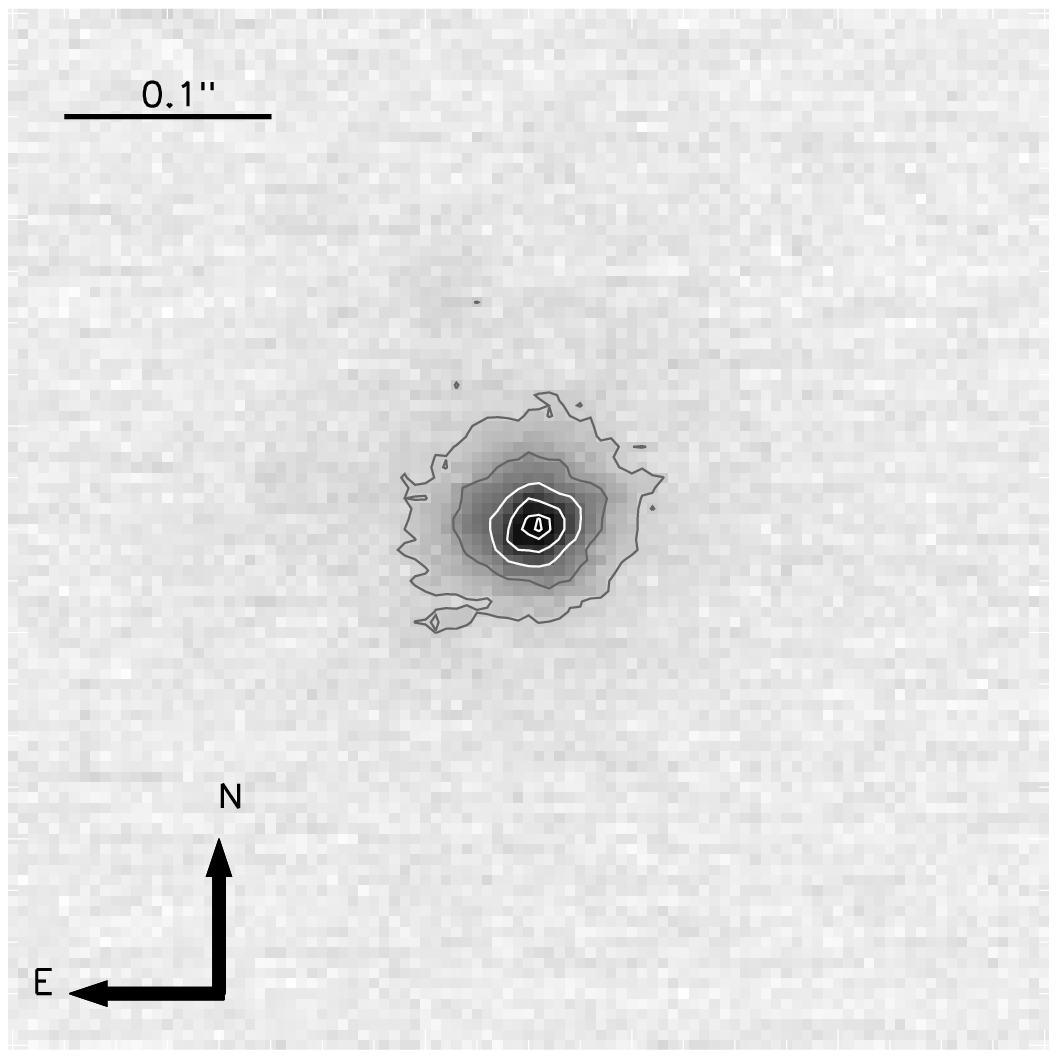}} \quad
\subfloat[][\emph{2MASS J1428+5923}.]
{\includegraphics[trim=0 0 140 0, clip, width=.3\textwidth]{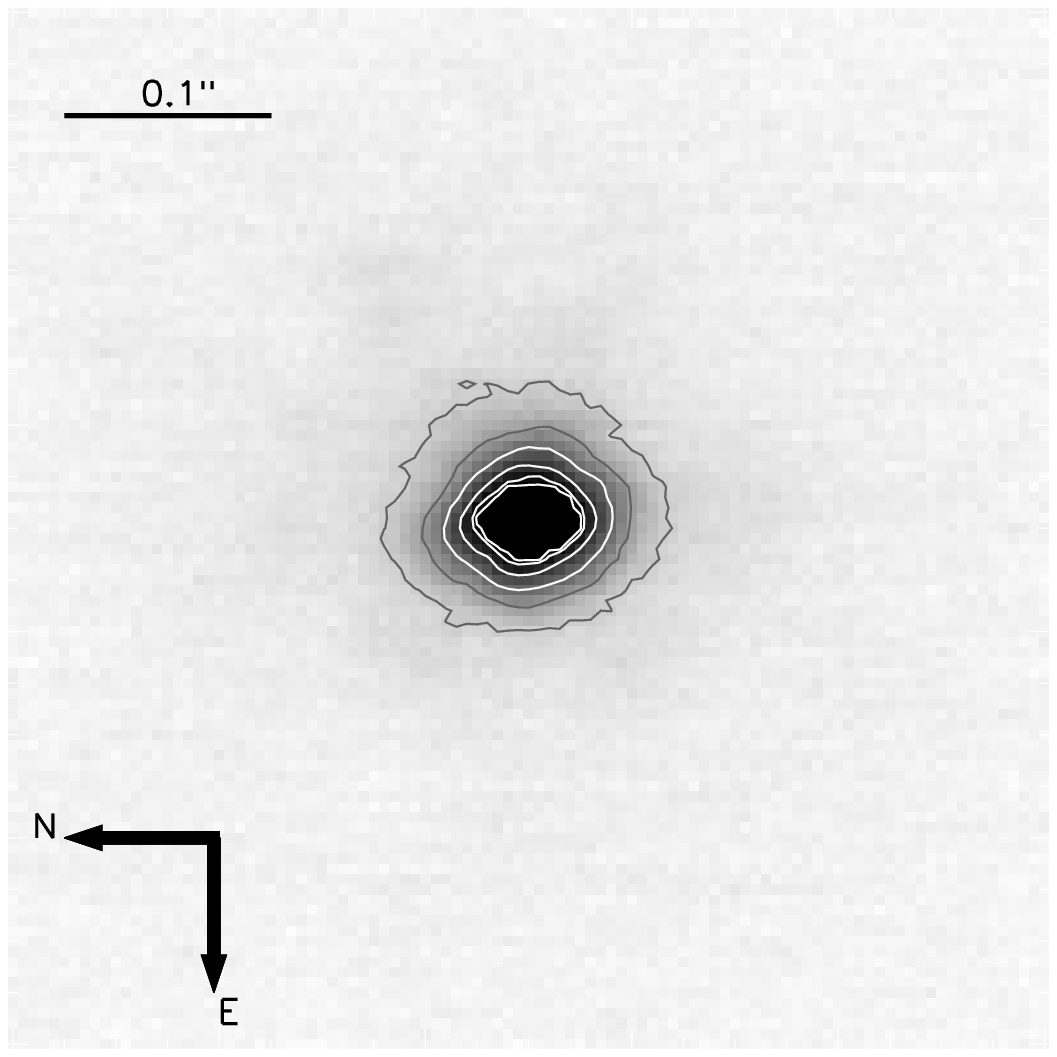}} \quad
\subfloat[][\emph{SDSS J1435+1129}.]
{\includegraphics[trim=0 0 140 0, clip, width=.3\textwidth]{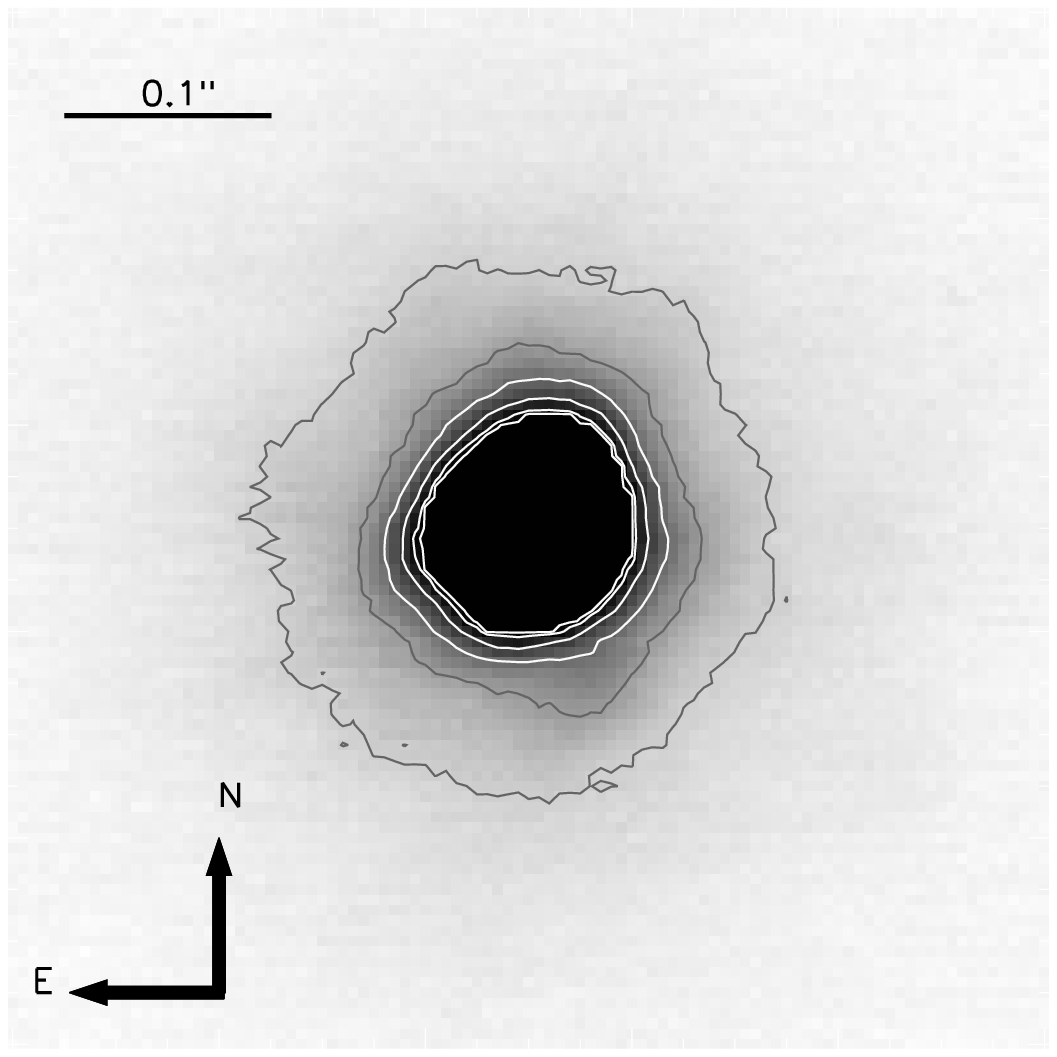}} \quad
\subfloat[][\emph{SDSS J1516+3053}.]
{\includegraphics[trim=0 0 140 0, clip, width=.3\textwidth]{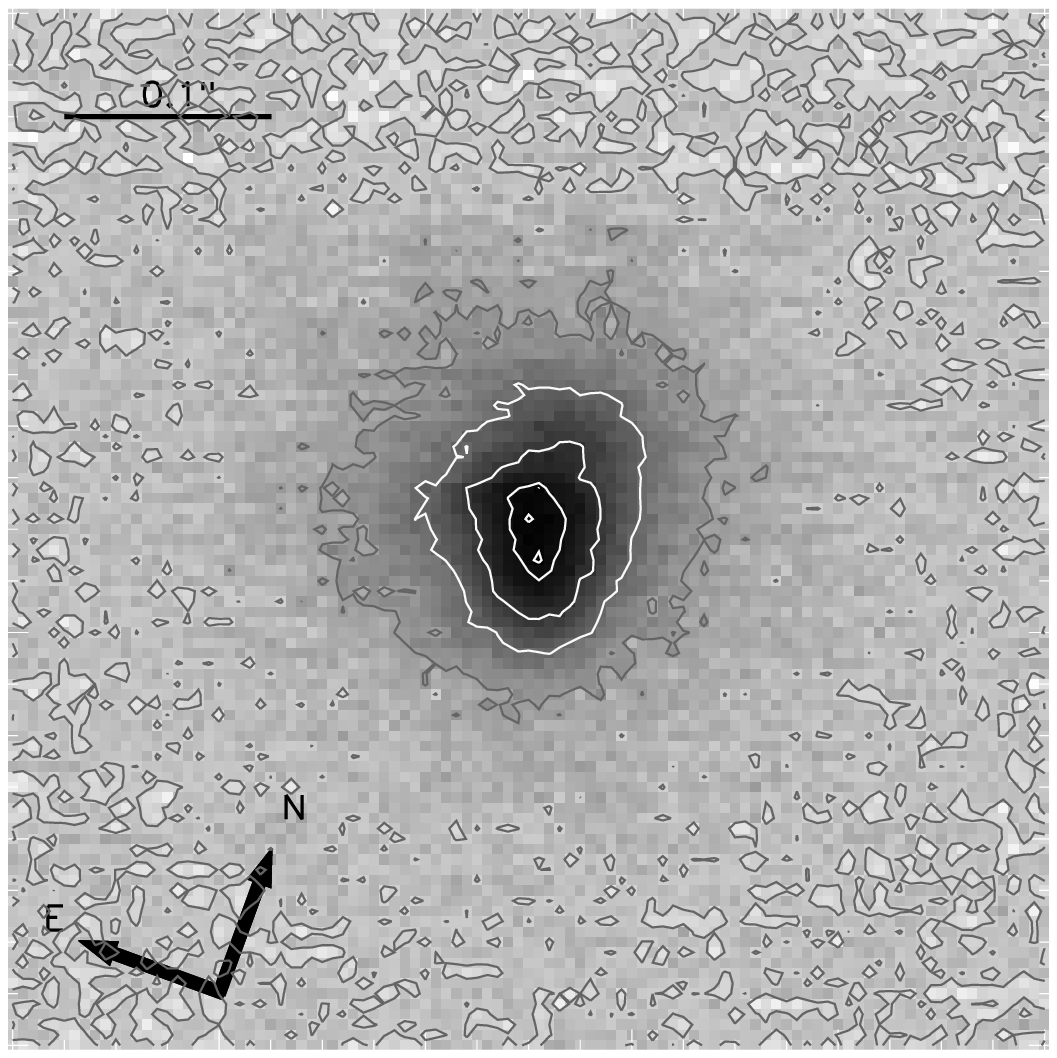}} \quad
\subfloat[][\emph{SDSS J1547+0336}.]
{\includegraphics[trim=0 0 140 0, clip, width=.3\textwidth]{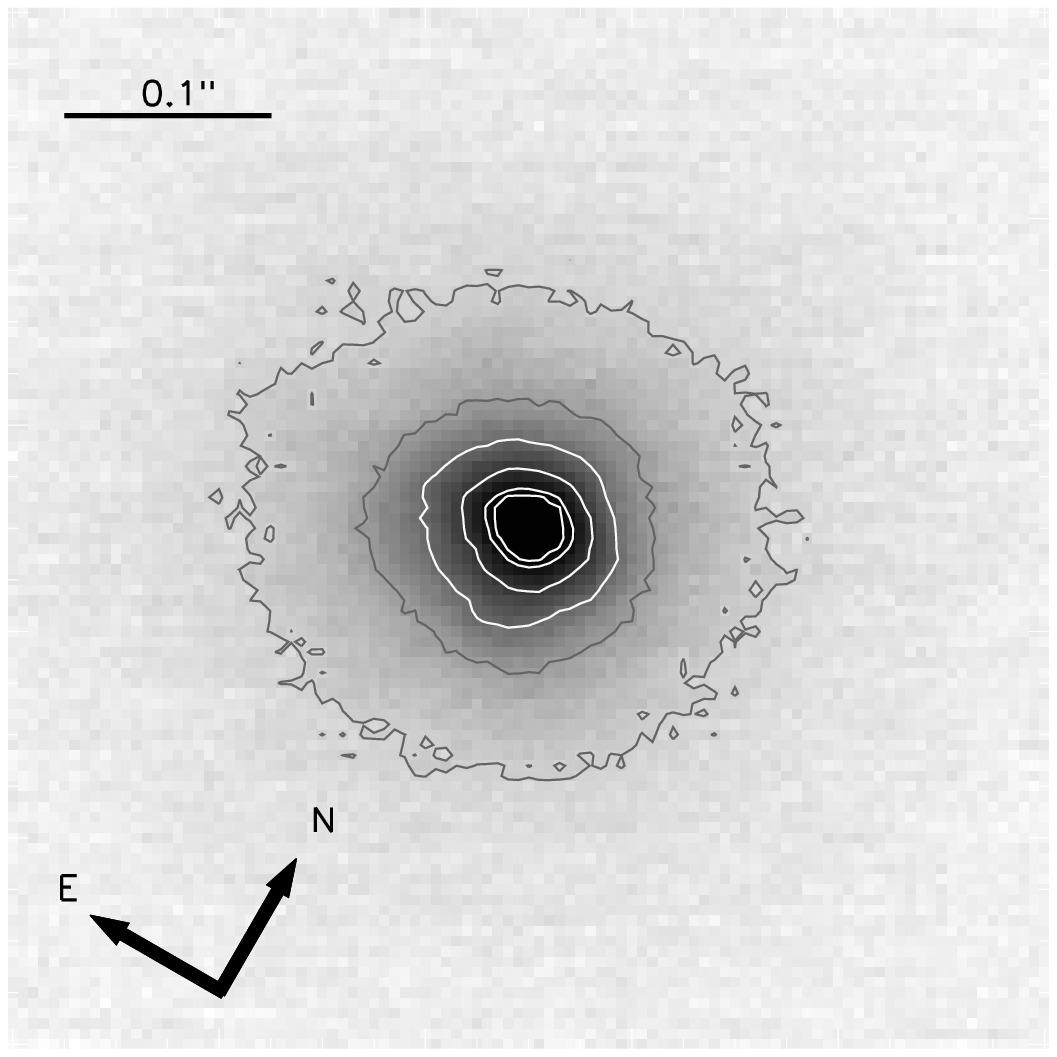}} \quad
\subfloat[][\emph{2MASS J1707+4301}.]
{\includegraphics[trim=0 0 140 0, clip, width=.3\textwidth]{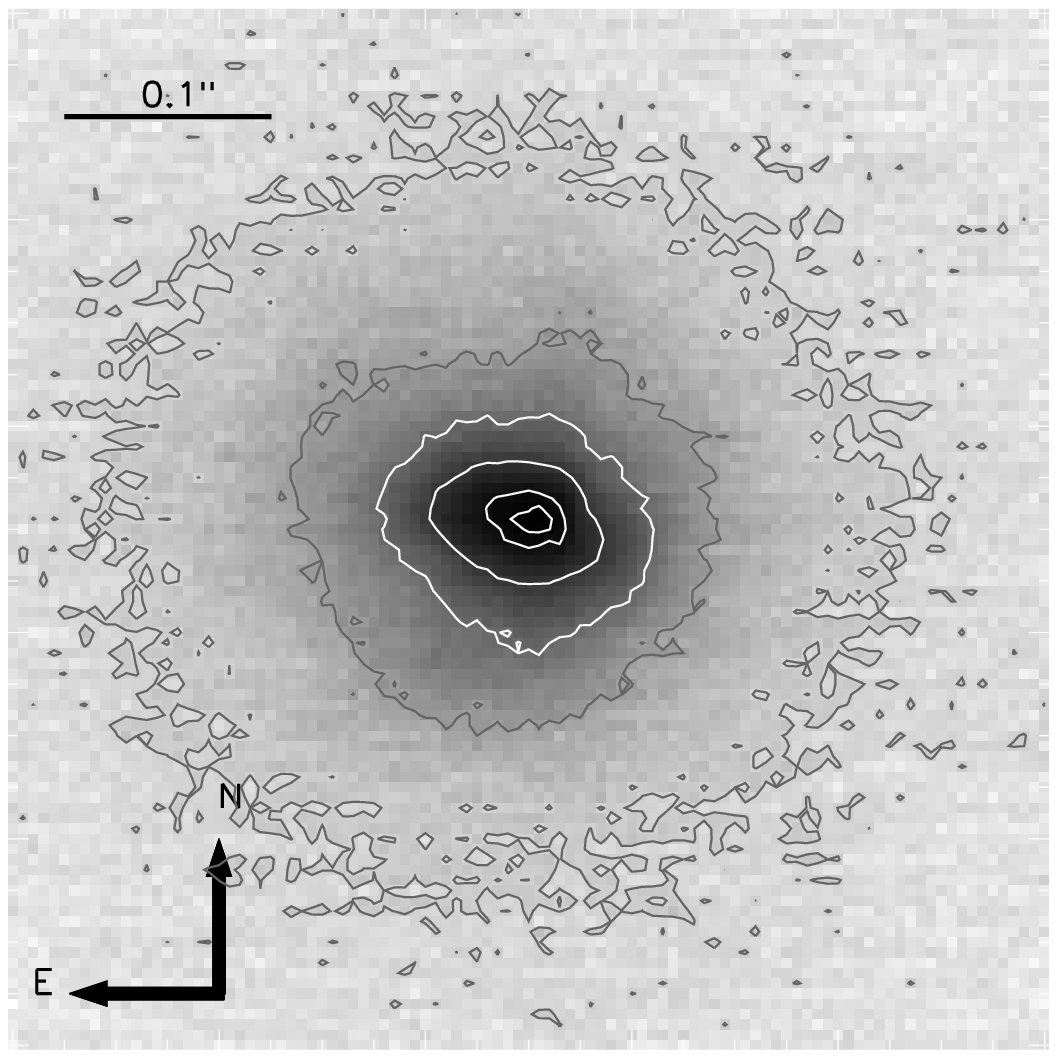}} \quad
\subfloat[][\emph{2MASS J1711+2232}.]
{\includegraphics[trim=0 0 140 0, clip, width=.3\textwidth]{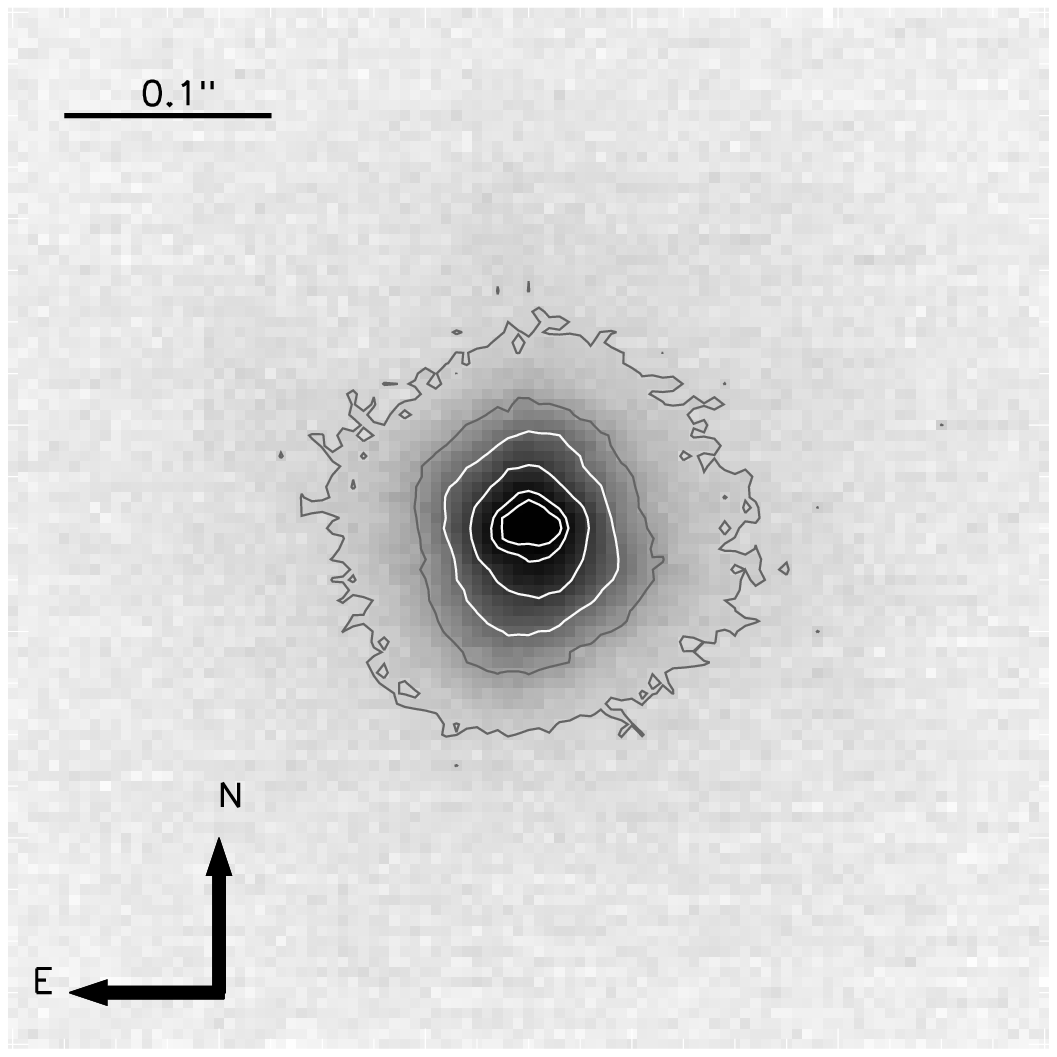}} \quad
\subfloat[][\emph{2MASS J1721+3344}.]
{\includegraphics[trim=0 0 140 0, clip, width=.3\textwidth]{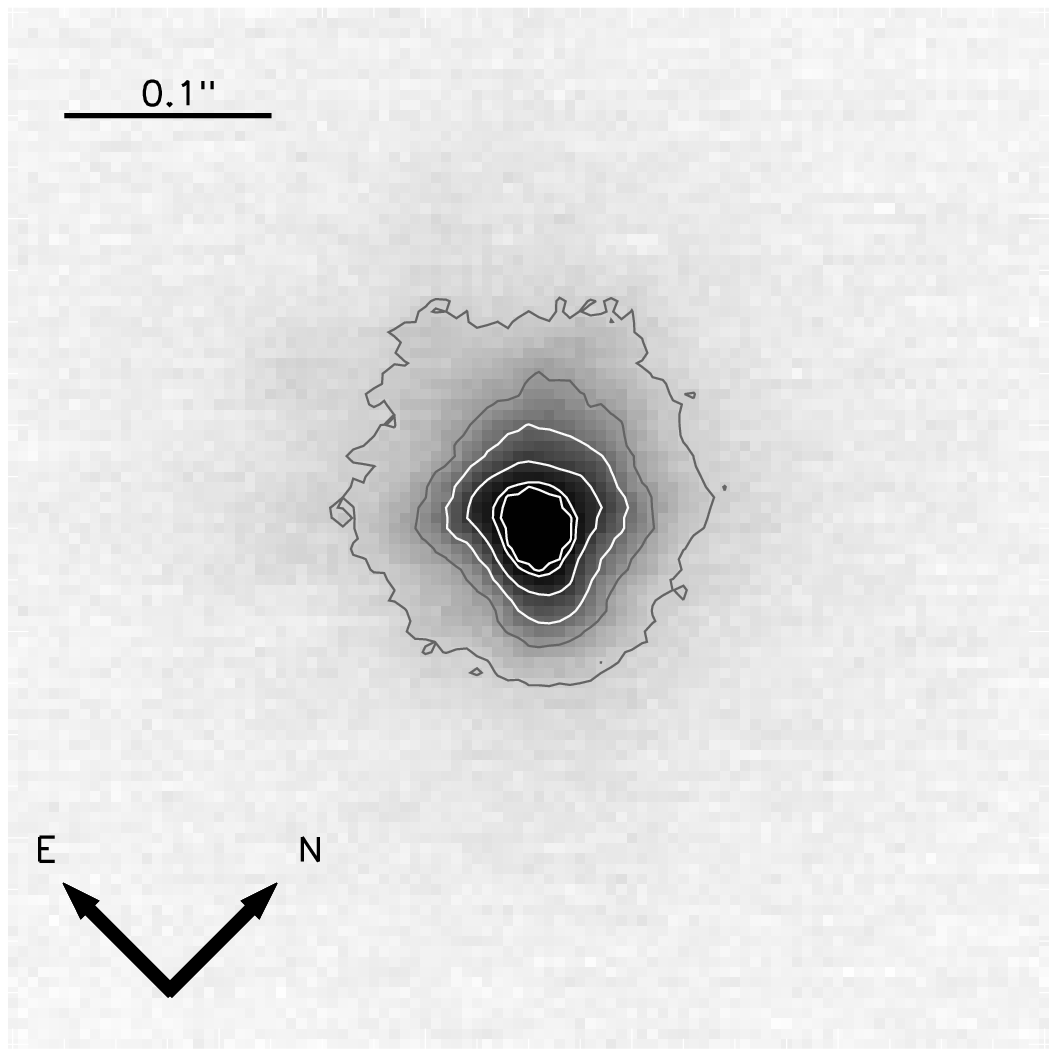}} \quad
\caption{\emph{(Continued.)}\label{fig:unresbin}}
\ContinuedFloat
\end{figure}

\begin{figure}[htp]
\renewcommand{\thesubfigure}{\roman{subfigure}}
\centering
\subfloat[][\emph{2MASS J1733$-$1654}.]
{\includegraphics[trim=0 0 140 0, clip, width=.3\textwidth]{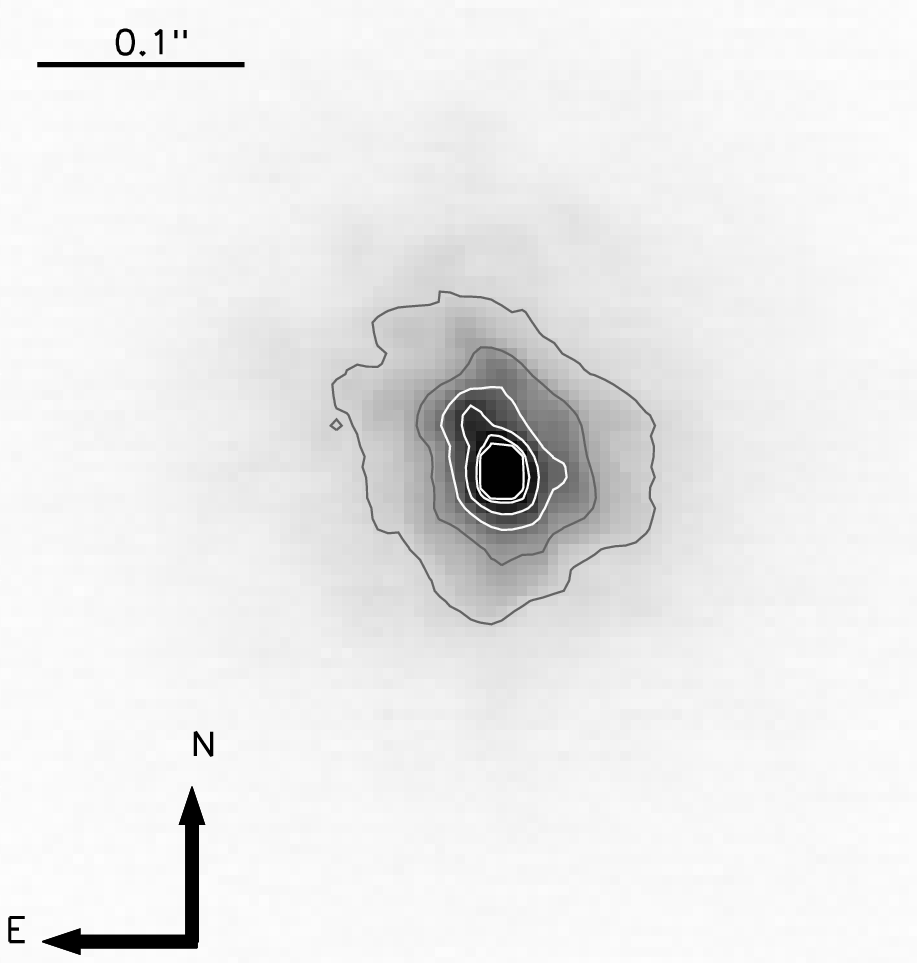}} \quad
\subfloat[][\emph{2MASS J2002-0521}.]
{\includegraphics[trim=0 0 140 0, clip, width=.3\textwidth]{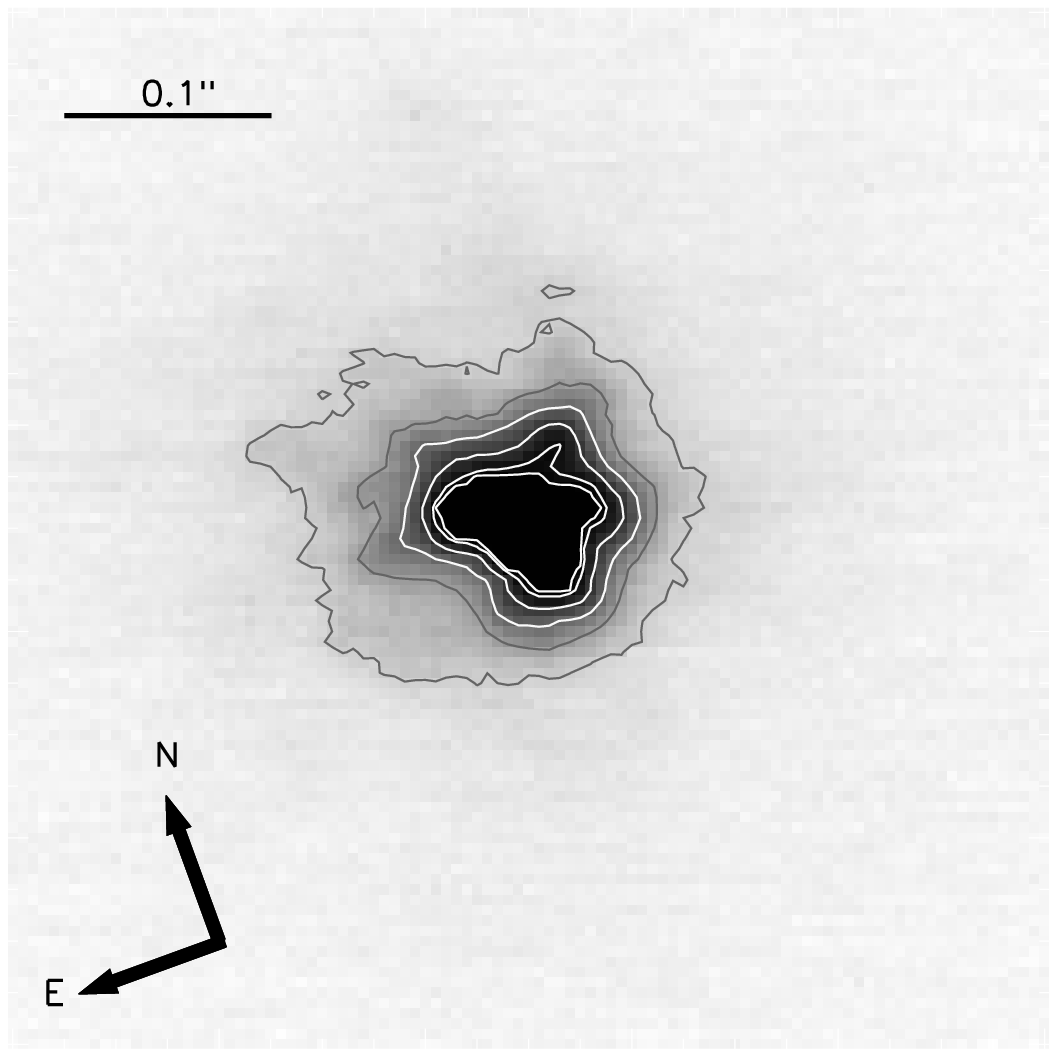}} \quad
\subfloat[][\emph{2MASS J2126+7617}.]
{\includegraphics[trim=0 0 140 0, clip, width=.3\textwidth]{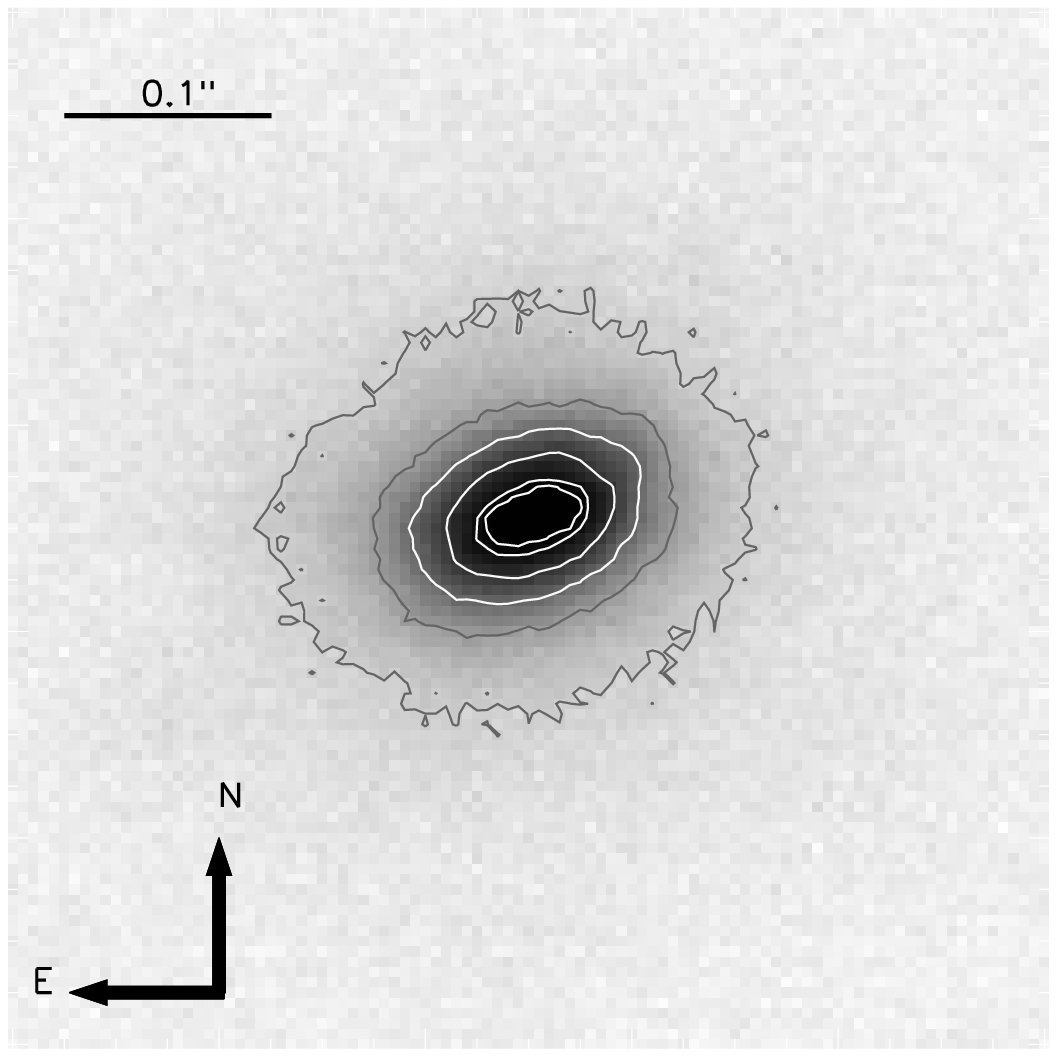}} \quad
\subfloat[][\emph{2MASS J2149+0603}.]
{\includegraphics[trim=0 0 140 0, clip, width=.3\textwidth]{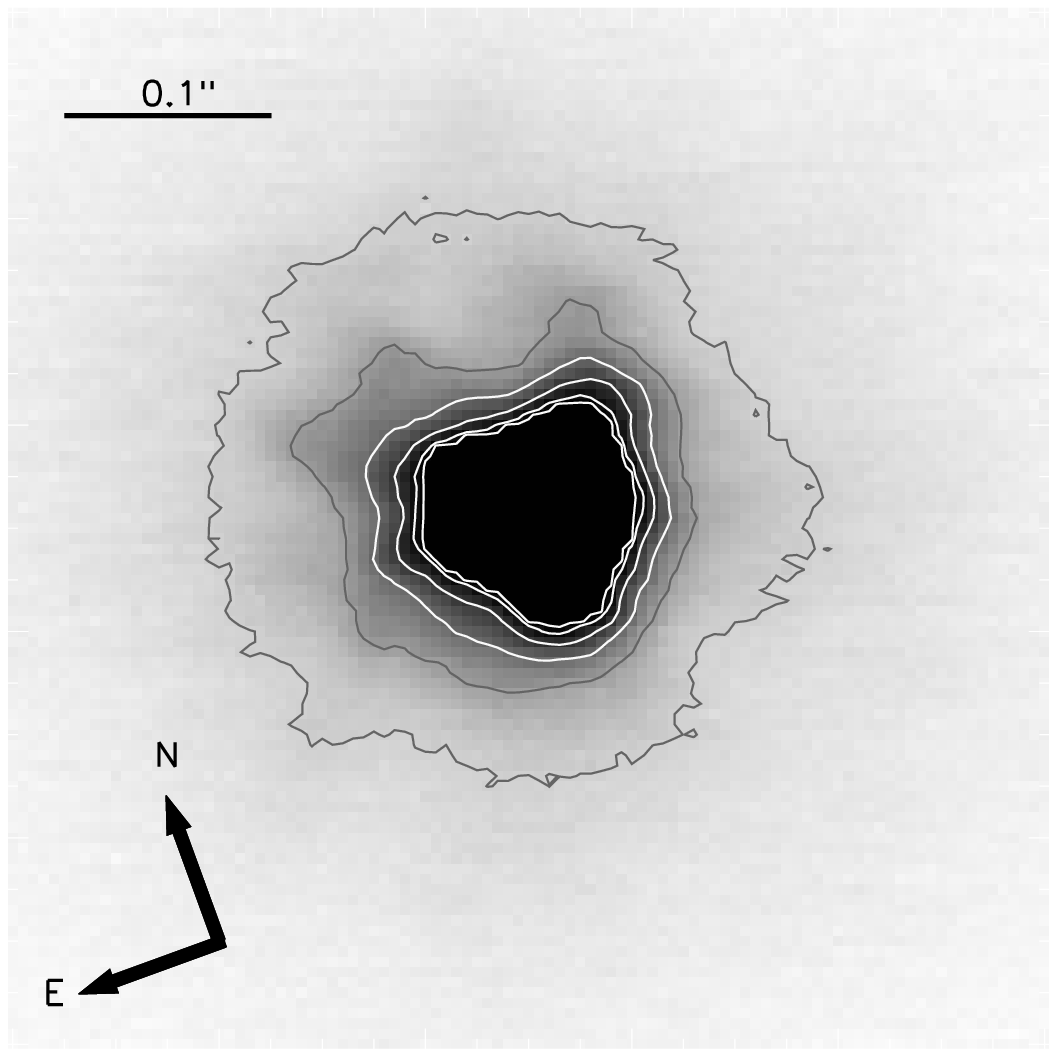}} \quad
\caption{\emph{(Continued.)}\label{fig:unresbin}}
\end{figure}

\clearpage

\end{document}